">arXiv:physics/0603035v1  [physics.bio-ph]  6 Mar 2006# SIMULATIONS IN STATISTICAL PHYSICS AND BIOLOGY: SOME APPLICATIONS

**María del Pilar Monsiváis-Alonso**

# M.Sc. Thesis

**Supervisors:**

Dr. Román López-Sandoval

Dr. Haret-Codratian Rosu

```
Division of Advanced Materials
for Modern Technology
DMATM -IPICyT
```

San Luis Potosí, S.L.P., Mexico
January 20, 2006

**INSTITUTO POTOSINO DE INVESTIGACIÓN CIENTÍFICA Y TECNOLÓGICA, A.C.**

**POSGRADO EN CIENCIAS APLICADAS**

# SIMULATIONS IN STATISTICAL PHYSICS AND BIOLOGY: SOME APPLICATIONS

Tesis que presenta

**María del Pilar Monsiváis-Alonso**

Para obtener el grado de

**Maestro en Ciencias Aplicadas**

En la opción de

**Nanociencias y Nanotecnología**

**Codirectores de la Tesis:**
Dr. Román López-Sandoval
Dr. Haret-Codratian Rosu Barbus

San Luis Potosí, S.L.P., 20 de Enero de 2006

# Acknowledgments

First of all, I would like to thank my advisor Dr. Román López Sandoval for his dedication, guidance and constant support during the development of this thesis. In the same spirit, I would like to thank my advisor Dr. Haret Codratian Rosu Barbus for his suggestions.

I also want to acknowledge the PhD student Vrani Ibarra for his important collaboration referring to chapter 3 of this thesis and I am also grateful to Dr. José Luis Rodríguez, Dra. Yadira Vega and Dr. Raúl Balderas, who read the document and provided helpful corrections.

I would like to thank in a special way to my parents, who always have been a support for me in everything, as well as, to Jorge and all my friends, in particular José Miguel, Víctor Hugo, Andrea, Gerardo, Pedro and Vianney.

My final thanks go to CONACyT for the master fellowship (no. 182493) during the years 2003-2005.

THANKS ALL OF YOU!

<div style="text-align: right">Pily Monsiváis</div>

# Contents

# Abstract

One of the most active areas of physics in the last decades has been that of critical phenomena, and Monte Carlo simulations have played an important role as a guide for the validation and prediction of system properties close to the critical points. The kind of phase transitions occurring for the Betts lattice (lattice constructed removing $1/7$ of the sites from the triangular lattice) have been studied before with the Potts model for the values $q = 3$, ferromagnetic and antiferromagnetic regime. Here, we add up to this research line the ferromagnetic case for $q = 4$ and 5. In the first case, the critical exponents are estimated for the second order transition, whereas for the latter case the histogram method is applied for the occurring first order transition. Additionally, Domany's Monte Carlo based clustering technique mainly used to group genes similar in their expression levels is reviewed. Finally, a control theory tool –an adaptive observer– is applied to estimate the exponent parameter involved in the well-known Gompertz curve. By treating all these subjects our aim is to stress the importance of cooperation between distinct disciplines in addressing the complex problems arising in biology.

# Introduction

> "Minerals grow, plants grow and live,
> animals grow, live and have feeling."
> Linnaeus, "Systema Naturae", 1735

Monte Carlo simulations have been used for many years to study the properties of physical models, and have also played a significant role in statistics, biology, computer science and other fields, demonstrating its versality and powerful approach. Furthermore, many advances in computation algorithms and computer technology have made possible to study systems which would be impossible to examine only a few years ago. The first part of this thesis aims to give a brief explanation of the Monte Carlo method, a review of the principal algorithms used, the study of phase transitions, finite size scaling theory and finally, some results obtained with the Potts model for a recently proposed lattice named Betts or Maple Leaf lattice.

Since the discovery of the helical structure of DNA and various complete genome sequences, biology has seen also an enormous advance. However, it seems that the only way to solve the complex problems raised in the study of biological systems is to share the challenge with other scientific disciplines such as chemistry, physics, and computer science. Research on cancer is one of the most important and interesting subjects in Biology. This terrible disease has received tremendous attention in the last part of the XX century, because of the huge amount of cases and the technological advances in analysis and medical treatment of tumours. Despite the efforts of the international scientific community, there are many unanswered questions related to the evolution of the cancer diseases, the causes that trigger them, the prediction of drugs and treatments effects, and the development of an effective cure. The introduction of the Monte Carlo method into biological problems has brought interesting results including the modeling of the structure and evolution of a epidermis cell nuclei, reproducing cancer growth.

The second chapter reviews the clustering techniques commonly used to group genes with similar behaviour in their expressions across various experiments, which helps in the construction of genetic networks and targeting of genes involved in diseases like cancer. The superparamagnetic gene clustering algorithm is also explained as an example of a clustering technique that employs the Monte Carlo method and is based on a physical phenomenom, leaving the subject to future implementation.

On the other hand, mathematical procedures, in particular models based on differential equations whose terms can represent not only the growth rate of a tumour, but also the growth or inhibition rates of substances existing in the medium or cell-cell interactions, provide an excellent tool to describe biological processes. There also exist empirical models that have proved to be very useful in fitting the experimental growth curves of tumours. The Gompertz model is a famous one, although there is not a convincing explanation of why it works so well. The Gompertz growth law has been introduced by Benjamin Gompertz in 1825 in his demographical studies, and in mathematical terms is written:

$$\lambda(a) = h_0 e^{\gamma a}, \tag{1}$$

where $\lambda(a)$ is the mortality rate.

The main problem is that the biological interpretation of its characteristic parameters is not very well settled. A link of these parameters with the biological phenomenology, if found, would make the Gompertz model extremely valuable as a predictive tool. The third part of this thesis discusses some of the most important models based on differential equations and gives a more complete idea about the formulation and applications of the Gompertz model, and finally presents a method based on control theory capable of accurately predict the first stages of Gompertz growth.

The main purpose of this work is to emphasize the importance of an interdisciplinary research. Nowadays, it is clear that many problems inherent to the biology field need to be adressed with tools coming from areas such as computational physics and applied mathematics.

# Chapter 1

# Monte Carlo Simulations in Statistical Physics

## Contents

## 1.1  Brief History of the Monte Carlo Method

The first electronic computer, ENIAC, was developed during the World War II period by a group of scientists working at the University of Pennsylvania in Philadelphia. They had realized that if electronic circuits could be made to count, then they could do arithmetic and hence, solve difference equations at incredible speeds. This would lead to a scientific revolution because it would give the possibility to study problems unsolved before due to the large amount of calculations needed.

In 1946, Stanislaw Ulam, a mathematician working in Los Alamos, attended a conference about a preliminary computational model of a thermonuclear reaction probed in ENIAC as a test for the computer. Like other scientists, he was impressed by the speed and versatility of the ENIAC. Additionally, Ulam's extensive mathematical background made him aware that statistical sampling techniques that had fallen into disuse because of tediousness of calculations, could be resuscitated with ENIAC. The basis of the Monte Carlo method has been proposed later by him as a consequence of his interest in random processes. As Stan Ulam mentioned in 1983, his first thoughts and attempts to practice the Monte Carlo method were suggested by a question that occurred to him in 1946 as he was playing solitaires. The question was what were the chances that a Canfield solitaire laid out with 52 cards will come out succcessfully? [1]. After spending a lot of time trying to estimate them by pure combinatorial calculations, he wondered whether a more practical method might not be to lay it out say one hundred times and simply observe and count the number of successful plays. He immediately thought about how to change processes described by certain differential equations into an equivalent form interpretable as a succession of random operations [1]. Ulam discussed his ideas with John von Neumann, Professor of Mathematics at the Institute for Advanced Study at Princeton, who was also a consultant to Los Alamos and one of the principals participating in the ENIAC probe conference in 1946. Von Neumann saw the importance of Ulam's approach and thought that it seemed especially suitable for exploring the behaviour of neutron chain reactions in fission devices. In March 1947, von Neumann wrote to Robert Richtmyer, the Leader of the Theoretical Division at Los Alamos, describing a possible statistical method to solve the problem of neutron diffusion in fissionable material using the newly developed electronic computing techniques. It was at that time when Nicholas Metropolis suggested the name Monte Carlo for this statistical method. It was related to the fact that Stan had an uncle who would borrow money from relatives because he "just had to go to Monte Carlo" [2] and also because of the similarities between the method and the games of chance abundant in the capital of Monaco, the european center of gambling.

Very similar methods, not fully developed, had been used earlier. An example is Buffon's needle problem, an experiment performed in the of the eighteenth century, which represents one of the first problems in geometric probability. It consists in throwing a needle randomly on a board with parallel lines, and inferring the value of $\pi$ from the number of times the needle intersects a line [3]; nowadays, Buffon's needle problem is practically solved by Monte Carlo integration. Descriptions of several modern Monte Carlo techniques appear in a paper by Kelvin [4], written nearly one hundred years ago, in the context of a discussion on the Boltzmann equation. In the 1940's, Enrico Fermi also used Monte Carlo in the calculation of neutron diffusion, and later designed the Fermiac, a Monte Carlo mechanical device used in the calculation of criticality in nuclear reactors [5]. Ulam's contribution was to recognize the potential for the newly invented electronic computer to automate such sampling.

The approach proposed by von Neumann in his letter was the first formulation of a Monte Carlo computation for an electronic machine. Von Neumann considered a spherical core of fissionable material surrounded by a shell of normal material, and the idea was to trace out the development of neutrons using random digits to select the outcomes of the various interactions along the way, such as scattering, absorption and fission. For example, once a neutron is selected to have an initial position with certain velocity, you have to decide the position of the first collision and the nature of the collision. If you select a fission to occur, then the number of emerging neutrons must be chosen, and each of the new neutrons is followed too. On the other hand, if you decide that the outcome of the collision is scattering, the new momentum of the neutron must be determined.

---

[1] Today is quite well known that the chance of winning is low: 3.3% (www.games.solitaire.com)

If the neutron crosses a material boundary, the characteristics of the new medium must be taken into account. At the end, a genealogical history of a neutron emerges. The same procedure is carried out for other neutrons until a statistically valid picture is obtained. Each neutron history is analogous to a single game of solitaire, and the use of random numbers to make the choices along the way is analogous to the random turn of the card.

To take decisions, the computer must have an algorithm for generating a uniformly distributed set of random numbers and these numbers must be transformed into the nonuniform distribution, say $g$, desired for the property of interest. In a 1947 letter, von Neumann discussed two techniques for using uniform distributions of random numbers to generate $g$. The first technique, which had already been proposed by Ulam, shown that the function $f$ needed to achieve this transformation is just the inverse of the nonuniform distribution function, that is, $f = g^{-1}$. For example, in the case of neutron physics, the distribution of free paths (how far neutrons of a given energy in a given material go before colliding with a nucleus) decreases exponentially with distance. If $x$ is uniformly distributed in the open interval $(0,1)$, then $f = -\ln x$ will give us a nonuniform distribution $g$ with just those properties. The rest of von Neumann letter describes an alternative technique that works when it is difficult or computationally expensive to form the inverse function, which is frequently true when the desired function is empirical. In this approach, two uniform and independent distributions $(x_i)$ and $(y_i)$ are used. If two numbers $x_i$ and $y_i$ are selected randomly from the domain and range, respectively, of the function $f$, then each such pair of numbers represents a point in the function's coordinate plane $(x_i, y_i)$. When $y_i > f(x_i)$ the point lies above the curve for $f(x)$, and $x_i$ is rejected; when $y_i \leq f(x_i)$ the point lies on or below the curve, and $x_i$ is accepted (see Fig. 1.1). Thus the fraction of accepted points is equal to the fraction of the area below the curve. In fact, the proportion of points selected that fall in a small interval along the x-axis will be proportional to the average height of the curve in that interval, ensuring generation of random numbers that mirror the desired distribution [1].

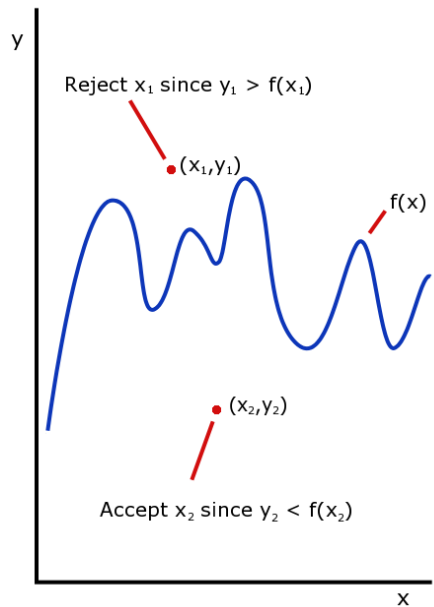

Figure 1.1: Generation of random numbers that mirror a given distribution $f(x)$ [1].

The first ambitious test of the Monte Carlo method consisted of nine problems in neutron transport, each one corresponding to various configurations of materials, initial distributions of neutrons, and running times. These problems did not include hydrodynamic and radiative effects, but complex geometries and realistic neutron-velocity spectra were handled easily. Neutron histories were checked with a variety of statistical analyzes and comparisons with other approaches. Conclusions about the efficiency of the method were quite favourable and gave rise to enthusiasm among scientists of distinct areas. At Los Alamos, the method was quickly adopted to study problems of thermonuclear and fission devices. Already in 1948, Ulam was able to report to the Atomic Energy Commission about the applicability of the method for cosmic rays and in the area of the Hamilton Jacobi partial differential equation. Other laboratory staff members started to run Monte Carlo codes in ENIAC. Among them, J. Calkin, C. Evans and F. Evans studied thermonuclear problems, and B. Suydam and R. Stark tested the concept of artificial viscosity for time-dependent shocks. By midyear 1949, Ulam and Metropolis published a paper describing the Monte Carlo method and its application to integro-differential equations [6] and the first symposium on the method was held in Los Angeles.

The construction of a new machine began later and N. Metropolis was the leader of the group in charge of it. He called the new machine MANIAC wishing to stop the use of acronyms for machine names, but contrary to what he sought, it only stimulated it. In early 1952, the MANIAC became operational at Los Alamos and soon after, Anthony Turkevich led a study of the nuclear cascades resulting from the collision of accelerated particles with atomic nuclei. Another computational problem run on the MANIAC was a study of equations of state based on the two-dimensional motion of hard spheres. The results were published in a famous paper in 1953 [7] and describes a strategy leading to greater computational efficiency for equilibrium systems obeying the Boltzmann distribution function. The idea developed in that is that if a move of a particle in the system causes a decrease in the total energy, the new configuration should be accepted. On the other hand, if there is an increase in energy, the new configuration is accepted only if it passes through a game of chances biased by a Boltzmann factor, otherwise, the old configuration is kept.

Since then, the Monte Carlo method has been proved to be a very powerful and useful tool. For example, deterministic methods for numerical integration of functions with many variables are very inefficient because with every additional dimension or variable, an exponential time increase takes place. The alternative way provided by the Monte Carlo method is the following: the function in question can be estimated by randomly selecting points in the many dimensional space and taking some kind of average of the values of the function at these points. This method will display $1/\sqrt{N}$ convergence, i.e. quadrupling the number of sampled points will halve the error, regardless of the number of dimensions. The use of Monte Carlo methods to model physical problems allows us to examine more complex systems that otherwise we are not able to handle. Solving equations which describe the interactions between two atoms is fairly simple but solving the same equations for hundreds or thousands of atoms is impossible. With Monte Carlo methods, a large system can be sampled in a number of random configurations, and those data can be used to describe the system as a whole. There are currently many applications of the Monte Carlo method: stellar evolution [8], reactor design [9], cancer therapy [10], traffic flow [11], finance [12], simulations of various systems of interacting particles (e.g. ferromagnetic materials), grain growth modeling in metallic alloys [13, 14], the behaviour of nanostructures and polymers [15], and protein structure predictions [16].

## 1.2 Basics of the Monte Carlo Method

In statistical mechanics, the partition function $Z(H,T)$ contains all the necessary information to calculate the thermodynamic properties of a system. The difficulty arise when the size of the system and the number of degrees of freedom for each particle is large, something that occurs in almost all cases. Then, summing over the large number of possible states to calculate $Z(H,T)$ is extremely expensive and almost impossible even in a computational way. The result is that, in general, the partition function can not be evaluated exactly [17].

The Monte Carlo approach consists of generating a series of possible states or configurations $X_1, X_2, ..., X_N$ of a system ($X_i = \{x_1, x_2...\}$ with $x_i$ being the position of the particles in the system), so that the probability $P_{X_i}$ of encountering the system in state $X_i$, is given by an appropriate probability density function. Averages over phase space may be constructed by considering a large number of identical systems which are held at the same fixed conditions. These are called **ensembles**, (Fig. 1.2), and depending on the parameters held fixed, one can have different types of ensembles. In the case in which $T$ is maintained constant, the set of systems obtained is said to belong to the canonical ensemble, in which the systems are allowed to have distinct energies. On the other hand, if the energy is fixed, the ensemble is called the microcanonical ensemble. In both cases the number of particles is also fixed, but if now we allow the number of particles to fluctuate, the ensemble is named the grand canonical ensemble [17].

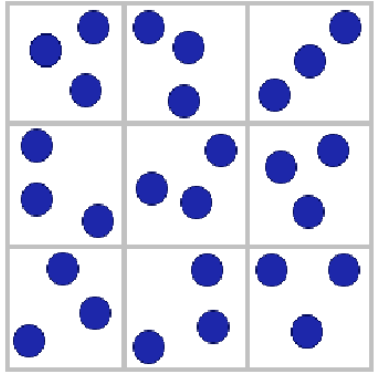

Figure 1.2: Graphical representation of a canonical ensemble: the positions of the particles and the energy can change in each system, but the number of particles and the temperature is fixed.

In the canonical approach, $Z(H,T)$ is calculated in the following way:

$$Z(H,T) = \sum_{all\,states} e^{-H/k_BT}, \quad (1.1)$$

where $k_B$ is the Boltzmann's constant, $H$ denotes the Hamiltonian and $T$ the temperature of the system. By all states we mean taking into the sum all the available configurations for the system. The probability distribution is called a canonical distribution if it is given according to the equation:

$$P_X = \frac{e^{-H(X)/k_BT}}{Z(H,T)}. \quad (1.2)$$

The general goal is to determine equilibrium properties of the canonical ensemble such as energy and magnetization. If $m(X)$ is the value of some physical property in a state $X$, and $H(X)$ the energy of this state, then the canonical ensemble average for the quantity $m$ is given by:

$$\langle m \rangle = \frac{\sum_{all\,states} m(X) e^{-H(X)/k_BT}}{Z(H,T)}. \quad (1.3)$$

As mentioned before, the problem is how to calculate $Z(H,T)$ in an efficient way.

If we have a finite state space $\mathscr{X}$, where $X_{(t)}$ is the state of the system at time $t$, that can only take $s$ discrete values $X_{(i)} \in \mathscr{X} = \{X_1, X_2, ..., X_s\}$, the stochastic process is called a Markov chain if the following condition is fulfilled:

$$P(X_{(t)}|X_{(t-1)},...,X_{(1)}) = T_M(X_{(t)}|X_{(t-1)}),$$

where $P(X_{(t)}|X_{(t-1)},...,X_{(1)})$ is the probability of the state $X_{(t)}$ to occur conditioned by the occurrence of the past states $X_{(t-1)},...,X_{(1)}$. $T_M$ is known as the transition probability matrix. The chain is homogeneous if the transition probability $T_M = T_M(X_{(t)}|X_{(t-1)})$ is constant for all $t$, with $\sum_{X_{(t)}} T_M(X_t|X_{(t-1)}) = 1$ for any $t$. That is, the evolution of the chain in the state space $\mathscr{X}$ depends solely on the current state of the chain and a fixed transition (probability) matrix [18].

For any starting point, the chain will converge to an invariant distribution $P(X)$, as long as $T_M$ is a stochastic transition matrix with the following properties:

1. *Irreducibility*: for any state of the Markov chain, there is a positive probability of visiting all other states. That is, the matrix $T_M$ can not be reduced to smaller matrices, which is also the same as stating that the transition graph is connected.

2. *Aperiodicity*: the chain should not get trapped in cycles [18], i.e., the system should not be limited to a subchain of states.

Consider now a large collection of copies of the same system in equilibrium. We allow each copy to evolve in time and, at any instant, we will find each different copy in one possible configuration, an all the copies will give a probability distribution over the configuration space. For each point $X_i$ in the configuration space, the probability $P$ of finding a copy in $X$ at time $t$ satisfies the equation:

$$\frac{d}{dt}P(X,t) = \sum_i [P(X_i,t)T_M(X_i \to X) - P(X,t)T_M(X \to X_i)]. \quad (1.4)$$

$T_M(X \to X_i)$ and $T_M(X_i \to X)$ are the probabilities of making a transition from the configuration $X$ to $X_i$ and viceversa. Because the collection is in equilibrium, the probability distribution is time invariant, and in the last equation we must have $dP(X,t)/dt = 0$ for all $t$. At any instant, there is an equal number of transitions to and from the configuration $X$. In fact, there exists an equation like (1.4) for each point in the configuration space, and the set of all such equations forms the master equation [19].

A sufficient (but not necessary) condition for an equilibrium (time independent) probability distribution needed to simulate equilibrium systems is the so-called **detailed balance condition** for the master equation that relates the transition between two configurations, $X_{n-1}$ and $X_n$ through:

$$P(X^n)T_M(X^{(n-1)}|X^{(n)}) = P(X^{(n-1)})T_M(X^{(n)}|X^{(n-1)}). \quad (1.5)$$

This method can be used for any probability distribution of configurations. If we choose the Boltzmann distribution, for which the probability of finding a configuration $X$ with energy $H$ at equilibrium is given by (1.2), and substitute it into (1.5), we get:

$$\frac{T_M(X^{(n-1)}|X^{(n)})}{T_M(X^{(n)}|X^{(n-1)})} = \frac{e^{-H(n-1)/k_B T}}{e^{-H(n)/k_B T}} = e^{\Delta E/k_B T}. \quad (1.6)$$

This is the detailed balance condition on the transition probabilities. It is very important to note that $Z(H,T)$ does not appear in this expression; it only involves quantities that we know $(k_B T)$ or that can be easily calculated $(E)$.

Thus, we have a valid Monte Carlo algorithm if we generate a new configuration $X_{(n)}$ from a previous one $X_{(n-1)}$ such that the transition probability satisfies the detailed balance condition, and the generation procedure is ergodic, i.e. every configuration can be reached from every other configuration in a finite number of iterations [20].

## 1.3 Measurements Using the Monte Carlo Method

Systems generated using a valid Monte Carlo algorithm are often held at fixed values of intensive variables, such as temperature, pressure, and so on. The corresponding conjugate extensive variables (energy, volume, etc.) will fluctuate in time; indeed these fluctuations will actually be observed during the Monte Carlo simulations and will help us to measure quantities of interest such as:

*Specific heat*

$$C_V = \frac{1}{V}\left(\frac{\partial E}{\partial T}\right)_V = \langle (E - \langle E \rangle)^2 \rangle = \langle E^2 \rangle - \langle E \rangle^2. \quad (1.7)$$

*Susceptibility*

$$\chi = \frac{1}{V}\left(\frac{\partial M}{\partial T}\right)_V = \langle (M - \langle M \rangle)^2 \rangle = \langle M^2 \rangle - \langle M \rangle^2. \quad (1.8)$$

These and other similar quantities are measured for each configuration and the averages and statistical errors calculated [17].

Summarizing, the idea of Monte Carlo simulations is to create an independently and identically distributed set of $N$ samples from a target density $P(X)$ distribution function defined on a high dimensional state space $\mathscr{X}$ (e.g., the set of possible configurations of a system). These $N$ samples can be used to approximate $P(X)$ [18].

When $P(X)$ has a standard form, e.g., Gaussian, it is straightforward to sample from it using easily available routines. However, when this is not the case, we need to introduce more sophisticated techniques such as Markov Chain Monte Carlo (MCMC) briefly presented above, which is a strategy of generating samples using a Markov chain mechanism while exploring the state space $\mathscr{X}$. This mechanism is constructed with the condition that the chain spends more time in the most important regions. In particular, it is constructed so that the samples mimic samples drawn from the target density distribution $P(X)$ [18].

## 1.4 Ising and Potts Models

The Ising model was proposed in 1925, in the doctoral thesis of Ernst Ising, a student of Wilhelm Lenz [21]. Using a model proposed by Lenz in 1920 [22], Ising tried to explain certain empirically observed facts about ferromagnetic materials in his thesis. The model was referred to in a paper by Heisenberg of 1928 in which he used the exchange mechanism to describe ferromagnetism [23]. After the publication of a paper by Peierls (1936) [24], in which he gave a non-rigorous proof that spontaneous magnetization must exist, the Ising model became a well-established paradigm. In 1941, Kramers and Wannier calculated the Curie temperature using a two-dimensional Ising model [25] and three years later Onsager gave a complete analytic solution of the model [26].

As a paradigm of statistical mechanics, the Ising model tries to imitate systems in which individual elements (e.g., atoms, animals, protein folds, biological membrane, social behaviour, etc.) modify their behaviour so as to conform to the dynamics of other elements in their neighbourhood [27].

In most specific terms, the Ising model in statistical mechanics considers a system with spins located at the sites of a D-dimensional lattice, where each spin can take the value +1, corresponding to spin up, or the value -1, corresponding to spin down. The Hamiltonian of such a spin lattice system is given by:

$$H_I = -J \sum_{\langle i,j \rangle} \sigma_i \sigma_j - B \sum_i \sigma_i, \tag{1.9}$$

where $J$ is the exchange constant, and $\sigma_i$ and $\sigma_j$ are the spins of the $i^{th}$ and $j^{th}$ sites respectively. The sites are usually a pair of nearest neighbours, though calculations for more distant neighbours can also be carried out. $B$ is an externally applied magnetic field with whom each spin interacts.

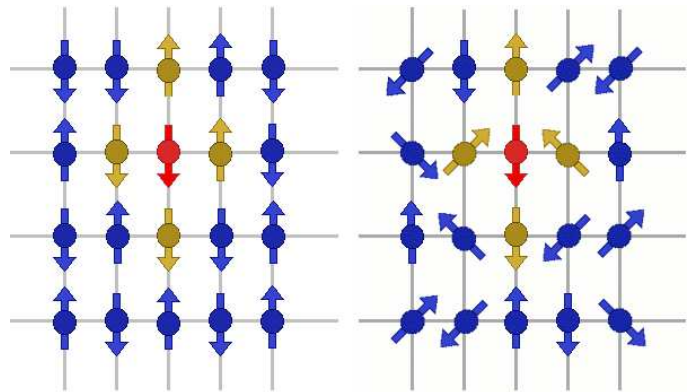

Figure 1.3: Lattice representations of Ising and Potts models. The red site interacts with his first neighbours (in yellow). Notice that in the Potts model, being a generalization of the Ising model, more than two possible directions for the spin are available.

When $J > 0$, the model describes a ferromagnetic system where parallel spins are favoured and antiparallel spins are discouraged.

In the case of $J < 0$, an antiferromagnetic system is modeled.

If $J$ is randomly chosen to be 1 or -1 for each pair of nearest neighbours and remain fixed during the course of observation, we obtain a model of a spin glass [28].

The energy associated with each state depends then on the exchange energy of the particles and the interaction of the particles with the external magnetic field. However, in the absence of the external field, the energy of the system depends only on the spin exchange energy:

$$H_I = -J \sum_{\langle i,j \rangle} \sigma_i \sigma_j. \tag{1.10}$$

The Potts model is a generalization of the Ising model, in which spins can choose its value from a discrete set of states (see Fig. 1.3). In 1952, C. Domb proposed it as a doctoral thesis for his student R. Potts [29]. Without the presence of an external field the Potts model is defined through the Hamiltonian:

$$H_P = -J \sum_{\langle i,j \rangle} \delta_{\sigma_i,\sigma_j} \sigma_i \sigma_j, \tag{1.11}$$

where $J$ denotes again the interaction exchange constant between nearest neighbours and the values $\sigma_i$ are characterized by an integer $\sigma_i = 1, 2, ..., q$. If two spins are parallel they contribute with energy $J$, otherwise their energy contribution is null.

## 1.5 Some Monte Carlo Algorithms: Metropolis, Swendsen-Wang and Wolff

The Metropolis [7], Swendsen-Wang [30] and Wolff [31] algorithms satisfy the master equation and the detailed balance condition for the Boltzmann distribution. Consequently, when the system reaches equilibrium, the probability distribution of all possible configurations will be the Boltzmann distribution.

The steps of the Metropolis algorithm for an Ising model are graphically represented in Fig. 1.7 and are the following:

1. Start with an arbitrary spin configuration $C_0$ of a lattice with $N$ sites.

2. Select a spin randomly and independently, and flip it.

3. Calculate the energy change $\Delta E$ which results if the spin is turned.

4. Generate a random number $r$ such that $0 < r < 1$.

5. If $\Delta E \leq 0$, accept the change; if $\Delta E > 0$, the configuration is accepted with a probability $e^{-\Delta E/k_B T}$. This is resumed as: if $r < e^{-\Delta E/k_B T}$ the spin is flipped. If not, the new configuration is rejected, and the system returns to the initial configuration $C_0$.

6. Choose randomly another spin to flip and go to (3).

It is important to discard some configurations at the beginning of the chain of configurations to ensure that the system forgets $C_0$ and that the configurations taken into account form a canonical ensemble. Then, after a considerable number of spins have been updated, the properties of the system are determined and added to the statistical average which is stored. The random number $r$ must be chosen uniformly in the interval $[0,1]$ and all the successive random numbers should be uncorrelated. Note that if a spin trial is rejected, the old state is counted again for the averages. For a $q$ state Potts model, the new value for the chosen spin is selected randomly among the other $q-1$ spin values [17].

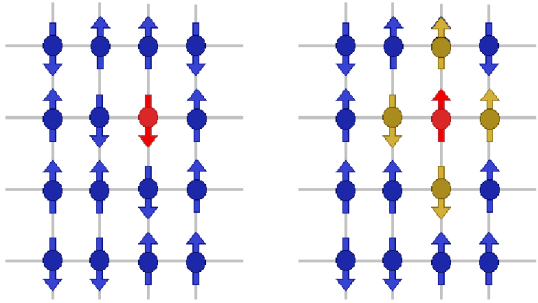

Figure 1.4: Metropolis algorithm: If the energy decreases with the spin flip, the new configuration remains. If not, is accepted or rejected with certain probability.

In the Metropolis algorithm, spins are updated one at a time and this single spin flip is the reason why this algorithm is inefficient at critical points where the phenomenon of slowing down occurs. The standard measure of Monte Carlo time is the Monte Carlo step per site (MCS/site), which corresponds to $N$ trial flips, regardless of whether the trial is successful or not ($N$ is the total number of spins in the system) [19].

The Swendsen-Wang and Wolff algorithms are cluster algorithms, where groups of spins are identified by establishing bonds between pairs of neighbouring spins. Once the clusters in the lattice are identified, a whole spin cluster is updated, and in this way these algorithms are more efficient near critical points.

The Swendsen Wang algorithm for a $q$ state Potts model is (Fig. 1.5):

1. Initialize the lattice of $N$ sites with an arbitrary spin configuration $C_0$.

2. Examine every pair of neighbouring spins in the system. If neighbouring spins are not parallel, nothing is done. If they are parallel, a bond is introduced between them with probability $p = 1 - e^{-K}$, where $K = J/k_BT$. (If $p < 1$, a random number $r$ is generated such that $0 < r < 1$, and if $r < p$ a bond is introduced between sites $i$ and $j$).

3. Once all clusters in the lattice have been formed, an arbitrary cluster is chosen.

4. Another random number $R$ is generated such that $1 \leq R \leq q$.

5. All spins in the chosen cluster are assigned $\sigma_i = R$.

6. Another cluster is selected randomly and return to (4).

7. When all clusters have been considered, erase the bonds, go to (2) and repeat the steps until the desired number of configurations has been obtained.

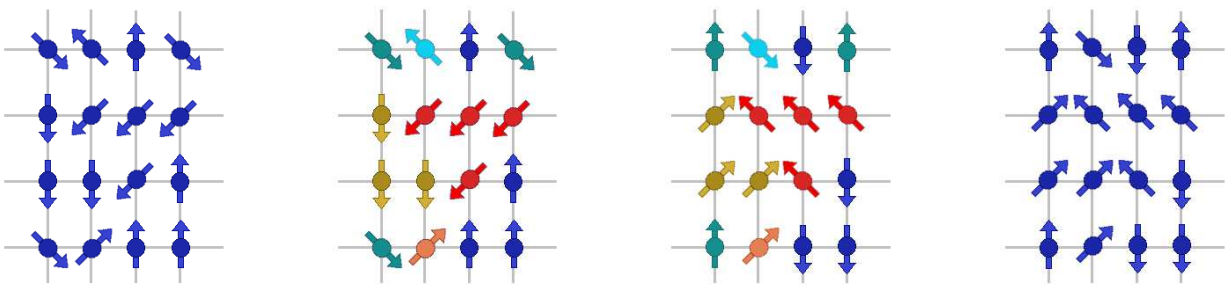

Figure 1.5: Swendsen-Wang algorithm: Once the clusters are formed (each one is represented by a diferent colour), their spin values are randomly modified. Some clusters maintain the same value (i.e., orange spin). After that, the cluster formation starts again.

One Monte Carlo cycle in the Swendsen-Wang algorithm is accomplished when all clusters have been updated (steps 2-6), and is equivalent to one Monte Carlo step per site (MCS/site) in the Metropolis algorithm [19].

The probability to set a bond between two sites depends on the temperature, which affects the resultant cluster distribution. At very high temperature, the clusters will tend to be quite small, whereas at very low temperature virtually all sites with nearest neighbours in the same state will belong to the same cluster and therefore there will be a tendency for the system to oscillate back and forth between quite structures. However, near a critical point, a quite rich array of clusters is produced and the net result is that each configuration differs substantially from its previous one. That is the main reason why the critical slowing down is reduced [17].

The Wolff algorithm is very similar to the Swendsen-Wang algorithm, the principal difference being that it flips the spins of one particular cluster with the maximum probability of 1 in each Wolff MC cycle. The Wolff algorithm was proposed to improve the Swendsen Wang algorithm in which significant effort is required in dealing with small clusters as well as large ones. However, the small clusters do not contribute to the critical slowing down [17] and can be disregarded. The Wolff algorithm is given by the following procedure (a graphical representation is provided in Fig. 1.6):

1. Start with an arbitrary spin configuration $C_0$ of a lattice with $N$ sites.
2. Randomly choose a spin to be the seed of a cluster.
3. Examine all its neighbours and draw bonds with probability $p = 1 - e^{-K\delta_i\delta_j}$.
4. If bonds have been drawn to any nearest neighbour site $j$, draw bonds to all nearest neighbours $k$ of site $j$ with probability $p = 1 - e^{-K\delta_j\delta_k}$.
5. Repeat step (4) until no more new bonds are created.
6. Flip all spins in the cluster to a different randomly chosen spin value.
7. Go back to (1).

The measurement of Monte Carlo time is more complicated. The natural unit of time is the number of cluster flips. However, in one cluster flip the number of spins visited is not equivalent to the total number of spins in the system and hence one Wolff cluster flip is not equivalent to one MC step per spin (MCS/site) or one MC cycle in the Metropolis and Swendsen-Wang algorithms. The generally accepted method of converting to MCS/site is to normalize the number of cluster flips by the mean fraction of sites $\langle c \rangle$ flipped at each step. The Monte Carlo time then becomes well defined if $\langle c \rangle$ is well defined, and this happens only after enough flips have occurred [17].

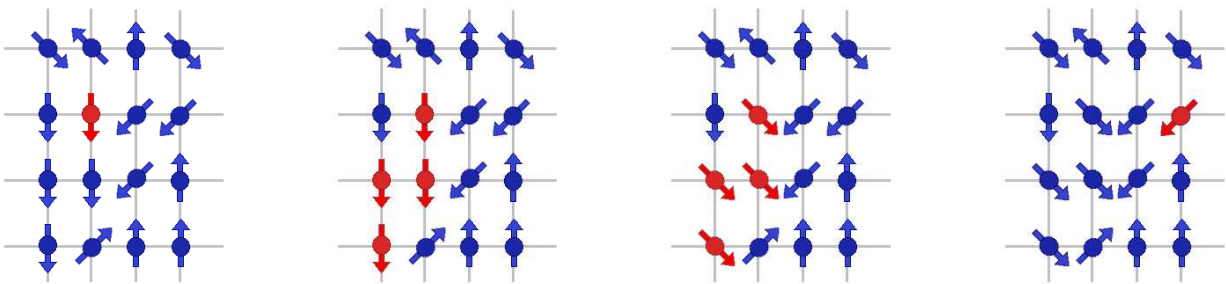

Figure 1.6: Wolff algorithm: A spin is chosen randomly, and the cluster is formed from it by introducing bonds to its neighbours and the neighbours of its neighbours with some given probability. The spin value of the cluster is changed and then another spin is selected to start a new cluster.

Although all these algorithms satisfy detailed balance, they do not give the same results for $M$ and $\chi$ in a simulation. This difference is due to the very small probability for $M$ to change sign using the Metropolis algorithm for large systems, at low temperatures. This corresponds to a physical situation, and one can calculate $\langle M \rangle$ and $\chi$ and obtain meaningful results. However in cluster algorithms, the clusters become very large at low temperatures, and by flipping them, we effectively flip the whole system, yielding $\langle M \rangle = 0$; the variance in $M$ is then simply $\langle M^2 \rangle$, a constant at low temperatures, which in turn gives a diverging $\chi_M$ as $T \to 0$. The solution is to use $|M|$ instead of $M$, and define $\chi_{|M|}$ just as we defined $\chi$ earlier. In this way, all three algorithms give the same results for $\langle M \rangle$ and $\chi_{|M|}$ at all temperatures [19].

Notice that cluster algorithms become inefficient at low temperatures, because in that situation, nearly all spins in the system are flipped when we flip the largest cluster, which is not helpful in achieving statistically independent configurations. In comparison, the Metropolis algorithm will be much more efficient [19].

Once an appropriate algorithm has been selected, one of the goals of Monte Carlo simulations is the study of the behaviour of systems in phase transitions.

## 1.6 Phase Transitions and Critical Exponents

One of the most common physical problems studied in simulations are phase transitions. A phase transition occurs when a thermodynamic system passes from one phase to another one with the change of some external variable, such as temperature or pressure. Some examples are the transitions between solid, liquid, and gaseous phases, the transition between the ferromagnetic and paramagnetic phases of magnetic materials, and the emergence superconductivity in certain metals when they are cooled below a critical temperature [32].

When a system goes from one phase to another, there will be in general a stage where the free energy is not analytic. Due to this, the free energies on either side of the transition are two different functions, so one or more thermodynamic properties will behave very differently after the transition. A system near or at the critical point of a phase transition presents peculiar behaviours that are universal, like divergence of some quantities and critical slowing down phenomena, which will be explained later. The most commonly examined property in this context is the heat capacity that in the transition region may become infinite, jump abruptly to a different value, or exhibit a discontinuity in its derivative [33]. This non-analytic behaviour stems generally from the interactions of an extremely large number of particles in a system, and does not show up with the same strength in systems that are too small [32].

Phase transitions are generally classified into first or second order transitions. A second order, or continuous phase transition, can be defined as a point at which a system changes from one state to another one without a discontinuity or jump in its density, internal energy, magnetization, or similar properties. In the case of a first order transition, the above mentioned properties jump discontinuously as the temperature or pressure passes through the transition point [34]. The name of different kind of phase transitions comes precisely from the number of derivatives of the free energy that we have to count before we can see a discontinuous behaviour. If the first derivative is discontinuous, we have a first order transition, if not, it is a second order one [17].

The first-order phase transitions involve a latent heat. During such transition, a system either absorbs or releases a fixed (typically large) amount of energy. Because energy can not be instantaneously transferred between the system and its environment, first-order transitions are associated with "mixed-phase regimes" in which some parts of the system have completed the transition and others have not. Continuous phase transitions, in many cases, are associated with a change of symmetry of the system and are easier to study than first-order transitions due to the absence of latent heat. They have shown many interesting properties. The phenomena associated with continuous phase transitions are called critical phenomena, because of their occurrence near critical points and because it turns out that continuous phase transitions can be characterized by parameters known as critical exponents [32].

In the case of many phase transitions a non-zero value of an order parameter appears, i.e., some property of the system which is non zero in one phase (usually called the ordered phase) but identically zero in the other phase (disordered phase). Thus, the order parameter can not be an analytic function at the transition point. The order parameter is defined differently in various kinds of physical systems [17]. For systems such as the ferromagnet, where there is a broken symmetry below critical temperature $T_c$, the order parameter is the magnetization. For systems without broken symmetry, one chooses some quantity that is very sensitive to the difference between the two phases, and measures the difference of this quantity from its value at the critical point and below it. For the liquid-vapor critical point, we may choose the order parameter as the difference between the actual density of the fluid and the density at the critical point. For liquid crystals the degree of orientational order is considered as the order parameter [34].

Another quantity of interest near a phase transition is the correlation function. In general, there will be microscopic regions in which the characteristics of the material are correlated. This is generally measured through the determination of a *two point correlation function*, which is the probability of finding that two sites separated by a distance $r$ have the same value of a certain given quantity $\rho$ [17]:

$$\Gamma_\rho(r) = \langle \rho(0)\rho(r) \rangle. \tag{1.12}$$

In the case of magnetic systems, the correlation function can be measured in neutron scattering experiments, whereas near the liquid vapour transition it can be measured by light scattering or small angle X-ray-scattering experiments [34].

If the correlation for the appropriate quantity decays to zero as the distance goes to infinity, then the order parameter is zero [17]. Close to the critical point, the correlation length $\xi$, which tells us how far correlations are still present, becomes extremely large. This is directly related to the large amount of long-wavelength fluctuations that occur in the system at the criticality [34]. The time taken for the system to change configuration near the critical point also increase significantly because of the divergence of the correlation length $\xi$. This phenomenon is called critical slowing down. For example, in the case of the Ising model, spins tend to align with their neighbours due to the exchange interaction, and regions or clusters of spins pointing in the same direction appear. These spins are said to be correlated, and, generally, there are clusters of various sizes. The span of the largest one is the correlation length $\xi$, while the time it takes to break up the existing conformation of spins and form another arrangement of clusters is called the decorrelation time $\tau$. At the critical point, there is a low probability for a spin in the middle of a spin cluster to change its direction, therefore spin regions are altered only at the boundary. This gives rise to a long decorrelation time which is related to the correlation length by a power law:

$$\tau \propto \xi^z, \tag{1.13}$$

where $z$ is the dynamical critical exponent [19]. For simulations of a finite lattice of linear dimension $L$, $\xi$ is naturally bounded by $L$ and then the basic assumption is that:

$$\tau \propto L^z. \tag{1.14}$$

These two equations describe the critical slowing down. In an infinite system, as the critical point is approached, the correlation length diverges (its value is $\infty$), and from (1.13), we see that the decorrelation time also diverges. In finite systems $\xi$ does not diverge as the critical point is approached, however, it reaches its peak with a sharp slope. Due to the power law dependence of $\tau$ on $\xi$, $\tau$ will also display a peak with a sharp slope, exhibiting critical slowing down [19].

Near the transition points, the critical slowing down phenomenon produces important effects that complicate the implementation of the Monte Carlo method. This is the main reason why the scientists introduced alternative approaches besides canonical Metropolis algorithm, such as Wolff and Swendsen-Wang algorithms. The computational effect of critical slowing down near a critical point can be understood in the following manner: when we simulate finite systems at the critical point, the decorrelation time depends on the linear dimension $L$ through a power law as $L$ approaches infinity. Take, for example, the 2D Ising model. The dynamical critical exponent $z$ is known to be approximately 2 using the Metropolis algorithm. If the time it takes to obtain 100 statistically independent configurations is $t$ in a system with $L = 32$, then if $L$ is increased by a factor of 2 to 64, the computational time needed to obtain 100 statistically independent configurations will increase to $4^2 t$. A factor of 4 is introduced because the number of spins is increased by 4, and another factor of 4 is due to the fact that $\tau \propto L^2$. In general, the amount of CPU time required to obtain a fixed number of statistically independent configurations for a system with linear dimension $L$ is proportional to $L^{d+z}$, where $d$ is the spatial dimension of the model, and $z$ is the corresponding dynamical critical exponent [19].

Data from experiments, as well as results for a number of exactly solvable models, show that in the vicinity of the critical point $T_c$, the thermodynamical properties can be described by a set of simple power laws [17]. For example, for the determination of the way which the magnitude of the order parameter approaches zero as the critical point is reached, we may write (according to the classical theories of phase transitions such as the van der Waals or mean field theories):

$$M = M_0 \, \varepsilon^\beta, \qquad (1.15)$$

where $M$ is the order parameter (i.e., the magnetization for a ferromagnet), $M_0$ is a constant that will vary from one system to another, $\varepsilon = |1 - T/T_c|$, and the exponent $\beta$ is called critical exponent [34].

The temperature variation of the order parameter is very important but not the only quantity of interest. Another key quantity is the specific heat, defined as the derivative of the internal energy with respect to the temperature. The specific heat is found to become infinite at the critical point in some systems but also one can have cases in which the specific heat is finite with only a sharp cusplike maximum at the critical point [34]. In either case, one may define an exponent $\alpha$ that characterizes the anomalous behaviour of the specific heat at the critical point:

$$C_V = C_0 \, \varepsilon^{-\alpha}. \qquad (1.16)$$

Susceptibility $\chi$ is another quantity of interest. It is defined as the derivative of the order parameter with respect to the applied field to which it is coupled, under constant temperature condition. For a magnetic system, this quantity is precisely the magnetic susceptibility. This quantity becomes extremely large near the critical point, and we may write the zero field magnetic susceptibility as [34]:

$$\chi = \chi_0 \, \varepsilon^{-\gamma}. \qquad (1.17)$$

Finally, the correlation length $\xi$ varies as:

$$\xi = \xi_0 \, \varepsilon^{-\nu}, \qquad (1.18)$$

where, again, $\nu$ is termed as critical exponent.

Note that the last equations represent asymptotic expressions which are only valid if $\varepsilon \to 0$ and more complete forms would include additional corrections to scaling terms which describe the deviations from the asymptotic behaviour. The exact values of these critical exponents are known exactly only for a small number of models, most notably for the 2D Ising square lattice [26], whose exact solution shows that $\alpha = 0$, $\beta = 1/8$, and $\gamma = 7/4$. Here, $\alpha = 0$ corresponds to a logarithmic divergence of the specific heat [17].

The power law behaviour near critical points is very general and many systems share the same critical exponents. In particular, the Ising universality class refers to the class of critical phenomena that share the same critical exponents as the Ising model [19].

Although the critical exponents, $\alpha$, $\beta$, and $\gamma$ defined above may be independent in principle, they were found empirically, in the 1960's, to be connected by the relationship:

$$\alpha = 2 - \gamma - 2\beta. \qquad (1.19)$$

This equality is known as the Rushbrooke relation, and the following three relations are also known [17], where $\eta$ and $\delta$ are two additional critical exponents:

$$\begin{aligned}
\text{Josephson:} \quad \nu D &= 2 - \alpha, \\
\text{Widom:} \quad \gamma &= \beta(\delta - 1), \\
\text{Fisher:} \quad \gamma &= \nu(2 - \eta).
\end{aligned}$$

In Table 1.1 we provide the theoretical values of the critical exponents for $q \leq 4$ 2D Potts model, which of course fulfills the latter relations.

|     | $\alpha$ | $\beta$ | $\gamma$ | $\nu$ | $\delta$ | $\eta$ |
|-----|----------|---------|----------|-------|----------|--------|
| q=0 | $-\infty$ | 1/6 | $\infty$ | $\infty$ | $\infty$ | 0 |
| q=1 | $-2/3$ | 5/36 | 27/18 | 4/3 | 18 1/5 | 5/24 |
| q=2 | 0 | 1/8 | 7/4 | 1 | 15 | 1/4 |
| q=3 | 1/3 | 1/9 | 13/9 | 5/6 | 14 | 4/15 |
| q=4 | 2/3 | 1/12 | 7/6 | 2/3 | 15 | 1/2 |

Table 1.1: Some theoretical critical exponents for the 2D Potts model [35].

The quantities discussed above are all equilibrium or static quantities; they can be measured in a time-independent experiment in thermal equilibrium conditions, and any involved correlation function refers to the correlation of fluctuations at a single instant of time. The majority of theoretical studies and experiments on critical phenomena are concerned with these static measurements. Thus, the usual division of systems into different universality classes is based on these static phenomena. There are other properties of systems, known as dynamical properties, which require a more detailed theoretical analysis. Moreover, they require a further subdivision of the universality classes. Two systems that belong to the same universality class for their static properties could show quite different behaviours in their dynamical properties. Some standard examples of dynamical properties are various relaxation rates of systems slightly disturbed from equilibrium, correlations involving fluctuations at two different time instants, and transport coefficients, e.g., thermal and electrical conductivities. Among the experiments used for studying dynamical properties we quote measurements of sound-wave attenuation and dispersion, widths of nuclear and electron magnetic resonance lines, and inelastic scattering experiments. Typically, one finds that the relaxation rate of the order parameter becomes anomalously slow at a critical point. However, some other relaxation rates are found to speed up and transport coefficients become large in a number of cases. In some cases, the results of a dynamical experiment may be interpreted as an indirect measurement of a static property of the system. As a matter of fact, some of the most precise measurements of static critical properties have been obtained by dynamical means. Examples are the measurements of the superfluid properties of liquid helium, the low-frequency sound velocity of a fluid, and the frequency of nuclear magnetic resonance in a magnetic system [34].

## 1.7 The Histogram Method

The canonical Metropolis algorithm yields mean values of various thermodynamical quantities, (energy, magnetization, etc) at particular values of the temperature $T$. Near a phase transition, many thermodynamical quantities change rapidly, and we need to determine these quantities at closely spaced values of $T$. If we use standard Monte Carlo methods, we will have to do many simulations to cover the desired $T$ range [36]. The use of histograms to overcome this problem became popular after the publication of a paper by Ferrenberg and Swendsen in 1988 [37]. However, the histogram technique is one of the oldest techniques proposed [38, 39]. Also often referred to as Ferrenberg-Swendsen reweighting technique, is used in almost all Monte Carlo calculations of statistical physics, especially when dealing with phase transition phenomena [40]. The idea is to use the knowledge of the equilibrium probability distribution at one value of $T$ (and other external parameters) to estimate the desired thermodynamical averages at neighbouring values.

A Monte Carlo simulation performed at $T = T_0$ generates configurations of the system with a frequency proportional to the Boltzmann weight, $e^{-\beta_0 H}$, where $\beta_0 = 1/k_B T_0$, and $H$ is the Hamiltonian of the system being studied. In the case of a magnetic system, the probability of simultaneously observing the system with

energy $E$ and magnetization $M$ is given by:

$$P_{\beta_0}(E,M) = \frac{1}{Z(\beta_0)} W(E,M) e^{-\beta_0 E}, \quad (1.20)$$

where $W(E,M)$ is the number of configurations (density of states) with energy $E$ and magnetization $M$, and $Z(\beta_0)$ is the partition function of the system. Because the simulation generates configurations according to the equilibrium probability distribution, a histogram $H(E,M)$ can be built during the simulation to provide an estimate for the equilibrium probability distribution that becomes exact in the limit of infinite-length run. For a finite length-simulation, the histogram will present statistical errors, but $H(E,M)/N$, where $N$ is the number of measurements, still provides an estimate of $P_{\beta_0}(E,M)$ over the $E$ and $M$ values generated during the simulation [41]. Keeping this in mind, we modify (1.20) as follows:

$$H(E,M) = \frac{N}{Z(\beta_0)} \tilde{W}(E,M) e^{-\beta_0 E}, \quad (1.21)$$

where $\tilde{W}(E,M)$ is an estimate of the true density of states, or number of configurations, $W(E,M)$.

The probability distribution for any value of $\beta$ has the same form as (1.20):

$$P_\beta(E,M) = \frac{1}{Z(\beta)} W(E,M) e^{-\beta E}. \quad (1.22)$$

Comparing (1.21) and (1.22), we can note that it is possible to determine $\tilde{W}(E,M)$ from (1.21):

$$\tilde{W}(E,M) = \frac{Z(\beta_0)}{N} H(E,M) e^{\beta_0 E}, \quad (1.23)$$

and replace $W(E,M)$ in (1.22) with it. After normalizing the distribution, we find that the relationship between the histogram measured at $\beta = \beta_0$ and the (estimated) probability distribution for an arbitrary $\beta$ is:

$$P_\beta(E,M) = \frac{H(E,M) e^{-(\beta-\beta_0)E}}{\sum_{E,M} H(E,M) e^{-(\beta-\beta_0)E}}. \quad (1.24)$$

From $P_\beta(E,M)$, the average value of any function $f(E,M)$ can be calculated as a continuous function of $\beta$:

$$\langle f(E,M) \rangle_\beta = \sum_{E,M} f(E,M) P_\beta(E,M). \quad (1.25)$$

The histogram method is useful only when the configurations relevant to the range of temperatures of interest occur with sufficient probability during the simulation at temperature $T_0$. For example, if we simulate an Ising model at low temperatures at which only ordered configurations occur (most spins aligned in the same direction), we can not use the histogram method to obtain meaningful thermodynamical averages at temperatures for which most configurations are disordered, and viceversa [36].

In the single histogram technique, the estimated $P(E,\beta)$ is accurate only for $\beta$ close to the reference value $\beta_0$. By generating many histograms that overlap each other we can widen the range of $\beta$. This is called the multiple histogram technique [42]. It is also clear that we can increase the range of $\beta$ by directly estimating the density of states $W(E,M)$. Multicanonical sampling [43] is an early technique proposed to do this. It is a very general and useful technique being often the method of first choice for a variety of problems that include critical slowing down near second order phase transition points, nucleation in first order phase transitions, and trapping in the metastable minima in systems with rugged energy landscapes.

## 1.8 Identifying the Nature of Transitions and Finite Size Scaling

The behaviour near phase transitions has been one of the main objectives of studies focusing on the properties of physical systems but a correlation length $\xi$ greater than the accessible size $L$ of the system may lead to many difficulties [44]. For systems close to a second order phase transition, finite-size scaling is routinely used to extract thermodynamic information from similar systems of fairly small size. An equivalent theory for first order phase transitions is clearly also of interest. A useful theory of finite-size scaling should allow us to extract the couplings at which the transition occurs, as well as other dimensional quantities like latent heat (or spontaneous magnetization) and specific heat (or magnetic susceptibility) [45].

First order transitions are characterized by a discontinuity in the order parameter and thermodynamic quantities, with an associated delta-peak behaviour in the susceptibility. As a matter of fact, the jump in the energy density is equivalent to the latent heat. However, at finite size, thermodynamic quantities become continuous and rounded. Instead of delta function behaviour in susceptibility there is only a hump. In simulations, this behaviour is visible only if the simulation time $\tau_s$ is larger than the decorrelation time $\tau$ at the transition point. $\tau_s$ is typically very large since $\tau \propto e^{-\sigma 2L^{D-1}}$, where $\sigma$ is the surface tension of the interface between the low temperature and high temperature phases [47]. It is the dimension $D$ that now plays the key role rather than the critical exponents as in the case of second order phase transitions [17].

At the transition temperature of a first-order phase transition, a mixed state can exist where two different bulk phases are separated by an interface. The free energy densities of the two bulk phases are equal and the free energy of the mixed state is higher than any of the coexisting pure phases by an amount $F_s = \sigma A$, where $A$ is the area of the interface and $\sigma$ is the interface tension [48]. In first order phase transitions, the correlation length remains finite in both the ordered and disordered phases, i.e., the correlation length does not diverge. Thus, a different approach to finite size scaling must be used [17].

From fairly general arguments about the nature of discontinuities at a first-order phase transition, Fisher and Berker [49] obtained the infinite volume limit approached by measurements performed at finite volumes. This conventional scenario is based on a smooth behaviour of the renormalization group flow and the existence of a discontinuity fixed point whose attraction domain contains the transition surface and has relevant exponents of the form $y = D$ [49]. The singularities associated with first order transitions are generated by infinite iterations of renormalization group transformations in the thermodynamic limit. Correction terms were later calculated in a particular phenomenological model called the double-Gaussian model, in which the peaks in the probability distribution for the coexisting phases were approximated by Gaussians [50, 51]. This model correctly predicts the first term in a series of corrections in inverse powers of the volume $V$, around the leading term obtained by Fisher and Berker [49].

More recent developments are due to Borgs, Kotecký and Miracle-Solé [52, 53]. The basic idea is to decompose the partition function into a sum of the contributions, each due to one of the coexisting phases, and to neglect contributions due to phase mixtures. Each of these contributions to the total partition function then yield quantities related to free energies in the pure phases. The analysis proceeds by power series expansions of these partial partition functions around the phase transition point, leading to moments expressed in inverse powers of the volume [45]. According to this theory, for periodic boundary conditions, the specific heats and Binder cumulants at the transition temperature can be represented by polynomials in $1/L^D$. If the $L >> \xi$, the contribution of the higher order terms are negligible [54, 55]. The difficulty arises when $\xi \geq L$. In this case, higher order corrections are necessary and deciding the order of the transition becomes difficult. Even when large lattices are used, higher order terms may create difficulties during the fitting procedure to the simulation data. Such difficulties may be reduced by choosing the quantities for which the correction terms play less important role. A good example for such quantity is the average energy measured at the infinite lattice transition point, which has exponentially small correction term enabling one to determine the infinite lattice critical point with great accuracy [53, 54, 56].

Finite size scaling ideas for first or second order transitions help to extract critical exponents and other information, but this requires prior knowledge of at least the nature of the transition. When the system undergoes a weak first order transition with $\xi \gg L$, it becomes very difficult to identify its nature even with large-scale computations. This problem is even worse when one encounters a system for which nothing is known [44, 57].

Lee and Kosterlitz [57] proposed a method which exploits the finite size scaling properties of the free energy $\Delta F(L)$. These properties are unambiguous even when $\xi \gg L$ and, more importantly, can be implemented with reasonable computational effort. This method depends on two key ideas: the identification of $\Delta F(L)$, which has a characteristic behaviour as a function of $L$ at a first or second order transition or in a single phase region, and the usage of histograms enabling this to be computed accurately. They have shown that the positions of the peak free energies in a histogram should scale as $1/L$ if the system is well into the first order region. The ratio of $P(E)$ at its peaks and minimum can be used to estimate an interface free energy $\Delta F(L)$, signaling a first order transition if it increases with system size $L$.

This method uses the Helmholtz free energy $F$ of a system. At low $T$, the low energy configurations dominate the contributions to the partition function $Z$, even though there are relatively few such configurations. At high $T$, the number of disordered configurations with high $E$ is large, and hence high energy configurations have a big contribution to $Z$. These considerations suggest that it is useful to define a restricted free energy $F_r(E)$ that includes only the main configurations at a particular energy $E$:

$$F_r(E) = -kT \left[ \ln g(E) \right] e^{-E/kT}. \tag{1.26}$$

For systems with a first-order phase transition, a plot of $F_r(E)$ versus $E$ will show two local minima corresponding to configurations that are characteristic of the high and low temperature phases. At low $T$, the minimum at the lower energy will be the absolute minimum, whereas at high T the higher energy minimum will be the absolute minimum of $F_rE$. At the transition temperature, the two minima will have the same value of $F_r(E)$. For systems with no transition in the thermodynamical limit, there will only be one minimum for all $T$. How will $F_r(E)$ behave for the relatively small lattices that we can simulate? In systems with first-order transitions, the difference between low and high temperature phases will become more pronounced as the system size is increased. If the transition is continuous, there are domains at all sizes, and we expect that the behaviour of $F_r(E)$ will not change significantly while increasing the size. If there is no transition, there might be a fake double minima for small systems that disappear for larger systems [36]. Lee and Kosterlitz proposed the following method to classify phase transitions:

1. Perform a simulation at a temperature close to the suspected transition temperature and calculate $H(E)$. Usually, the temperature at which the peak in the specific heat occurs is chosen as the simulation temperature.

2. Make use of the histogram method to calculate $F_r(E) \propto -\ln H_0(E) + (\beta - \beta_0)E$ at neighbouring values of $T$. If there are two minima in $F_r(E)$, vary $\beta$ until the values of $F_r(E)$ at the two minima are equal. The corresponding temperature is an estimate of the possible transition temperature $T_c$.

3. The difference between the maxima and the minimum between the two peaks is used to estimate the free energy barrier $\Delta F_r(E)$ at $T_c$.

4. Repeat steps (1-3) for larger systems. If $\Delta F_r(E)$ increases with size, the transition is first order. If $\Delta F_r(E)$ remains the same, the transition is continuous. If $\Delta F_r(E)$ decreases and goes to zero with size, there is no thermodynamic transition.

The above procedure is applicable when the phase transition occurs by varying the temperature. Transitions also can occur by varying the pressure or the magnetic field. These field-driven transitions can be tested by a similar method. For example, consider the Ising model in a magnetic field at temperatures below $T_c$. As we vary the magnetic field from positive to negative values, there is a transition from a phase with magnetization

$M > 0$ to a phase with $M < 0$. Is this a first-order or continuous transition? To answer this question, we can use the Lee-Kosterlitz method with a histogram $H(E,M)$ generated at zero magnetic field, and calculate $F_r(M)$ instead of $F_r(E)$. The quantity $F_r(M)$ is proportional to $-\ln \sum_E H(E,M) e^{-(\beta - \beta_0)E}$. Because the states with positive and negative magnetization are equally likely to occur for zero magnetic field, we should see a double minima structure for $F_r(M)$ with equal minima. As we increase the size of the system, $\Delta F_r$ should increase for a first order transition and remain the same for a continuous transition [36].

Another way to determine the nature of a first order phase transition is to use the Binder cumulant of energy defined by [58]:

$$U_L = 1 - \frac{\langle E^4 \rangle}{3 \langle E^2 \rangle^2}. \tag{1.27}$$

If various cumulants (each one corresponding to different lattice sizes) are plot in the same graph, a behaviour characteristic of a first order transition appears as will be discussed in the next section.
It can be shown that the minimum value of $U_L$ is

$$U_{L,min} = \frac{2}{3} - \frac{1}{3}\left(\frac{E_+^2 - E_-^2}{2 E_+ E_-}\right)^2 + O(L^{-d}), \tag{1.28}$$

where $E_+$ and $E_-$ are the energies of the two phases in a first order transition. These results are derived by considering the distribution of energy values to be a sum of Gaussians about each phase at the transition point, which become sharper and sharper as $L \to \infty$ [36].

On the other hand, equations (1.15) to (1.18) for second order transitions are valid only for infinite systems and, as a matter of fact, we can simulate only finite systems. Quantities that diverge in the infinite case now present peaks in the finite system. Furthermore, the peaks occur at a value $T_c(L)$, for a given linear dimension $L$, slightly different from the infinite-lattice critical temperature $T_c$. However, at a second order phase change, the critical behaviour of a system in the thermodynamical limit can be extracted from the properties of finite systems by examining the size dependence of the singular part of the free energy density. This finite size scaling approach was first developed by Fisher [59]. According to his theory, the free energy of a system of linear dimension $L$ is described by the scaling ansatz:

$$F(L,T,h) = L^{-(2-\alpha)/\nu} F^0(tL^{1/\nu}, \quad hL^{(\gamma+\beta)/\nu}), \tag{1.29}$$

where $t = (T - T_c)/T_c$, $h$ is the magnetic field and $F^0$ is a scaling function. The critical exponents $\alpha$, $\beta$, $\gamma$, and $\nu$ all correspond to the values for the infinite system. Appropriate differentiation of the free energy yields the various thermodynamic properties with their corresponding scaling forms:

$$\begin{aligned} m &= L^{-\beta/\nu} m^0 x_t, \\ C &= L^{\alpha/\nu} C^0 x_t, \\ \chi &= L^{\gamma/\nu} \chi^0 x_t, \end{aligned} \tag{1.30}$$

where $x_t = tL^{1/\nu}$ is the temperature scaling variable [41].

To determine the transition temperature accurately one find the location of the peak in a thermodynamic derivative, for example, specific heat. For a finite lattice the peak occurs at the temperature where the scaling function $Z^0(x_t)$ is maximum, i.e., when

$$\left. \frac{dZ^0(x_t)}{dx_t} \right|_{x_t = x_t^*} = 0.$$

This temperature is the finite lattice (or effective) transition temperature $T_c(L)$, defined through the condition $x_t = x_t^*$ to vary with the lattice size, asymptotically, as:

$$T_c(L) = T_c + T_c x_t^* L^{-1/\nu}.$$

These results for the scaling of thermodynamic quantities and $T_c(L)$ are valid only for sufficiently large $L$ and temperatures close to $T_c$. Corrections to finite size scaling must be taken into account for smaller systems. These are introduced as power law corrections with an exponent $-w$, such that, for example, the magnetization at $T_c$ would scale with system size like $L^{-\beta/\nu}(1+cL^{-w})$. As we move away from $T_c$, corrections to scaling due to irrelevant scaling fields, or nonlinearities in the scaling variables must be introduced. Corrections due to irrelevant fields are expressed in terms of an exponent $\theta$ leading to additional terms like $a_1 t^\theta + a_2 t^{2\theta} + ...$, while nonlinearities in the scaling variables give rise to corrections terms of the form $b_1 t^1 + b_2 t^2 + ...$, [41].

If we take one correction term into account, the estimate for $T_c(L)$ is then modified in terms of the coupling $K = J/k_B T$ as follows:

$$K_c(L) = K_c + \lambda L^{-1/\nu}(1+bL^{-w}).$$

Before this equation can be used to determine $K_c$, it is necessary to have an accurate estimate for $\nu$ and accurate values for $K_c(L)$.

It has traditionally been difficult to determine $\nu$ from Monte Carlo simulation data because of a lack of quantities which provide a direct measurement. This situation was greatly improved by Binder's introduction of the fourth order magnetization cumulant $U$ [58] defined by:

$$U = 1 - \frac{\langle m^4 \rangle}{3\langle m^2 \rangle^2}, \quad (1.31)$$

where $m$ is the magnetization per spin. Binder showed that the slope of the cumulant at $K_c$, or anywhere in the finite size scaling region, varies with system size like $L^{1/\nu}$. In particular, the maximum value of the slope scales as $L^{1/\nu}$. If we take into account a correction to scaling term, the size dependence of the peak becomes:

$$\frac{dU}{dK}\bigg|_{max} = aL^{1/\nu}(1+bL^{-w}).$$

The location of the maximum slope of $U$ also serves as an estimate for an effective transition coupling which can be used to determine $K_c$. In the same paper, Binder introduced the cumulant crossing method which extracts a transition temperature by examining the behaviour of the magnetization cumulant for different lattice sizes.

Additional estimates for $\nu$ can also be obtained by considering the logarithmic derivative of any power of the magnetization, which has the same scaling properties as the cumulant slope. The location of the maximum slope also provides an additional $K_c(L)$:

$$\begin{aligned} \frac{\partial}{\partial K} \ln \langle m^n \rangle &= \frac{1}{\langle m^n \rangle} \frac{\partial}{\partial K} \langle m^n \rangle \\ &= \frac{\langle m^n E \rangle}{\langle m^n \rangle} - \langle E \rangle. \end{aligned} \quad (1.32)$$

To this end, the methods of finite size scaling are very helpful to determine the behaviour of infinite systems from data obtained on finite systems.

## 1.9 Monte Carlo Simulations on the Betts Lattice

Research of properties of lattices distinct from the commonly studied ones (square, triangular lattice) is a key step in the development and prediction of the behaviour of possible new materials. A different lattice proposed by Donald Betts is constructed removing 1\7 of the sites in a two dimensional triangular lattice [65], accomplishing that each vertex has a coordination number of five and yielding another translationally invariant lattice (see Fig.1.7). This structure is known as Betts or Maple Leaf lattice, and lies between the kagomé and triangular ones, which have coordination numbers of four and six, respectively. It has a hexagonal unit cell of six sites and fifteen bonds, it is invariant under rotations through multiples of $60°$, and, contrary to the kagomé and honeycomb lattices, it has no inversion symmetry [66]. To study the critical behaviour of this lattice, we performed Monte Carlo simulations using the Potts model for $q=3$, $q=4$ and $q=5$.

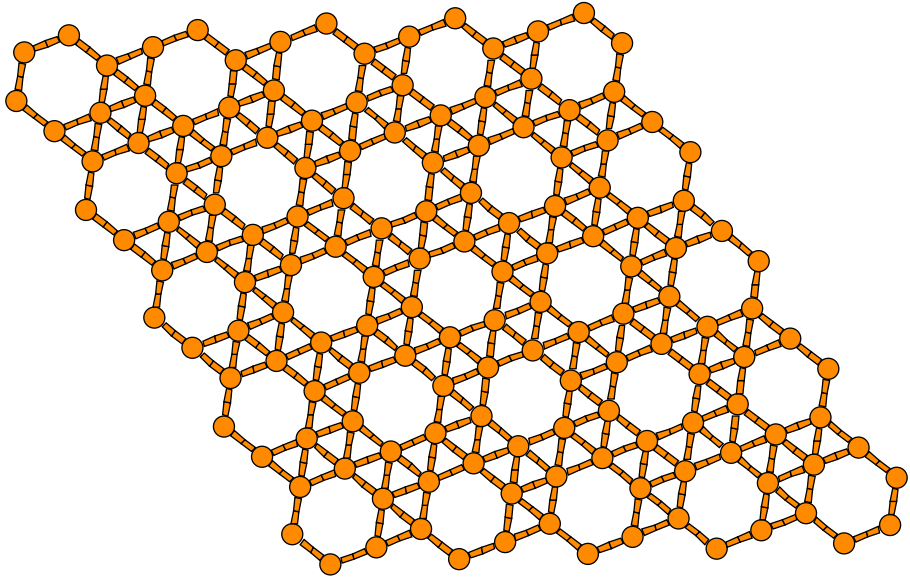

Figure 1.7: Maple Leaf lattice

For the q-Potts model, the magnetization is defined as follows:

$$m = \frac{N_{max} - 1/q}{1 - 1/q}, \qquad (1.33)$$

where $N_{max}$ is the maximum number of equally oriented spins for certain configuration. We denote the lineal size of the system studied as $L$, and this is related to the number of sites as $nsit = L \times L \times 6$. Earlier work has been already done on this lattice for $q = 3$, using the Metropolis algorithm, by Wang and Southern [67]. We applied Wolff algorithm instead, due to its proved better performance, and obtained similar results for ferromagnetic and antiferromagnetic cases. As predicted, calculations shown a second order transition for the ferromagnetic case and a first order transition for the antiferromagnetic case. For $q = 4$ and $q = 5$ there is no published work. We focus on the ferromagnetic regime in which the transition is found to be of second order for $q = 4$ and of first order for $q = 5$. In the latter case, the transition is very weak and more calculations are needed to obtain better results.

### 1.9.1 $q=3$, $J<0$: Antiferromagnetic Case

We selected four lattice sizes $L = 12$, 18, 24 and 36 to perform Monte Carlo simulations. The number of Monte Carlo steps used to equilibrate the system before making the average was of the order of $2 \times 10^5$, and the number of steps used for averaging was $6 \times 10^5$. Binder cumulants of the order parameter $E$ as a function of temperature for all lattice sizes demonstrate that the system undergoes a first order transition, as each curve shown a deep minima whose value moves to lower temperature regions (Fig. 1.8). The critical temperature is obtained from the deep minimums showed by all curves, and its near $T_c = 0.444$.

In Fig. 1.9, specific heats for each lattice size are plotted. There, the lattice size effect on the results can bee clearly seen: the peaks are sharper and moves toward smaller temperatures at larger lattice sizes. The transition temperature can be estimated as the temperature where the peaks have their maximum values, and obviously, the best approximation is obtained for the largest lattice size.

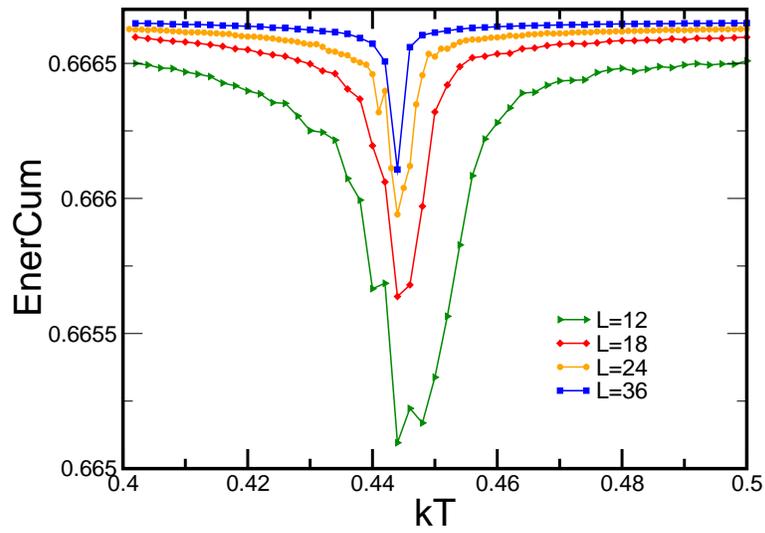

Figure 1.8: Energy cumulants suggesting a first order phase transition for $q = 3$, $J < 0$, $T_c = 0.444$ and four lattice sizes.

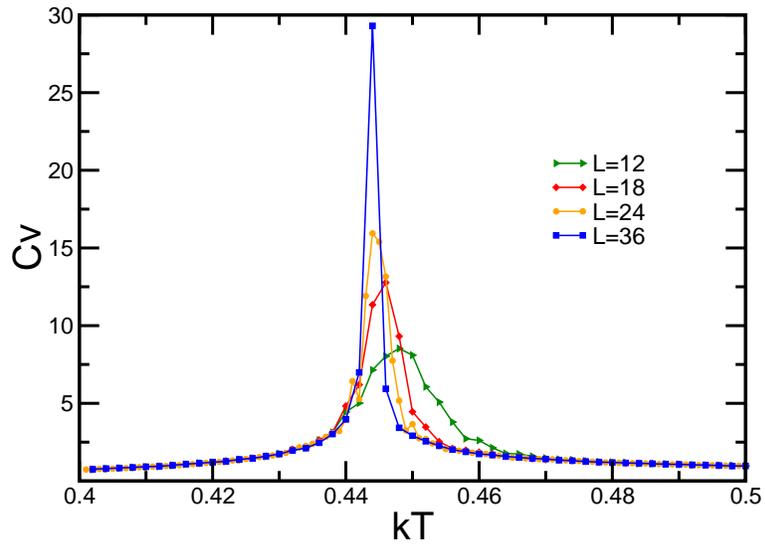

Figure 1.9: Specific heats for $q = 3$, $J < 0$ and the same four lattice sizes.

Realizing that the phase transition appears to be of first order, the next step is to calculate the energy distribution histograms $P(E)$ for various lattice sizes near the estimated critical temperature. We used $1 \times 10^6$ steps to equilibrate the system and $4 \times 10^6$ steps for averaging. The histograms always present two well-defined peaks, and while increasing $L$, the minimum between the peaks becomes deeper. Moreover, the histograms are sharper when more sites are taken into account (see Fig. 1.10). As explained in section 1.8, this is typical for first order phase transition, confirming the nature of the transition for this case.

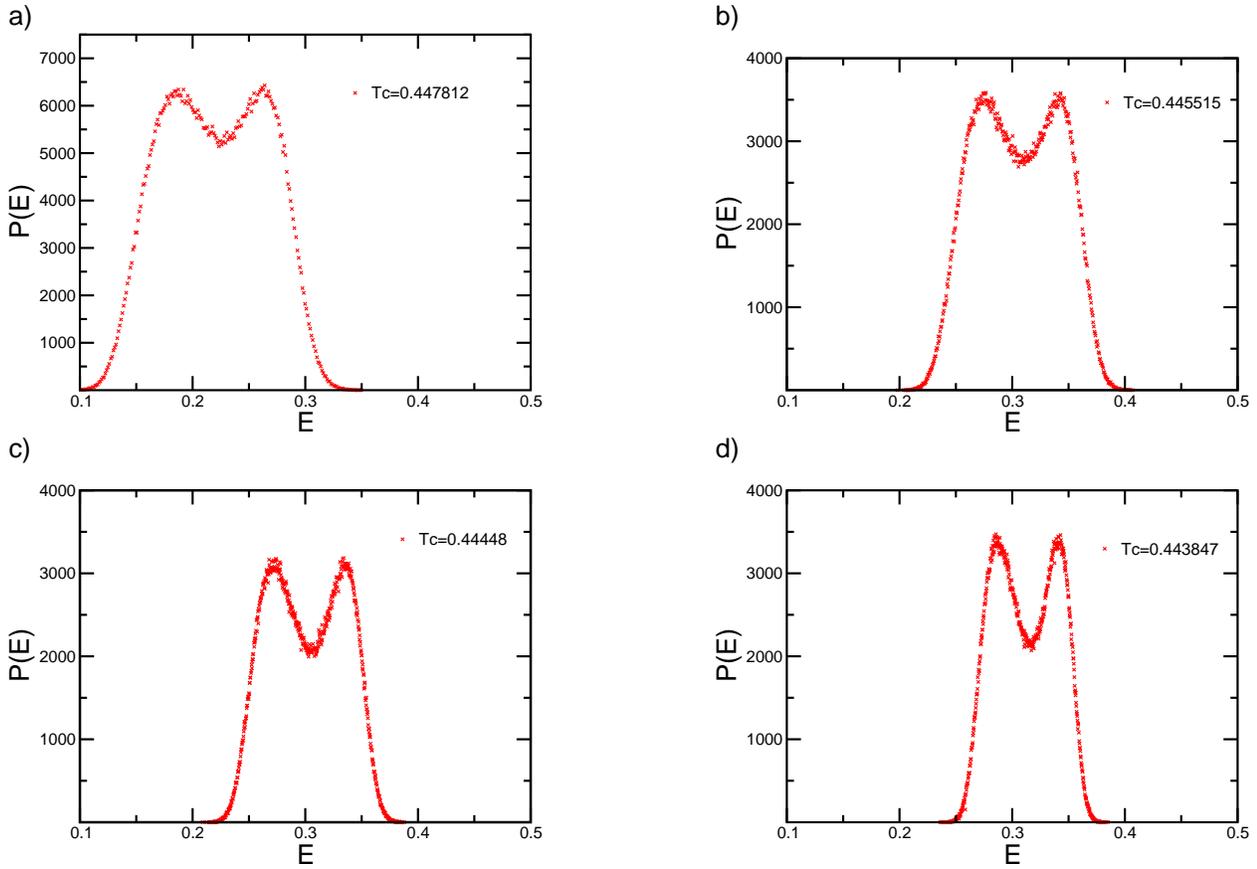

Figure 1.10: Energy histograms for lattice sizes a) $L = 12$, b) $L = 18$, c) $L = 24$, and d) $L = 36$ for $q = 3$, $J < 0$.

The results shown in the present subsection correspond well with the values reported by Wang and Southern [67]. The transition temperature reported by them is $T_c = 0.445$ and their histograms present a behaviour identical to ours.

## 1.9.2 $q=3, J>0$: Ferromagnetic Case

In this case, the used lattice sizes are $L = 18, 24, 30, 36, 48, 54$ and $60$. We considered a larger number of lattice sizes in order to have more points available to estimate the critical exponents. The number of Monte Carlo steps used to thermalize was $2 \times 10^5$, and the number of steps for averaging was $6 \times 10^5$. Binder cumulants of the order parameter $m$ as a function of temperature for the various $L$ values demonstrated that the system undergoes a second order transition. This is presented in Fig. 1.14. The critical temperature is obtained from the intersection of all curves, each curve corresponding to a distinct lattice size. The obtained value for the critical temperature is $T_c = 1.2275$. In Fig. 1.15, specific heats for different values of $L$ are shown.

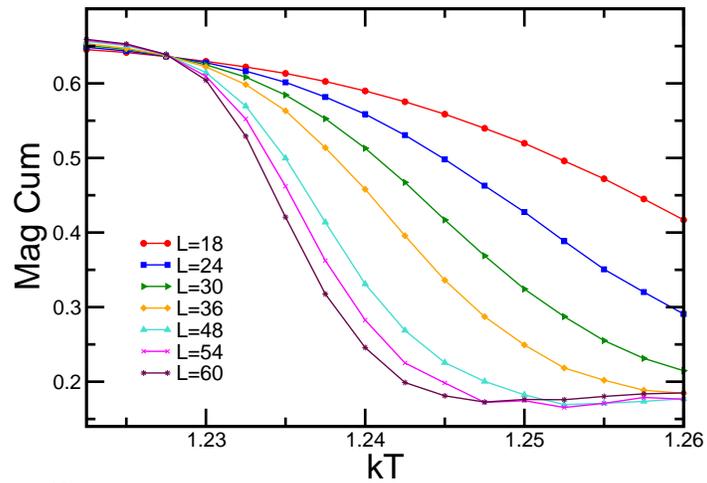

Figure 1.11: Magnetization cumulant showing second order transition at $T_c = 1.2275$. ($q=3, J>0$).

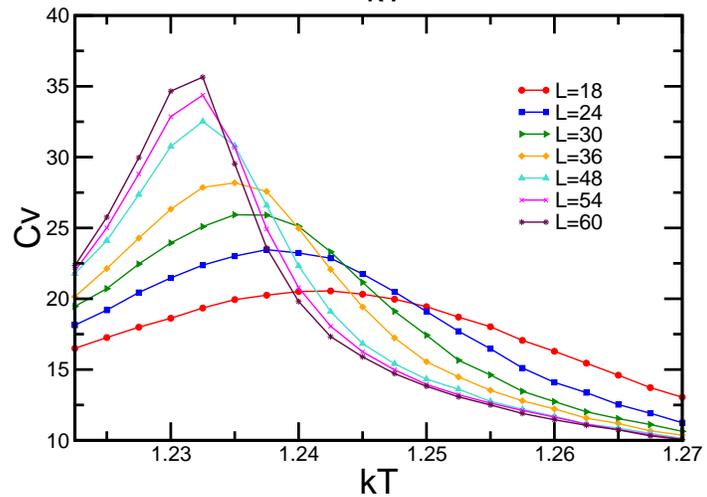

Figure 1.12: Specific heats for distinct lattice sizes for $q = 3$, $J > 0$.

We used finite size scaling techniques (see section 1.9) to calculate the critical exponents. To obtain $\nu$, for example, we calculated the logarithmic derivative of the magnetization in a range near the critical temperature for all lattice sizes selected, and the maximum value obtained for each curve was plotted against lattice size in a log-log plot. A line was fitted to these points, and its slope gave an estimate of the value of $1/\nu$. Fig. 1.16 illustrates the procedure. It is important to note that logarithmic derivatives of higher orders of magnetization can be also used to obtain estimations of $1/\nu$.

To calculate $\alpha$, the quantities plotted as functions of lattice sizes are the maximum values of specific heat $C_v$. Again, a linear fit gives the value of $\alpha/\nu$, from which $\alpha$ can be estimated using the value of $\nu$ obtained earlier.

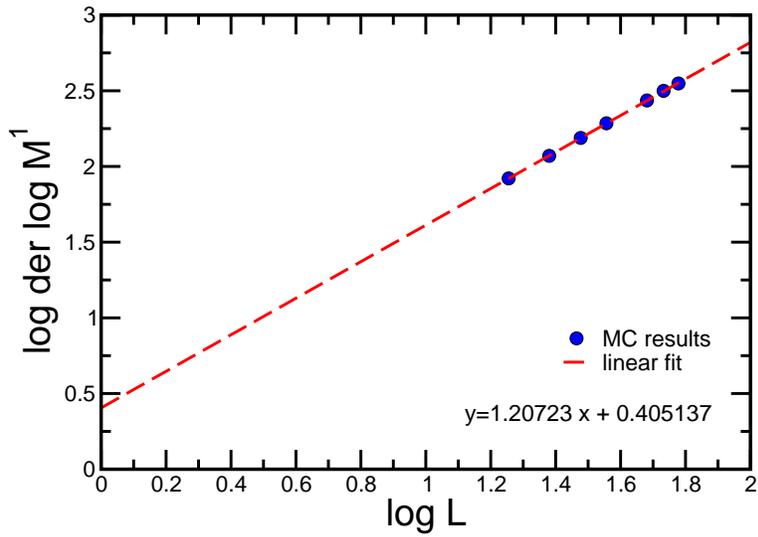

Figure 1.13: Values of the logarithmic derivatives of the magnetization for different sizes of Betts lattice versus the logarithm of $L$. The slope of the fitted line $y$ gives the value of $\nu$ for $q = 3$.

The data and the linear fit are shown in Fig. 1.17.

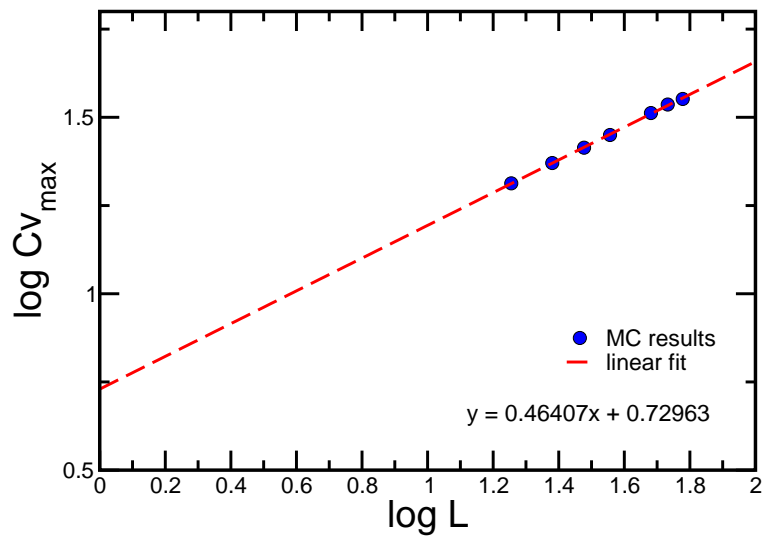

Figure 1.14: Log-log plot of the maximum values of $C_V$ for distinct sizes of Betts lattices. The fit gives the value of $\alpha/\nu$ for $q = 3$.

The critical exponent $\beta$ is extracted from the magnetization values at the critical temperature suggested by the Binder cumulant of magnetization. The logarithm of these values (remember that each value corresponds to a lattice size) are plotted versus the logarithm of $L$, and the slope of the line fitting the data corresponds to $-\beta/\nu$. If instead, the maximum values of the susceptibility are plotted versus $L$, the critical exponent $\gamma$ is obtained using the same procedure. (see Figs. 1.18 and 1.19).

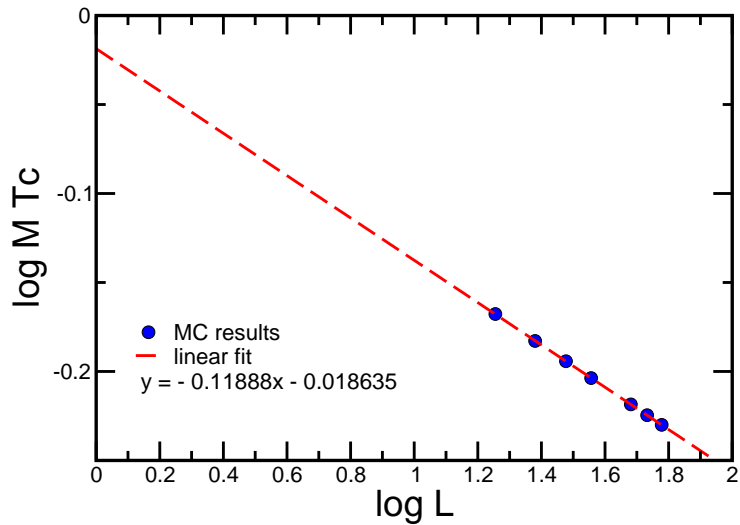

Figure 1.15: Logarithms of magnetization at the $T_c$ value suggested by the magnetization cumulant versus logarithms of $L$ values. The fit gives $-\beta/\nu$ for $q = 3$.

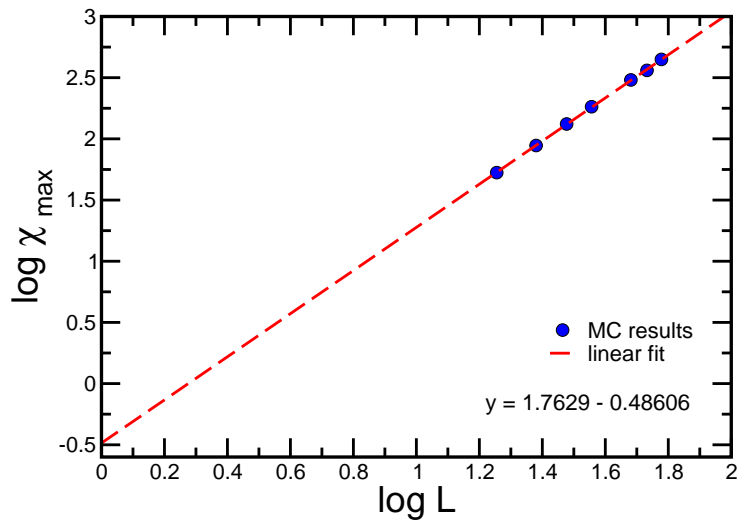

Figure 1.16: Logarithms of susceptibility at the critical temperature versus logarithms of different linear sizes $L$. The fit gives $\gamma/\nu$ for $q = 3$.

One of the common procedures used to obtain the transition temperature consists in plotting the temperature at which the logarithmic derivatives of the magnetization for each lattice size have their maxima versus $L^{-1/\nu}$. A line is fit to the data and the intersection of this line with the y-axis gives an approximate of the true transition temperature. This can be seen in Fig. 1.17 from which one gets $T_c = 1.22676$.

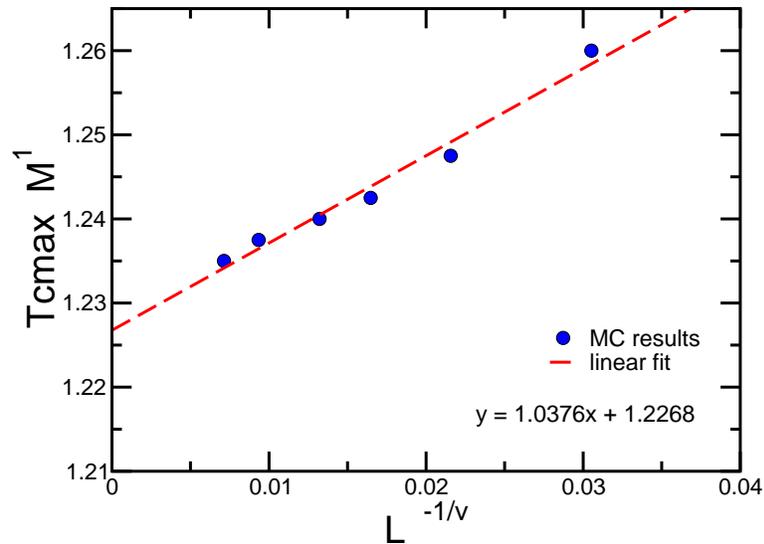

Figure 1.17: Estimation of the transition temperature $T_c$ ($q = 3$, $J > 0$).

In the next table, the critical exponents calculated by Wang and Southern [67], the values obtained in this work, and the theoretical values are summarized. The values obtained with the Monte Carlo simulations agree well to the 2D Potts classical values, but are not perfectly equal. This can be due to numerical errors, lattice size effects and also because the Betts lattice can be seen as a triangular lattice with a large number of defects. Something that is not so clear to us is why values obtained with the Wolff algorithm are less similar to the universal values than those calculated by Wang and Southern.

|  | Wang & Southern Results | Our Results | Theory |
|---|---|---|---|
| $\alpha/\nu$ | $0.42 \pm 0.04$ | $0.464068 \pm 0.00479$ | $0.4$ |
| $\beta/\nu$ | $0.132 \pm 0.002$ | $0.118885 \pm 0.000203$ | $0.13333$ |
| $\gamma/\nu$ | $1.74 \pm 0.05$ | $1.76294 \pm 0.01072$ | $1.73333$ |
| $1/\nu$ | $1.19 \pm 0.02$ | $1.20723 \pm 0.004891$ | $1.2$ |

Table 1.2: Comparison of the reported critical exponent values with the universal values predicted for the $q = 3$ 2D-Potts model.

### 1.9.3 $q = 4, J > 0$: Ferromagnetic Case

The used lattice sizes are once again in the range $L = 18$ to $L = 60$. The number of Monte Carlo steps used to thermalize is $2 \times 10^5$, and the number of steps for averaging is $6 \times 10^5$. The Binder cumulant of the order parameter $m$ as a function of temperature shows that the system undergoes a second order transition, and it is displayed in Fig. 1.21. The critical temperature is obtained from the intersection of all curves, each curve corresponding to a distinct lattice size, and is near $T_c = 1.126$. In Figs. 1.22 and 1.23, specific heats and susceptibilities for different values of $L$ are shown.

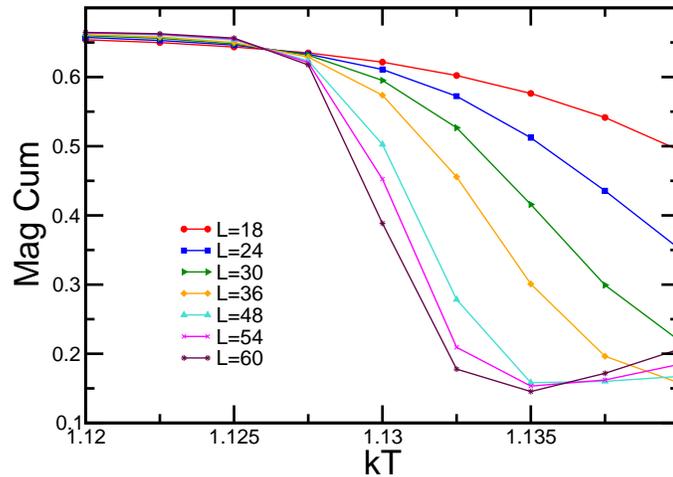

Figure 1.18: The Binder magnetization cumulant for $q = 4, J > 0$.

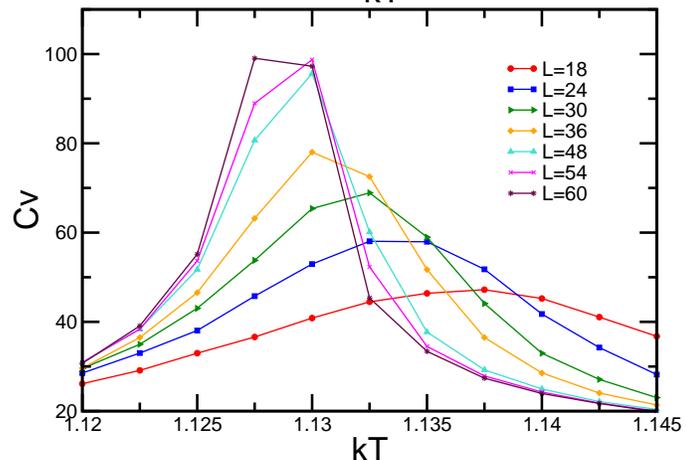

Figure 1.19: Specific heats for different lattice sizes ($q = 4, J > 0$).

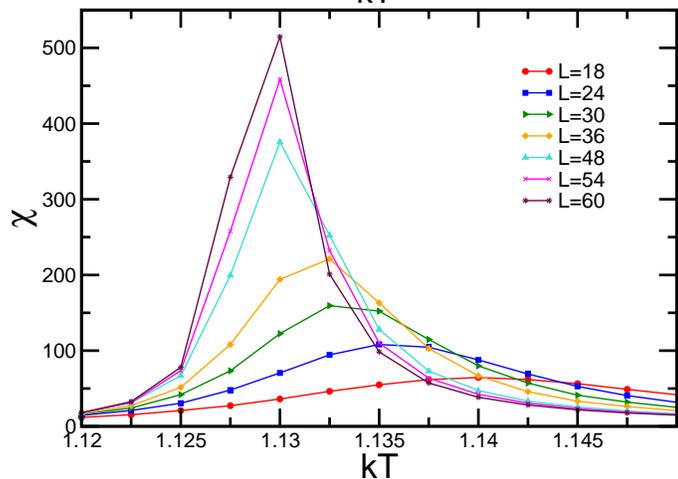

Figure 1.20: Susceptibilities for different lattice sizes ($q = 4, J > 0$).

The critical exponents were obtained with the same procedures explained for $q = 3$. Different thermodynamic quantities are calculated for each lattice size, and the values near critical temperature are plotted against linear size in various log-log plots. A line is fit to the data and its slope is representative of some critical exponent, depending on which thermodynamic quantity was selected to be plotted (Fig. 1.24 to Fig. 1.27). The critical temperature is estimated in the same way explained earlier, and is shown in Fig. 1.28.

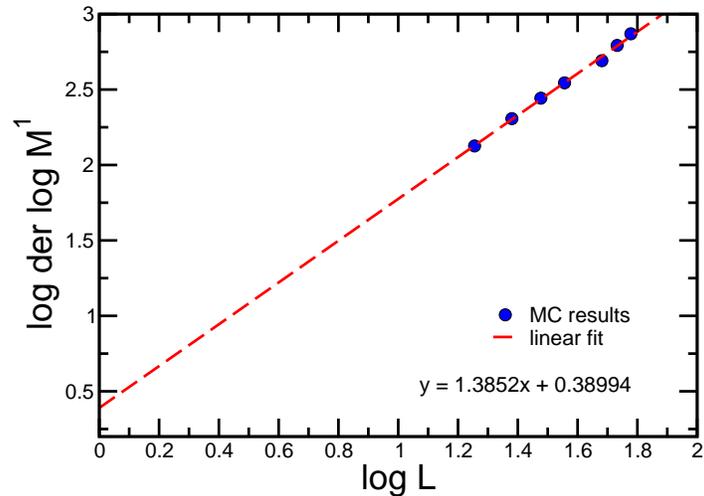

Figure 1.21: Values of the logarithmic derivatives of magnetization for different sizes of Betts lattice, versus logarithm of $L$. The fit gives the value of $1/\nu$ for $q = 4$, $J > 0$.

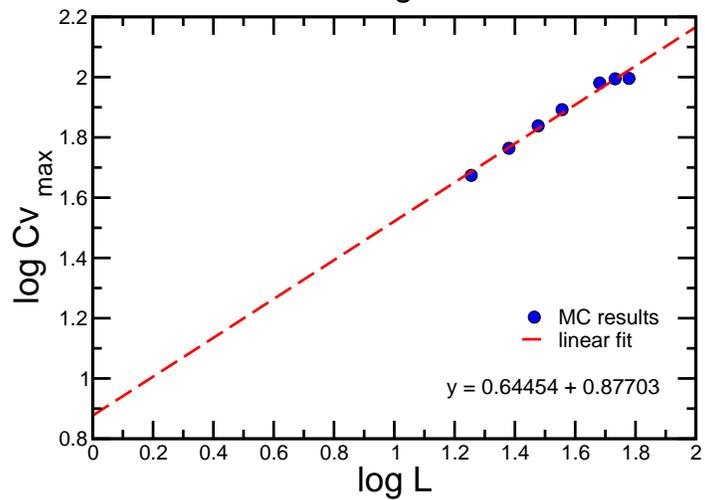

Figure 1.22: Logarithm of the maximum value of $C_v$ for different sizes of Betts lattice, versus logarithm of $L$. The fit gives the value of $\alpha/\nu$ for $q = 4$, $J > 0$.

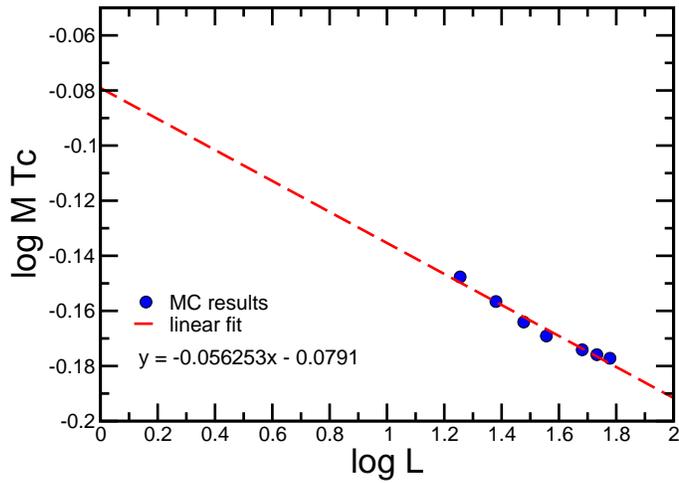

Figure 1.23: Logarithms of the magnetization values at the critical temperature of various lattice sizes versus logarithms of $L$. The fit gives $-\beta/\nu$ for $q=4$, $J>0$.

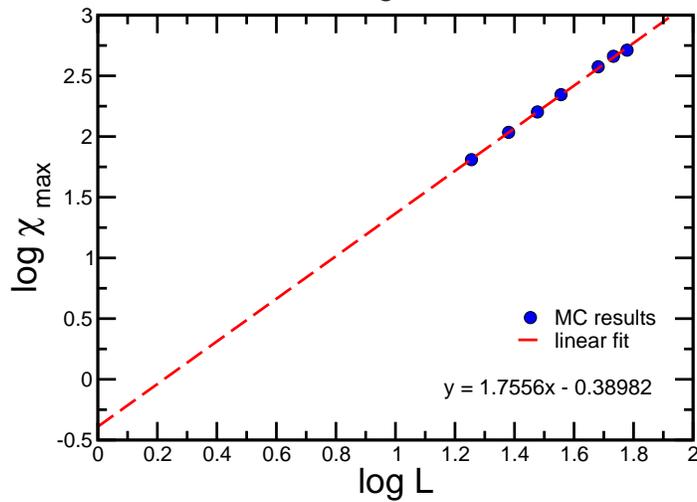

Figure 1.24: Logarithms of the susceptibilities at the critical temperature of various lattice sizes versus logarithms of $L$. The fit gives $\gamma/\nu$ for $q=4$, $J>0$

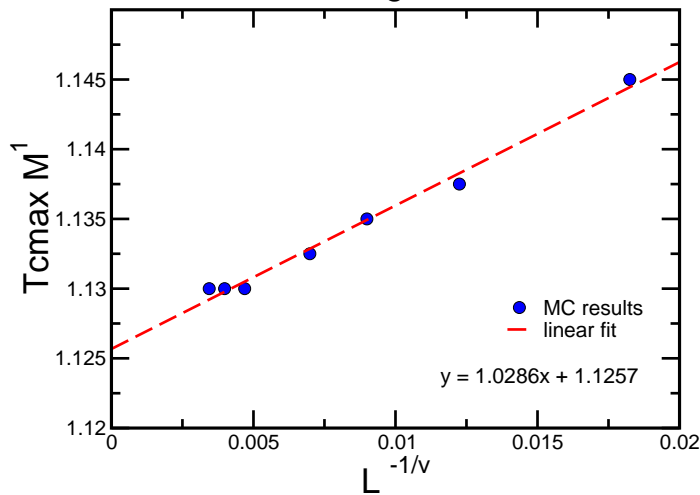

Figure 1.25: Estimation of the critical temperature for $q=4$, $J>0$.

The values obtained for the critical exponents are summarized in the next table, along with the theoretical values expected. The value obtained for β is not near the expected value, and this can be due to an effect of magnetic frustration on the lattice. More calculations need to be done in the future, with larger lattice sizes, however it could be that this lattice does not belong to an universal group.

|  | Our Results | Theory |
|---|---|---|
| $\alpha/\nu$ | $0.644537 \pm 0.03695$ | 1 |
| $\beta/\nu$ | $0.0562526 \pm 0.004001$ | 0.125 |
| $\gamma/\nu$ | $1.75564 \pm 0.02506$ | 1.75 |
| $1/\nu$ | $1.38519 \pm 0.03172$ | 1.5 |

Table 1.3: Comparison of critical exponent values obtained by us with the universal values predicted theoretically for $q = 4$ 2D-Potts model.

### 1.9.4 $q = 5, J > 0$: Ferromagnetic Case

The lattice sizes considered once again are in the range $L = 18$ to $L = 60$. The number of Monte Carlo steps used to thermalize is $2 \times 10^5$, and the number of steps for averaging is $6 \times 10^5$. The Binder cumulant of the order parameter $E$ as a function of temperature suggests that the system undergoes a first order transition, and it is presented in Fig. 1.29. The critical temperature is obtained from the deep of all curves, each curve corresponding to a different lattice size, and is near $T_c = 1.0575$. Specific heats are shown in Fig. 1.30.

For the calculation of energy histograms, the number of Monte Carlo steps used for thermalization is $1 \times 10^6$, and the number of steps for averaging varies from $3 \times 10^6$ for lattice sizes until $L = 36$, and $4 \times 10^6$ for the next lattice sizes. The results are shown in Fig. 1.31.

Energy histograms confirm that the transition is first order. The histograms are narrow at increased lattice size, and the valleys also become deeper.

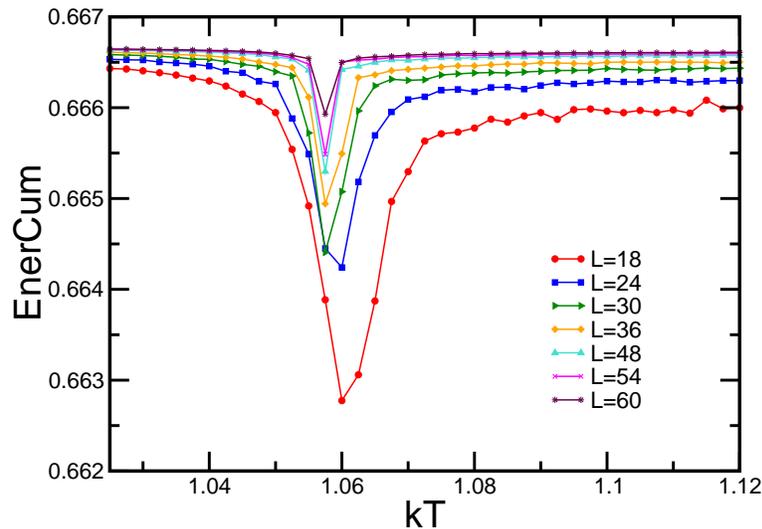

Figure 1.26: Energy cumulant for $q = 5, J > 0$. The transition temperature is near 1.0575.

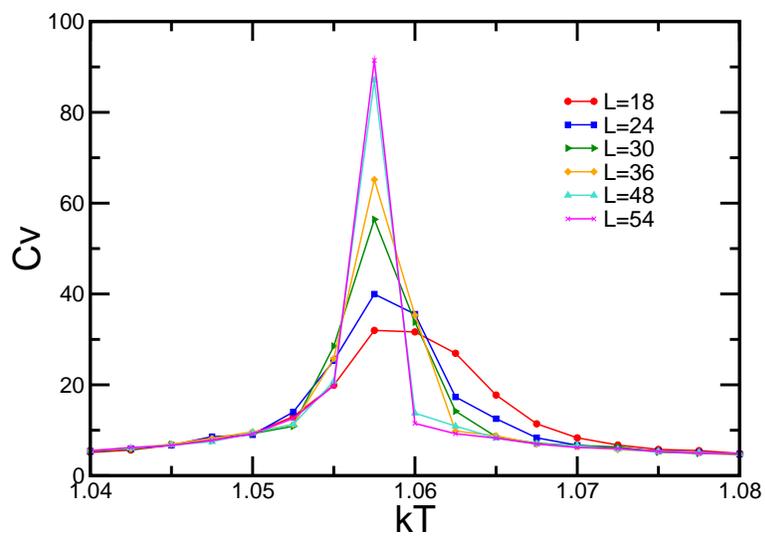

Figure 1.27: Specific heats for $q = 5$, $J > 0$.

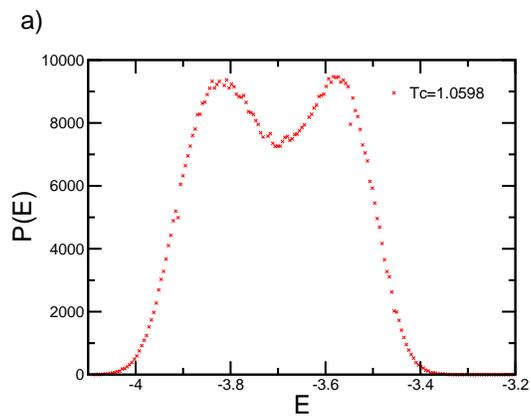
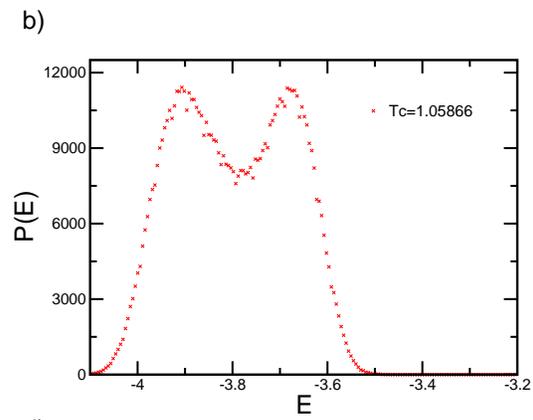
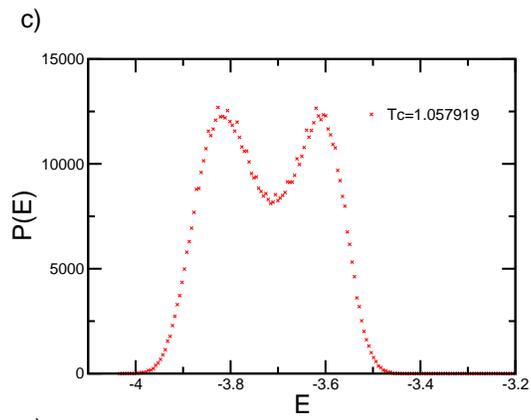
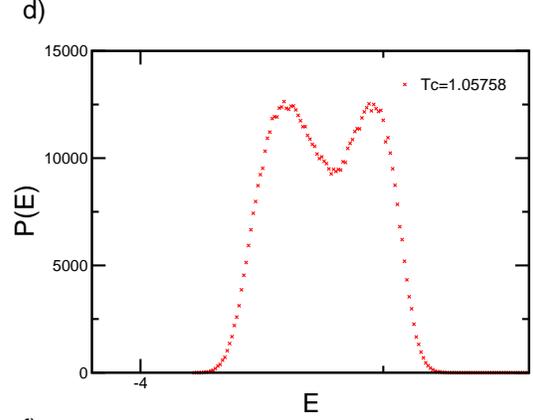
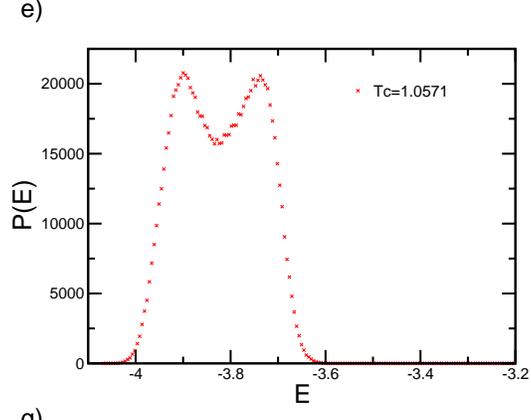
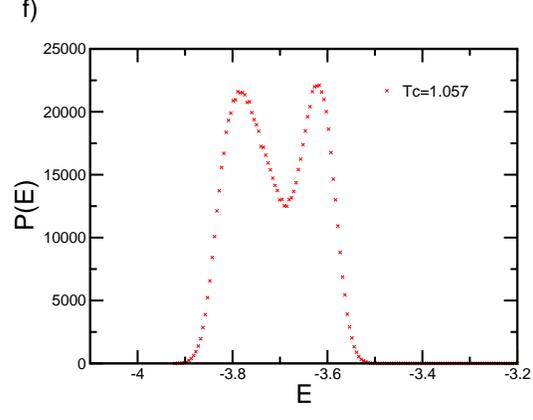
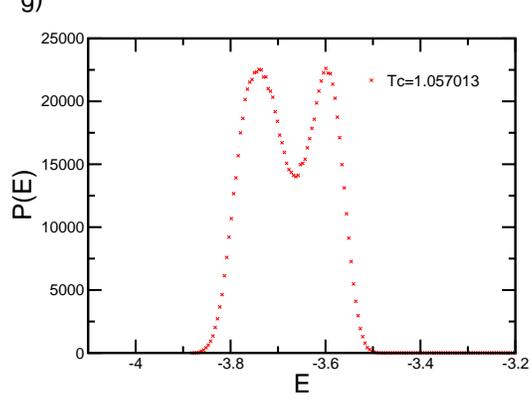

Figure 1.28: Energy histograms for lattice sizes a) $L = 18$, b) $L = 24$, c) $L = 30$, d) $L = 36$, e) $L = 48$, f) $L = 54$ and g) $L = 60$.

### 1.9.5 Conclusion

The results obtained with Monte Carlo simulations and finite size scaling techniques show clearly the kind of transition for each of the cases presented. The calculated critical exponents were near the theoretical values for second order phase transitions, except for the exponent $\beta$ in the case $q = 4$ that requires a more detailed analysis. Remember that for $J > 0$ the transition is supposed to be of second order for $q \leq 4$ and of first order for $q > 4$. As it lies at the border between the two, the case $q = 4$ is difficult to assess. Another aspect that must be taken into account for further analysis is the type of lattice, because it is quite probable that magnetic frustration effects could modify the magnetization-related critical coefficients. For first order transitions, the values of the free energy barriers could be estimated from the difference between the two peaks and the valley.

As we mentioned at the end of section 1.1, Monte Carlo simulations are a helpful tool in other areas. Thus, in the next chapter we will move on and review a cluster identification technique that involves Monte Carlo calculations for the analysis of biological data.

# Chapter 2

# Monte Carlo Simulations in Biology

**Contents**

## 2.1 Proteins, DNA and Gene Expression

Proteins are the complex molecules that make life possible. Keratin, essential in the structural conformation of our hair and nails, is one among the many proteins used as supporting materials in biology. Cells are also made of proteins, and enzymes, which are responsible for all chemical reactions inside living organisms, are proteins too. The information to produce the sequence of amino acids conforming certain protein resides in the DNA molecule, making it of great importance for life.

A single strand of DNA is formed by unities named nucleotides. Each nucleotide is composed by deoxyribose (a sugar formed by five carbons), a phosphate group, and one of the four possible nitrogenous bases: adenine (A), cytosine (C), guanine (G) and thymine (G). Phosphodiester bridges link the phosphate group of a nucleotide and the sugar of the next one, building in this way a chain of nucleotides. The DNA molecule is formed when two of these linear chains are joined by hydrogen bonds connecting the nitrogenous bases standing out of the sugar-phosphate backbone of each chain. All this chemical construction has a double helix structure envisioned in 1952 by Cricks and Watson [1]. The diameter of the helix is of 2 nm, and adjacent bases are separated by 0.34 nm along the helix axis. Hence, the helix repeats itself every 10 residues on each chain at intervals of 3.4 nm. Only specific pairs of bases can form hydrogen bonds: the purine base A always pairs with the pyrimidine base T through two hydrogen bonds, and the other pyrimidine base C always pairs with the purine base G this time by three hydrogen bonds [2]. A possible explanation for this situation is that two purines require more than 2 nm for connection, which does not fit within the diameter of the helix. On the other hand, there is too much space for two pyrimidines to get close enough to form hydrogen bonds between them [3]. The rules of base pairing tell us that if we can "read" the sequence of nucleotides on one strand of DNA, we can immediately infer the complementary sequence on the other strand. Thus, DNA looks like a chemical code based on four letters, each one corresponding to the four nitrogenous bases, aligned along a double-helicoidal chain. At the same time, the genetic code is a universal translation table formed by three-bases words called codons. Each codon codifies for a specific amino acid, so different sequences of codons build distinct proteins, and they can also build RNA molecules with a functional role. One can obtain 64 distinct combinations mixing the four nitrogenous bases in clusters of three letters. This quantity is enough to code for the 20 amino acids forming all proteins, and, as a matter of fact, almost all amino acids have more than one codon to codify them. Three of these triplets are left apart to code for chain termination to release proteins at the end of their production process. Moreover, one triplet is left as a signal to start the synthesis process. A segment of DNA sequence with the instructions to codify for a functional product, protein or RNA, is named a gene, and the genome is the collection of all the "recipes" for the products that an organism needs. The genome of a simple organism such as yeast has around 7,000 genes and the latest estimate for the human genome is of 25,000 genes [4]. It is important to stress that not all chain segments codify for proteins or RNA molecules: an overwhelming majority of human DNA (98%) contains non-coding regions (introns) that do not represent any particular functional product [5], although it is believed that they help to protect the genes.

Every cell of multicellular organisms has the entire set of information needed, but only some genes are expressed depending on the function of the cell. For example, cells in our retina need photosensitive molecules, whereas our liver do not. **A gene is expressed in a cell when the protein or RNA it codes for is synthesized**. The large majority of abundantly expressed genes are associated with common functions, such as metabolism, and hence are expressed in all cells. However, there will be differences in the expression profiles of different cells, and even in a single cell, expression will vary with time, in a manner dictated by external and internal signals that reflect the state of the organism and the cell itself [5].

Although DNA molecule contains all the instructions to make a huge amount of diverse proteins, DNA is not able to come out of the eucaryotic cell nucleus. Therefore, when certain protein is needed, another molecule called messenger RNA (m-RNA) is formed from DNA in a fundamental process called transcription, and is this molecule who travels outside the nucleus carrying the information. RNA is also a nucleic acid but is formed by a single of nucleotides, and its sugar (ribose) is slightly different from deoxyribose. Besides, RNA has uracil U instead of the base thymine T. In cells, one can find three important types of RNA: messenger

RNA (m-RNA), which transports the instructions to make a protein from the nucleus to the ribosomes, transfer RNA (t-RNA) which carries the amino acids to the ribosomes where the proteins are assembled and is found in the cytoplasm, and, finally, the r-RNA, or ribosomal RNA, which is one of the substances from what the ribosomes are made out [6].

In the transcription process, the portion of DNA containing the sequence for the needed product splits off, and then the free RNA nucleotides existing in the nucleus are attached to the exposed DNA nucleotides forming a complementary chain of RNA. This m-RNA chain comes out of the nucleus carrying a complementary sequence and arrives finally to the ribosome, where the sequence is translated into a protein(translation process). A cell may need a large number of some proteins and a small number of others, i.e. every gene may be expressed at a different level. The manner in which the instructions to start and stop transcription are given for a certain gene is governed by regulatory networks. Transcription is regulated by special proteins, called transcription factors, which bind to specific locations on the DNA, upstream from the coding region. Their presence at the right site initiates or suppresses transcription. The basic paradigm of gene expression analysis states that the biological state of a cell is reflected by its expression profile: the expression levels of all the genes of the genome. These in turn are reflected by the concentrations of the corresponding m-RNA molecules [5].

Several genomes of diverse organisms have been completed in the last years, the human genome, published on 2002, among them. Now, the main focus is to understand the underlying function and mechanisms behind these genomes. Some of the questions that remain unanswered included what are the functional roles of different genes, how genes are regulated, how do genes and gene products interact, how does gene expression level differ in various cell types and states, and how is gene expression changed by various diseases or compound treatments[7].

## 2.2 DNA Microarrays

Although m-RNA is not the ultimate product of a gene, transcription is the first step in gene regulation, and information about the transcript levels is needed as a first approach for understanding gene regulatory networks. DNA microarrays or DNA chips are one of the latest breakthroughs in experimental molecular biology precisely because they allow to monitor the expression of thousands of genes at the same time. The potential this technology is tremendous: monitoring gene expression levels in different developmental stages, tissue types, clinical conditions and different organisms can help understanding gene function and gene networks, assist in the diagnosis of disease conditions and reveal effects of medical treatments.

There are currently two main technologies that generate large-scale gene expression data: cDNA and oligonucleotide microarrays. CDNA microarrays contain large sets of complementary DNA sequences several hundred bases long, each set representing a gene, immobilized on a solid substrate. In oligonucleotide microarrays, each gene is represented on the array by a set of 15-20 different oligonucleotide probes designed to hybridize perfectly to some particular sequence, and some mismatch control oligonucleotides, identical to the perfect match probes except for a single base-pair mismatch. These mismatch control oligonucleotides allow estimation of cross-hybridization, improving reproducibility and accuracy of RNA quantification, and reducing the rate of false-positives. In general, oligonucleotides used consist about 20-25 nucleotides, and are synthesized in situ with photolithography techniques [8], [9]. In brief, functioning of microarrays is based on the preferential binding of complementary single stranded nucleic acid sequences, and a single microarray may contain tens of thousands of spots. One of the most popular experiments involving cDNA microarrays consists in compare m-RNA abundance in two samples. The m-RNA molecules are extracted from cells taken from two tissues of interest (e.g. tumour and normal tissues) [10]. They are reverse transcribed from RNA to DNA and their concentration is enhanced. The resulting DNA is transcribed back into fluorescently marked single strand RNA. For example, tumour tissue is labeled with a red dye and normal tissue with a green one. The solution of marked and enhanced m-RNA molecules ("targets") that are copies of the m-RNA molecules originally extracted from the tissue, is placed on the chip and diffuses over the collection of single strand DNA probes.

When an m-RNA encounters a part of the gene of which it is a perfect copy, it hybridizes to it with a high affinity (considerably higher than with a bit of DNA of which it is not a perfect copy) and when the m-RNA solution is washed off, only those molecules that found their perfect match remain stuck to the chip. Next, the chip is illuminated with a laser, and the stuck targets fluoresce. If RNA from tumour tissue is in abundance, the spot will emit red light, but if instead RNA from normal tissue is in abundance, it will appear green. If tumour and normal RNA bind equally, the spot will be yellow, while if neither binds, it will not fluoresce and appear black [7]. Therefore, by measuring the light intensity emanating from each spot, one obtains a measure of the number of targets that stuck, which, in turn, is proportional to the concentration of these m-RNA in the investigated tissues. CDNA microarrays are a differential technique because only ratios between both fluorescence wavelengths give meaningful information and hence, only relative expressions levels are obtained. On the other hand, with oligonucleotide arrays, absolute expression levels are measured [11]. The characterization of genes expressed differently in normal and their corresponding tumour cells has been particularly important [12]. Arrays have also being used to discover transcribed regions in genomic DNA [13]; to detect polymorphism in copy number of regions of the genome[14], which may be a new and important class of mutation; and to analyze amplifications and deletions that are associated with oncogenic transformation and some inherited conditions [15], [16]. A number of diagnostic applications for arrays have been suggested. The first to be granted approval by the US Food and Drug Administration is the Roche AmpliChip for cytochrome P450. This test will help doctors determine an individual's genotype to determine appropriate drugs and doses to prescribe, minimizing harmful drug reactions [17].

## 2.3 Gene Clustering

Microarray data analysis can be divided into two general classes: supervised and unsupervised analysis. The supervised approach assumes that for some (or all) profiles we have additional information, such as functional classes for the genes, or diseased/normal states attributed to the samples. We can viewed this additional information as labels attached to the rows or columns. Having this information, a typical task is to build a classifier able to predict the labels from the expression profile. If a clasiffier that is able to distinguish between two different, but morphologically closely related tumour tissues, can be constructed, such a classifier can be used for diagnosis. Classifiers are trained on a subset of data with a priori given classification and tested on another subset with known classification. After assessing the quality of the prediction they can be applied to data with unknown classification [7]. Unsupervised data analysis consists on clustering expression profiles to find groups of co-regulated genes or related samples. An example of these two kind of clustering combined to predict clinical outcome of breast cancer can be seen at [18].

Some short DNA sequences in or around the gene, specifically in the promoter region, serve as switches that control gene expression. Special proteins (transcription factors) interact with these binding sites, and represses or activates the transcription process of the gene [11]. Various genes that share common functions or the same regulatory mechanism at the sequence level, can have the same binding site sequence. As a result, similar expression patterns can correspond to similar binding site patterns. Then, a key step in the analysis of gene expression data is the detection of groups of genes that exhibit similar behaviour of expression patterns (i.e. are coexpressed), based on the idea that these genes share common regulatory or functional roles, assumption that has proved right in many experiments (see for example [19]).

The challenge is then transformed into the problem of clustering genes into groups based on their similarity in expression profiles. Instead of clustering genes, experimental conditions can also be clustered, the task being now to find groups of experimental conditions (which can be, for example, tumour samples) across which all the genes behave similarly. This type of clustering can be helpful for diagnosis. In gene expression, elements to be clustered are usually genes and the vector of each gene is its expression pattern; similarity between genes can be measured in various ways that are problem dependent, for example by the correlation coefficient between vectors. The goal is to partition the elements into subsets, i.e. clusters, so that elements in the same cluster are highly similar to each other and elements from different clusters have low similarity to each other.

An essential step to obtain an effective cluster analysis is the preprocessing of the initial data. Although this work does not attempt to give a detailed explanation of the various preprocessing steps, the most common ones are mentioned in the following. The first step is the normalization of the hybridization intensities within a single experiment or across experiments. Besides, expression ratios are not symmetrical in the sense that upregulated genes have expression ratios between one and infinity, while downregulated genes have expression ratios squashed between one and zero. Taking the logarithms of these expression ratios results in symmetry between expression values of up- and downregulated genes. This preprocessing step is called nonlinear transformation of the data. Other, but uncommonly used transformations, include square, square root, and inverse transformations. Missing value replacement is another step, and is made only in the case of using a clustering algorithm that are not able to handle missing values due to technical reasons in the data. Some genes do not really contribute to the biological process because their expression values show little variation over the different experiments, and another problem are expression profiles with a considerable number of missing values. This non desirable data is removed in the filtering process. Biologists are mainly interested in grouping gene expression profiles that have the same relative behaviour, i.e. genes that are up- and downregulated together. Genes showing the same relative behaviour but with diverging absolute behaviour will have a relatively high Euclidean distance, and cluster algorithms based on this distance measure will therefore wrongfully assign the genes to different clusters. This can be prevented by applying standardization or rescaling to the gene expression profiles so that they have zero mean and unit standard deviation. ([11]).

In a typical experiment to monitor gene expression levels several DNA chips are used, and since each DNA chip contains thousands of spots, a huge amount of information is obtained. These results are summarized in a $G \times S$ expression table, in which $G$ represents the number of genes placed on every chip and $S$ is the number of DNA chips used (each chip accounting for different conditions, experiments, time points or samples). Therefore, each row on this matrix corresponds to one particular gene and each column to a different sample. Each element $E_{gs}$ of the matrix represents the expression level of gene $g$ in sample or condition $s$. Each column is called the profile of that condition, and each row vector is the expression pattern of a gene across all the conditions, commonly named expression profile. If the input data for a clustering problem is given in this form, it said to be as fingerprint data. Other type of input data is similarity data, where pairwise similarity values between elements are used. These values can be computed from fingerprint data. Alternatively, the data can represent pairwise dissimilarity. Fingerprints contain more information than similarity values, but the latter can be used to represent the input to clustering in any application. Moreover, the fingerprint matrix is of order $G \times S$ while the similarity matrix is of order $G \times G$. It is important to note that in gene expression applications typically $G >> S$, while in tissue classification applications often $G << S$ [20].

We need a way to measure the similarity (or distance) between the genes or samples being compared and clustered. We can regard these rows or columns in the matrix as points in $n$-dimensional space or as $n$-dimensional vectors, where $n$ is the number of samples for gene comparison, or number of genes for sample comparison. The natural, so called Euclidean distance between these points in the $n$-dimensional space may be the most obvious, but not necessarily the best choice. There is no theory how to choose the best distance measure. Possibly one right distance measure in the expression profile space does not exist, and the choice should depend on the problem studied [7]. Some distance metrics commonly applied are the following:

1. Pearson correlation. The Pearson correlation $r$ is the dot product of two normalized vectors (i.e. the cosine between two vectors). It measures the similarity in the shapes of two profiles, while not taking the magnitude of the profiles into account, and therefore suits well the biological intuition of coexpression.

2. Squared Pearson correlation. This is the square of the Pearson correlation, which considers two vectors pointing in the exact opposite directions to be perfectly similar, which might also be interesting for biologists (because repression is a form of coexpression).

3. Euclidean distance. Euclidean distance measures the length of the straight line connecting the two points. It measures the similarity between the absolute behaviours of genes, while the biologists are more interested in their relative behaviours. Thus, a standardization procedure is needed before clustering using Euclidean distance. Importantly, after standardization, the Euclidean distance between two points $x$ and $y$ is related to the Pearson correlation by $|x-y|^2 = 2(1-|r|)$.

4. Jackknife correlation. The jackknife correlation is an improvement for the Pearson correlation (which is not robust to outliers). Jackknife correlation increases the robustness to single outliers by computing a collection of all the possible leave-one-(experiment-)out Pearson correlations between two genes and then selecting the minimum of the collection as the final measure for the correlation.

The first generation of cluster algorithms used for gene expression profiles were developed for purposes did not related with biological research (e.g. hierarchical clustering, K-means and self organizing maps(SOM)). Although it is possible to obtain biologically meaningful results with these algorithms, some of their characteristics often complicate their use for clustering expression data. More recently, new algorithms have been developed specifically for gene expression profile clustering to overcome some of the limitations of earlier methods. These algorithms include, among others, model-based algorithms, the self-organizing tree algorithm (SOMA), quality-based algorithms, and biclustering algorithms. Also, some procedures have been developed to help biologists estimate some of the parameters needed for the first generation of algorithms, such as the number of clusters present in the data. While it is impossible to give an exhaustive description of all clustering algorithms developed for gene expression data, we try here to illustrate some of them.

### 2.3.1 Hierarchical Clustering

Agglomerative or hierarchical clustering algorithms ([21]) are among the oldest and most widely used clustering methods applied to gene expression data. Typically, the algorithm takes each expression profile as one cluster at the beginning. Then computes the distance between every pair of clusters, and the pair of clusters with the minimum distance is merged; the procedure is carried on iteratively until all elements ends into one single cluster. The whole clustering process is presented as a tree called a dendrogram and the original data are often reorganized in a heat map demonstrating the relationships between genes or conditions. After the full tree is obtained, the determination of the final clusters is achieved by cutting the tree at a certain level or height, which is equivalent to putting a threshold on the pairwise distance between clusters. The decision of the final clusters is arbitrary, because it is difficult to predict which level will give the best biological results. Note that the memory complexity of hierarchical clustering is quadratic in the number of gene expression profiles, which can be a problem due to the large number of genes involved in experiments.

As in every step of agglomerative clustering, the two subsets that are closest or more similar to each other are merged, the distance between two clusters has to be defined. There are four common options:

1. Single linkage. The distance between two clusters is taken as the distance between the two closest data points, each point belonging to a different cluster.

2. Complete linkage. The distance between the two furthest data points, each one in a different cluster.

3. Average linkage. Both single linkage and complete linkage are sensitive to outliers. Average linkage provides an improvement by defining the distance between two clusters as the average of the distances between all pairs of points in the two clusters.

4. Ward's method. At each step of agglomerative clustering, instead of merging the two clusters that minimize the pairwise distance between clusters, Ward's method ([22]) merges the two clusters that minimize the "information loss" for the step. The "information loss" is measured by the change in the sum of squared error of the clusters before and after the merge. In this way, Ward's method assesses the quality of the merged cluster at each step of the agglomerative procedure.

These methods yield similar results if the data consist of compact and well separated clusters. However, if some of the clusters are close to each other or if the data have a dispersed nature, the results can be quite different. Ward's method, although less well known, often produces the most satisfactory results [23].

Eisen et al. developed a clustering software package based on average linkage hierarchical clustering [19]. The clustering program is called Cluster, and the accompanying visualization program is called TreeView. Both programs are available at http://rana.lbl.gov/EisenSoftware.htm.

### 2.3.2 K-Means Clustering

K-means clustering ([24], [25]) is a simple and popular partitioning method for data analysis. The number of clusters K in the data is needed as an input for the algorithm. K-means starts by assigning at random all gene expression profiles to one of the K clusters. Iteratively, the center, which is nothing more than the average expression vector of each cluster, is calculated and then the gene expression vectors are reassigned to the cluster with the closest cluster center. The initial mean vectors are called the seeds. The iterative procedure converges when all the mean vectors of the clusters remain stationary or the given number of iterations is exceeded. Since it is difficult to predict the number of clusters in advance, the predefinition of the number of clusters by the user is arbitrary. In practice, this implies the use of a trial-and-error approach where a comparison and biological validations of several runs of the algorithm with different parameter settings are necessary ([11]). Another parameter that influence the result of K-means clustering is the choice of the seeds. The algorithm suffers from the problem that with different seeds the algorithm can yield very different result.

### 2.3.3 Self-Organizing Maps

SOM ([26]) is a technique developed by Kohonen for fitting a number of reference vectors to the distribution of gene data, by means of a set of nodes. The nodes are the intersections of a two-dimensional grid (usually of hexagonal or rectangular geometry). In the high-dimensional input space (with the gene expression vectors), each node represents a reference vector (similar to the mean vectors in the K-means algorithm). The dimension of the grid (e.g. lattice of 6x5 nodes) needs to be specified a priori. The initial position of the reference vectors is randomly chosen, and then the algorithm selects a random data vector $p$, identifies the node $n_p$ whose reference vector is closest to $p$, and updates the position of all reference vectors towards $p$ according to a predefined learning function. The amount of position adjustment determined by the learning function decreases as the distance between $n$ and $n_p$ (in the grid) and the iteration number grow. The intuition for this learning process is that the reference vectors that are close enough to $p$ will be pulled towards it, and the stiffness of the grid structure will propagate some of impact to neighbouring nodes. As a result, a reference vector is pulled more towards input vectors that are closer to the reference vector itself and is less influenced by the input vectors located further away. In the meantime, this adaptation procedure of reference vectors is reflected on the output nodes (nodes associated with similar reference vectors are pulled closer together on the output grid). The algorithm terminates when convergence of the reference vectors is achieved or after completing a pre-defined number of training iterations.

Because of the advantage in visualization, choosing the geometry of the output grid is not as crucial a problem as the choice of the number of clusters for a K-means method. Like the K-means method, the initial choice of reference vectors remains a problem that influences the final clustering result of SOM clustering. A good way to seed the reference vectors is to use the result from a principal component analysis(PCA) [23].

Tamayo et al. [27] devised a gene expression clustering software, GeneCluster, which uses the SOM algorithm. The software is available at
http://www.broad.mit.edu/cancer/software/genecluster2/gc2.html.

### 2.3.4 Self-Organizing Tree Algorithm

SOTA combines SOM and hierarchical clustering techniques. As in SOM, SOTA maps the input gene profiles to an output space of nodes. However, the nodes in SOTA, instead of being in a two-dimensional grid, are in the topology (or geometry) of a binary tree. The number of nodes in SOTA is not fixed from the beginning (in contrast to SOM) because the tree structure of the nodes grows during the clustering procedure.

The initial system is composed of two external elements, denoted as cells, connected by an internal element that its called node, like a tree with two leaves. Each cell (or node) is a reference vector with the same size as the gene profiles. In the beginning, the entries of the two cells and the node are randomly initialized. The series of operations performed until a cell generates two descendants is called a cycle. During a cycle, cells and nodes are repeatedly adapted by the input gene profiles. Adaptation in each cycle consists on the presentation of all expression profiles to the network, and this implies two steps: first, each gene profile is associated with the cell whose reference vector is located closest to it, and second, the reference vector of this cell and its neighbouring nodes, including its parent node and its sister cell, are updated based on some neighbourhood weighting parameters (which perform the same role as the neighbourhood function in SOM). Thus, a cell is moved into the direction of the expression profiles that are associated with it. The network follows its growing process by replicating the cell whose associated profiles exhibits the highest heterogeneity, i.e., the largest variability (defined by the maximal distance between two profiles that are associated with the same cell). This cell gives rise to two new descendants cells and become a node. The values of the two new cells are identical to the node that generate them and the whole procedure starts again. The growing process ends when the heterogeneity of the system falls below a threshold. This threshold can be set to zero for a fully resolved dendogram similar to that provided by hierarchical clustering. If the threshold is obtained from the randomized distribution of data, SOTA will provide the cluster hierarchy that minimizes the probability of having missassigned genes to them [28].

SOTA has two crucial advantages: the topology is that of a hierarchical tree, and the clustering obtained is proportional to the heterogeneity of the data, instead of the number of items (this is due to the fact that SOTA is distribution preserving while SOM is topology preserving). In both SOM and SOTA, the training process changes the vectors in the nodes to weighted averages of the gene expression patterns associated to them. The advantage in the case of SOTA is that the binary topology produces a nested structure in which nodes at each level are averages of items below them (items that can be nodes or in the case of terminal nodes, genes). This makes it straightforward to compare average patterns of gene expression at different hierarchical levels even for large data sets [28].

### 2.3.5 Model Based Clustering

Model Based Clustering assumes that the data are generated by a finite mixture of underlying probability distributions such as multivariate normal distributions. In this case, each cluster $C_i$ is represented by a multivariate Gaussian model $p_i$ in $d$ dimensions:

$$p_i(y_j|\mu_i, \sum_i) = \frac{1}{(2\pi)^{d/2}|\Sigma_i|^{d/2}} e^{-1/2(y_j-\mu_i)^T (\Sigma_i)^{-1}(y_j-\mu_i)}, \qquad (2.1)$$

where $y_j$ is an expression profile and $\mu_i$ and $\Sigma_i$ the mean and covariance matrix of the multivariate normal distribution respectively [11].

The covariance matrix $\Sigma_i$ can be represented by its eigenvalues decomposition, which in general takes the following structure:

$$\sum_i = \lambda_i D_i A_i D_i^T, \qquad (2.2)$$

where $D_i$ is the orthogonal matrix of the eigenvectors of $\Sigma_i$, $A_i$ is a diagonal matrix whose elements are proportional to the eigenvalues of $\Sigma_i$, and $\lambda_i$ is the constant of proportionality. This decomposition implies a nice geometric interpretation of the clusters: $D_i$ controls the orientation, $A_i$ controls the shape, and $\lambda_i$ controls the volume of the cluster. Note that simpler forms of the covariance structure can be used (e.g., by having some of the parameters take the same values across clusters), thereby decreasing the number of parameters that have to be estimated but also decreasing the model flexibility (capacity to model more complex data structures).

The mixture model $p$ itself takes then the following form:

$$p(y_j) = \sum_{i=1}^{K} \pi_i p_i(y_j|\mu_i, \Sigma_i), \qquad (2.3)$$

where $K$ is the number of clusters and $\pi_i$ is the prior probability that an expression profile belongs to cluster $C_i$ so that:

$$\sum_{i=1}^{K} \pi_i = 1. \qquad (2.4)$$

In practice we would like, given a collection of expression profiles $y_j (j = 1,...,n)$, to estimate all the parameters $(\pi_i, \mu_i, \Sigma_i (i = 1,...,K)$, and $K$ itself) of this mixture model. In a first step $\pi_i, \mu_i, \Sigma_i (i = 1,...,K)$ are estimated with an EM algorithm using a fixed value for $K$ and a fixed covariance structure [11]. In the EM algorithm, the Expectation steps and Maximization steps alternate. In the E step, the probability of each observation belonging to each cluster is estimated conditionally on the current parameter estimates. In the M step the model parameters are estimated given the current group membership probabilities. When the EM algorithm converges, each observation is assigned to the group with the maximum conditional probability [29]. This parameter estimation is then repeated for different values for $K$ and different covariance structures. The result is thus a collection of different models fitted to the data and all having a specific value for $K$ and a specific covariance structure. In a second step the best model in this group of models is selected (i.e., the most appropriate number of parameters and a covariance structure is chosen here). This model selection step involves the calculation of the Bayesian Information Criterion (BIC) for each model [30], which is not further discussed here.

A good implementation for model based clustering (called MCLUST) is available at www.stat.washington.edu/fraley/mclust. Yeung et al. reported good results using this software on several synthetic data sets and real expression data sets. McLachlan et al. [31] have also implemented model-based clustering in a Fortran program called EMMIX, which is also freely available from the web at http://www.maths.uq.edu.au/g̃jm/emmix/EMMIX.f.

### 2.3.6 Quality-Based Algorithms

Quality-based algorithms produces clusters with a quality guarantee that ensures that all members of a cluster are coexpressed (this property is called transitivity). This concept was introduced by Heyer, Kruglyak and Yooseph, ([32]) and their implementation is called QT _ Clust. The quality of a cluster $C$ is defined as a diameter (equal to $1 - min_i, j \in s_{ij}$, where $s_{ij}$ is the jackknife correlation between expression profiles $i$ and $j$), but the method can be easily extended to other definitions.

The algorithm considers every expression profile in the data set as a cluster seed (one could also call this a cluster center) and iteratively assigns the expression profiles to these clusters that cause a minimal increase in diameter until the diameter threshold, i.e., quality guarantee, is reached. Note that at this stage every expression profile is made available to every candidate cluster and that there are as many candidate clusters as there are expression profiles. At this point, the candidate cluster that contains the most expression profiles is selected as a valid cluster and removed from the data set where after the whole process starts again. The algorithm stops when the number of points in the largest remaining cluster falls below a threshold. Note that this stop criterion implies that the algorithm will terminate before all expression profiles are assigned to a cluster.

This approach has some advantages, for example it is possible to find clusters containing highly coexpressed genes, and these clusters might therefore be good seeds for further analysis. Moreover, genes not really coexpressed with other members of the data set are not included in any of the clusters. Some disadvantages are that the quality guarantee of the clusters is a user defined parameter hard to estimate, it is hard to use by biologists, needs extensive parameter fine-tuning, and produces clusters all having the same fixed diameter not optimally adapted to the local data structure [11].

### 2.3.7 Adaptive Quality-Based Clustering

Adaptive quality-based clustering ([33]) consist of a two-step approach. In the first step, a quality-based approach is performed to locate a cluster center in an area where the density of gene expression profiles is locally maximal using a preliminary estimate of the radius (i.e. the quality) of the cluster. In the second step, called adaptive step, the algorithm re-estimates the radius of the cluster so that the genes belonging to it are, in a statistical sense, significantly coexpressed. To this end, a bimodal and one-dimensional probability distribution (the distribution consists of two terms: one for the cluster and one for the rest of the data) describing the Euclidean distance between the data points and the cluster center is fitted to the data using an expectation-maximization (EM) algorithm.

Finally, step one and two are repeated, using the re-estimation of the quality as the initial estimate needed in the first step, until the relative difference between the initial and re-estimated quality is sufficiently small. The cluster is subsequently removed from the data and the whole procedure is restarted. Note that only clusters whose size exceeds a predefined number are presented to the user.

In adaptive quality-based clustering, users have to specify a threshold for quality control. This parameter has a strict statistical meaning and is therefore much less arbitrary (in contrast to the case in QT_Clust). It can be chosen independently of a specific data set or cluster and it allows for a meaningful default value (95%) that in general gives good results. This makes the approach user friendly without the need for extensive parameter fine-tuning. Furthermore, with the ability to allow the clusters to have different radius, adaptive quality-based clustering produces clusters adapted to the local data structure[11]. An application of Adaptive Quality- Based Clustering to nervous system is found in [34].

However, the method has some limitations like it does not converge in every situation. A server running the program is available at http://homes.esat.kuleuven.be/˜thijs/Work/Clustering.html

### 2.3.8 Biclustering and Some Physics Related Algorithms

Clustering can be applied to either the rows or the columns of the data matrix, separately. Biclustering, on the other hand, performs clustering in these two dimensions simultaneously. This means that clustering derives a global model while biclustering produces a local model[35]. The term biclustering was first used by Cheng and Church [36] in gene expression data analysis. It refers to a distinct class of clustering algorithms that perform simultaneous row-column clustering. One of the earliest biclustering formulations is the direct clustering algorithm introduced by Hartigan [24], also known as block clustering.

The goal of biclustering techniques is thus to identify subgroups of genes and subgroups of conditions, by performing simultaneous clustering of both rows and columns of the gene expression matrix, instead of clustering these two dimensions separately. We can then conclude that, unlike clustering algorithms, biclustering algorithms identify groups of genes that show similar activity patterns under a specific subset of experimental conditions [35].

There are also several physics related clustering algorithms, e.g. Deterministic Annealing [37] and Coupled Mass [38]. Deterministic Annealing uses the same cost function as K-means, but rather than minimizing it for a fixed value of clusters K, it performs a statistical mechanics type analysis, using a maximum entropy principle as its starting point. The resulting free energy is a complex function of the number of centroids and their locations, which are calculated by a minimization process. This minimization is done by lowering the temperature variable slowly and following minima that move and every now and then split (corresponding to a second order phase transition). Since it has been proven [39] that in the generic case the free energy function exhibits first order phase transitions, the deterministic annealing procedure is likely to follow one of its local minima [5].

Finally, it is important to stress that clustering methods have been used in a large variety of scientific disciplines and applications that include pattern recognition [40], learning theory [41], astrophysics [42], medical images and data processing [43], machine translation of text [44], satellite data analysis [45], as well as speech recognition [46].

## 2.4 Superparamagnetic Gene Clustering: Monte Carlo Simulations

This method takes the data points generated by gene expression profiles as sites of an inhomogeneous Potts ferromagnet, and was first proposed by Eytan Domany et al. [47]. The presence of clusters in the data gives rise to magnetic grains, and working in the superparamagnetic phase, the SPC algorithm decides if a data point belong to the same grain using the pair correlation function of the Potts spins. Additionally, temperature controls the level of resolution obtained.

A Potts system is said to be homogeneous when its spins are on a lattice and all nearest neighbour couplings are equal, $J_{ij} = J$. This system exhibits two phases, at high temperatures is paramagnetic or disordered, and at low temperatures is ordered. In the disordered phase the correlation function $G_{ij}$ decays to $1/q$ when the distance between points $v_i$ and $v_j$ is large (remember from last chapter, that $q$ is the number of possible states in the Potts model). This is the probability to find two completely independent Potts spins in the same state. At very high temperatures even neighbouring sites have $G_{ij} \approx 1/q$. As the temperature is lowered, the system undergoes a sharp transition to an ordered, ferromagnetic phase, meaning that one Potts state dominates the system. At very low temperatures $G_{ij} \approx 1$ for all pairs $v_i, v_j$, i.e. all spins have the same $q$ [48].

In strongly inhomogeneous Potts models, spins form magnetic grains with very strong couplings between neighbours that belong to the same grain, and very weak interactions between all other pairs. At low temperatures such a system is also ferromagnetic, but as the temperature is raised the system may exhibit an intermediate, super-paramagnetic phase. In this phase strongly coupled grains are aligned (i.e. are in their respective ferromagnetic phases), while there is no relative ordering of different grains. This is illustrated in Fig. 2.1.

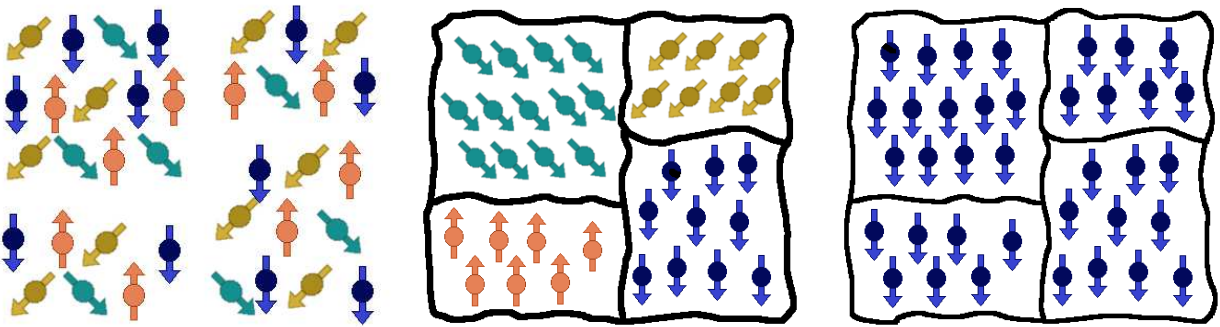

Figure 2.1: At high T all sites have different spin values, but as T is lowered, regions of aligned spins appears (superparamagnetic phase). At low T, the system is completely ordered.

At the transition temperature from the ferromagnetic to super-paramagnetic phase a pronounced peak of $\chi$ is observed [47]. As the temperature is further raised, the super-paramagnetic to paramagnetic transition is reached; each grain disorders and $\chi$ abruptly diminishes by a factor that is roughly the size of the largest cluster. Thus the temperatures where a peak of the susceptibility occurs and the temperatures at which $\chi$ decreases abruptly indicate the range of temperatures in which the system is in its super-paramagnetic phase. In principle, one can have a sequence of several transitions in the super-paramagnetic phase: as the temperature is raised the system may break first into two clusters, each of them in turn breaks into more (macroscopic) sub-clusters and so on. Such a hierarchical structure of the magnetic clusters reflects a hierarchical organization of the data into categories and sub-categories [49].

In concreteness, SPC method consists on three stages. First, to specify the Hamiltonian which governs the system. Second, find the temperature range where the superparamagnetic phase take place, taking into acount the susceptibility behaviour. Finally, the correlation of neighbouring pairs of spins, $G_{ij}$ is measured and, taking into account these values, the clusters are formed.

### 2.4.1 Detailed Description of SPC

Each expression profile is represented as a point in a $D$ dimensional space, and a random spin value $\sigma_i$, $i = 1, 2, ..., q$ is assigned to it. A small value $q$ hinders the identification of the SPM clusters since different clusters are then forced to point into the same Potts direction. Too large $q$ makes the calculations more cumbersome. However, the results depend only weakly on the value of $q$. In the next step, the neighbours of each spin $v_i$ are calculated using the $K$ mutual neighbour criterion. This criterion initially calculates the $K$ nearest points of each site. If $v_i$ has $v_j$ among its $K$ nearest points, and $v_j$, in turn, has $v_i$ as one of its $K$ nearest points, then $v_i$ and $v_j$ are considered as neighbours.

The average number of neighbours $\hat{K}$ and the average of all distances $a$ between neighbouring pairs $v_i$ and $v_j$ are then computed, and finally the interaction couplings which will appear in the Hamiltonian will be calculated as follows:

$$J_{ij} = \begin{cases} \frac{1}{\hat{K}} e^{-\frac{d_{ij}^2}{2a^2}} & \text{if } v_i \text{ and } v_j \text{ are neighbours} \\ 0 & \text{otherwise} \end{cases} \quad (2.5)$$

Choosing $J_{ij}$ in this way creates strong interactions between spins associated with the data from high density regions, and weak interactions between neighbours that are in low density regions [50].

Any different assignment of spins to data points $S$ has an energy cost given by:

$$H(S) = \sum_{i,j} J_{ij} \delta_{\sigma_i, \sigma_j}, \quad (2.6)$$

where the sum is over neighbouring sites. The function $\delta_{\sigma_i, \sigma_j}$ is the Kroenecker symbol taking the value 1 when $\sigma_i = \sigma_j$ and 0 otherwise. The lowest possible energy cost, $H(S) = 0$ is attained when we assign the same spin to all points, which corresponds to all data points being assigned to the same cluster. Moreover, as one chooses interactions that are a decreasing function of the distance $d_{ij}$, then the closer two points are to each other, the more likely is for them to be in the same state. In summary, this Hamiltonian procedure penalizes placing spins at points $i, j$ in different clusters, and this penalty decreases with the distance between the points [49].

The next step is the calculation of magnetization, susceptibility and correlation function for pairs of neighbours $G_{ij}$ over a range of temperatures using Monte Carlo technique. The original creators of SPC used the Swendsen Wang algorithm.

As the temperature increases, $M$ varies from 1 to 0 via sharp phase transitions. At low temperatures the system is fully magnetized and the fluctuations in $m$ are negligible. As $T$ increases to the point where the single cluster breaks into subclusters (or become completely disordered), fluctuations become very large. Hence, one expect to identify the transitions at which clusters break up by the sharp peaks of the susceptibility [51].

The strategy is to vary $T$ and measure $\chi(T)$. Transitions show up as peaks of $\chi$. At temperatures between transitions, we expect to observe relatively stable phases that correspond to some clusters being ordered internally and uncorrelated with other clusters. Within each such phase, $G_{ij}$ is measured. The value of $G_{ij}$ is the probability to find the two Potts spins $\sigma_i$ and $\sigma_j$ in the same state, i.e. the probability to find them in the same cluster. By the relation to granular ferromagnets we expect that the distribution of $G_{ij}$ is bimodal; if both spins belong to the same ordered grain (cluster), their correlation is close to 1; if they belong to two clusters that are not relatively ordered, the correlation is close to 0. Rather than thresholding the distances between pairs of points to decide their assignment to clusters, we use the pair correlations, which reflect a collective aspect of the data's distribution [49].

Clusters are identified in three steps:

1. Build the cores of the clusters using a thresholding procedure. If $G_{ij} > 0.5$, a link is set between the neighbour data points $v_i$ and $v_j$. The resulting connected graph depends weakly in the value used in this thresholding, as long as it is bigger than $1/q$ and less than $1 - 2/q$ [49]. The reason is that the distribution of the correlations between two neighbouring spins peaks strongly at these two values and is very small between them.

2. Capture points lying in the periphery by linking each point to its neighbour of maximal correlation. Of course, some points were already linked in step one.

3. Data clusters are identified as the linked components of the graph obtained in the previous steps.

The temperature controls the resolution at which the data are clustered.

It is intuitively clear that if a set of data points form a dense cloud, isolated from the rest of the data, the corresponding spins will form a ferromagnetic domain at some low temperature, which will become paramagnetic and lose its correlations only at a high temperature. Hence the size of the temperature interval $dT$ in which such a ferromagnetic domain exists can be used as a measure of the stability and significance of the corresponding data cluster.

Some of the demonstrated useful properties of SPC are the following: (a) the number of clusters is determined by the algorithm itself and not externally prescribed (as is done by SOM and K-means); (b) presents stability against noise; (c) generates a hierarchy (dendrogram) and provides a mechanism to identify in it robust, stable clusters (by the value of dT ); (d) ability to identify a dense set of points forming a cloud of an irregular (non-spherical shape) as a cluster [5].

The SPC method has been used in various contexts, like computer vision [52], speech recognition [49] and identification of clusters of companies in stock indices [53]. Its first direct application to gene expression data has been for analysis of the temporal dependence of the expression levels in a synchronized yeast culture [54], identifying gene clusters whose variation reflects the cell cycle. Subsequently, the SPC was used to identify primary targets of p53 [55], the most important tumour suppressor that acts as a transcription factor of central importance in human cancer. SPC has been used also to cluster protein sequences [56], and to classify or identify new genes associated with colon and skin cancer [57].

### 2.4.2 Future Directions

The location of the superparamagnetic phase in the SPC algorithm is closely related to the phase transitions occurring in the system. The introduction of the Wolff algorithm instead of the originally used Swendsen-Wang algorithm will probably improve the efficiency of the method, and this is left for future investigations, as well as a comparison of different methods with the SPC algorithm.

# Chapter 3

# Gompertz Equation

**Contents**

## 3.1 History of Gompertz Equation

In 1724 Moivre presented his hypothesis of uniform decrement, summarized in the expression $y(x) = K(w-x)$, where $y(x)$ represents the surviving persons with age $x$, $K$ is the slope or velocity with which the population diminishes in the mortality table, and $w$ is the maximum survival age for the population. The Moivre straight line was recommended for an age range between 12 and 86, in which it adjusted better. This linear hypothesis was exceeded by Benjamin Gompertz, who believed in the existence of two general causes of mortality: chance and the increasing inability of men to avoid death. Gompertz took into account only the biological causes, and his hypothesis was based on the following idea: "Men resistance to death diminish with time in a proportional rate" [1].

Benjamin Gompertz was a British mathematician interested, besides other subjects as astronomy, in the problem of life insurances and mortality rates in the nineteenth century. He worked with death and population records of people in England, Sweden and France between ages 20 and 60 and noted that the arithmetic increases in age were consistently accompanied by geometric increases in mortality, and that this law of geometrical progression appeared in large portions of the different tables of mortality. Nowadays, the simple formula describing the exponential rise in death rates between sexual maturity and extreme old age, $\gamma(t) = e^{\gamma t}$ is better known as Gompertz equation. In his first paper about this subject published in 1820 [2], Gompertz identified this peculiar pattern among different european populations for a limited portion of the age range. For his second paper, Gompertz used equal intervals of longer periods of time than in his previous work and found, for example, that the differences in the natural logarithm between successive 10 year age intervals between ages 15 and 55 in a mortality table for Deparceaux, France, were all nearly identical. Gompertz believed he had discovered a general law of mortality after observing similar patterns of geometrical progression in other tables of mortality, and published it in 1825 in the Philosophical Transactions of the Royal Society, in a paper whose title was "On the Nature of the Function Expressive of the Law of Human Mortality" [3]. In his third paper he improved his original notation and finally presented the last one in 1860, published after his death, where he noted that in his primary equation for geometric progression, the parameters were supposed to represent constant quantities for a very long term of years [4].

From 1825 to 1862 Gompertz was involved on the subject of what was called *vital statistics* in an effort to understand why there were consistent age patterns of death among people. Gompertz assumed that human beings have certain powers of integration and that those powers could be divided into a principal or fundamental part and an auxiliary part responsible for the maintenance of the principal power of integration. This auxilar force is some kind of recuperative force, a power to oppose destruction that the organisms lose in equal proportions in equal small intervals of time. Gompertz also believed on the presence of powers destroying this auxiliary force and multiplied this hypothetical force against life by the population alive to estimate the number of deaths in the age interval. Gompertz realized that if the force to destroy life operated equally on everyone, then all individuals should have the same length of life, something he knew could not be true. As a possible explanation, Gompertz emphasized the importance of chance in the timing with which death occurs. At that time the concept of genetic heterogeneity was not known, instead, Gompertz invoked chance to explain why members of a presumed homogeneous cohort die at different times [5].

After Gompertz death, the subject remained mostly unknown in the scientific community. In 1860, W. M. Makeham improved Gompertz law of mortality incorporating a term due to chance in the equation. He noted that the logarithms of the probabilities of living from Gompertz's formula increased at a faster pace at higher age than at younger ages, so he developed a theory of partial forces of mortality that intended to explain this. Makeham linked the diseases associated with the diminution of the vital power to specific organ systems -the lungs, heart, kidneys, stomach and liver, and brain. These diseases represented a significant portion of total mortality at that time and worked well in solving the observed problem of greater increased forces of mortality at older ages than at younger ages. His formula accurately portrayed the mortality experience of various human populations between ages 10 and 95 [6].

Gompertz and Makeham recognized that the original Gompertz equation did not apply to the entire age range,

the formula was intended to apply only between the ages of 20 and 60. In fact, Gompertz suggested in his last paper that there are four distinct periods in the life span between which separate laws of mortality apply: birth to 12 months, 12 months to 20 years, 20 years to 60 years, and 60 years to 100 years. Even within this range he recognized that his formula worked best "provided the intervals be not greater than certain limits." The applicability of the Gompertz function to only a specified range within the life span have been recognized by many researchers but still nowadays some researchers reject the entire Gompertz paradigm after finding that it does not apply to older ages for some organisms [5].

Scientists started searching biological explanations for Gompertz's law of mortality until the first years of the twentieth century, motivated in part by the fact that increases in mortality among nonhuman species also followed Gompertz's law for a large portion of their life span. Differences among species were assumed to be just a matter of scale.

Brownlee (1919) suggested that mortality due to senescent causes should be expressed first at about age 12, become the dominant force of total mortality by age 30, and advance at an exponential rate from ages 12 to 85. He also recognized that a law of mortality was likely to be obscured by nonsenescent mortality. Brownlee identified a formula that accurately describes the rate of decay of substances subject to the action of organic ferments (i.e., bacteria exposed to a disinfecting solution) which he believed produced a time dependent decay analogous to the loss of vital power. He found that his formula corresponded to Makeham's adjustment of Gompertz's equation, leading him to the conclusion that life depends on the energy of certain substances in the body, an energy which is gradually being destroyed throughout life [7].

Wright (1926) [8] appears to have first suggested the use of the Gompertz curve for biological growth. Following Wright, Davidson (1928) used the Gompertz curve to represent the growth in body weight of cattle [10]. Weymouth, McMillin and Rich (1931) used the Gompertz curve to represent the growth in shell size of the razor clam [9]. They stated that the curve also gives good fits for the guinea pig and the rat. It must be noted that they have found necessary in their most extensive series, the use of two different curves to graduate the first and second halves of their data. Weymouth and Thompson (1931) also applied the Gompertz curve to the growth of the Pacific cockle [11]. Since then, a number of authors fitted Gompertz formula to growth data for animals and organisms with remarkable success.

Already in 1934, Casey fitted the Gompertz model to tumour growth data and was followed by numerous authors [12]. The general conclusion has been that the Gompertz law very well describes tumor growth, but a biological explanation for this success has not been found.

The first person who attempted to perform an interspecies comparison of mortality, in this case, the mortality schedules of *Drosophila* and humans, was Raymond Pearl. Pearl (1921) plotted the survival curves of US males in 1910 on a scale with those of the longwinged male *Drosophila* [13]. Although Pearl acknowledged the arbitrary nature of this comparison, particularly in the choice of the beginning age interval for both species, he demonstrated a remarkable similarity in the curves. In his second study (1922), Pearl refined his interspecies approach and found that the form of the distributions was fundamentally the same [14]. In addition, he found that humans had a higher life expectancy at every age relative to the *Drosophila*, a discovery that he attributed to humans' control over their environment. Pearl was the first to manipulate experimentally the living conditions of his study populations to test the importance of accidental deaths on the survival curves. He was convinced that his research would reveal a "fundamental biological law" of mortality for more than one species, but after two decades of research using this scaling approach on an expanded repertoire of species, Pearl and Minor [15] emphatically declared that a universal law of mortality did not exist. Pearl and Minor identified what Makeham had identified 68 years before as the main problem -the inability to partition total mortality into its intrinsic and extrinsic causes of death.

In the 50's, researchers turned to the use of radiation, which they thought was a method to accelerate senescence, for understanding aging and making interspecies extrapolation of mortality risks. George Sacher (1950), a pioneer in this field, assumed that the effects of radiation combined additively with natural aging, without introducing new pathology [16]. Under this assumption, the Sacher model accounted for natural aging by the inclusion of a simple linear time dependent term to the integral lethality function for radiation injury. He observed that at low daily dose rates, the reciprocal difference in mean survival times for a control and for an irradiated population was proportional to the intensity of exposure. In 1952 Austin Brues and George Sacher envisioned injury as a process that disrupts the normal physiological oscillations about a mean homeostatic state within an organism, and that there were lethal injuries that an organism could not tolerate. Brues and Sacher noted that this biological model of injury and failure lead directly to the formulation Gompertz derived to describe his law of human mortality [17]. Using mean survival times, Sacher estimated cumulant lethality functions to compare empirically the similarities and differences in species' responses to radiation injury within phases of the injury process. Sacher and Trucco, however, noted that they had insufficient knowledge about the fluctuation process in real systems and that the very fact of performing an observation introduced a disturbance in the study [18].

Like Brody before him, Failla (1958) defined *vitality* as the reciprocal of the age specific mortality rate [19]. After expressing the Gompertz function in terms of vitality, he suggested that the resulting equation described the loss of vitality from a one hit random process acting on the cell population. Failla concluded that the vitality curve must describe a deterioration in the function of cells with age. He attributed the deterioration of function to somatic mutations, and interpreted the Gompertz aging parameter (derived from mortality data) as an estimate of the spontaneous somatic gene mutation rate per cell per year. With some assumptions about generation length and the number of genes in diploid cells, Failla (1960) calculations suggested that the mutation rate per generation was similar across species [20]. This would imply that the somatic mutation rate per unit time is higher in short-lived animals than in animals with longer life span.

Szilard (1959) also developed a theory on the nature of the aging process based on the concept of accumulated somatic damage [21]. Inherited mutations in somatic genes whose function is critical late in the life span was viewed as the major explanation for the different lengths of human beings' life. Like Sacher's lethal bound, Szilard envisioned death occurring when the fraction of somatic cells unaffected by mutation reached a critical threshold. He suggested that the magnitude of life shortening following exposure to radiation should be inversely related to the square root of the number of chromosomes of a species. As such, mice and humans should experience a similar radiation-induced life shortening when expressed as a fraction of the life span.

The quantitative as well as the biological importance of the Gompertz distribution was further enhanced by the work of Bernard Strehler and Albert Mildvan (1960) [22], these investigators presented a Gompertz-based theory of mortality and aging that was based on disruptions of the homeostatic state of an organism. Their approach differed from that of Sacher in the functional form of the equations used to describe the disturbances of the " energetic environment" of an organism challenged by stress. Strehler also made several important observations of the biological effects of radiation compared to the effects of aging. He noted that (1) aging effects are typically associated with post-mitotic cell whereas radiation primarily affects dividing cells; (2) radiation damage is primarily genetic whereas the effects of aging appear to be more broad spectrum; (3) some species (e.g., Drosophila) do not exhibit life shortening even after large doses of radiation; and (4) the dose required to double the mortality rate (i.e., Gompertz slope) produces a much larger increase in the mutation rate. Based on this observations, Strehler rejected the notion that radiation acts through a general acceleration of the normal aging process.

Studies of radiation effects continued to make extensive use of the Gompertz distribution throughout the 1960's. Like Greenwood (1928) before him [23], Grahn (1970) proposed to use the ratio of Gompertz slopes to adjust for life span differences when making mortality comparison between species [24]. Grahn successfully used this scaling approach to predict reductions in human life expectancy following radiation exposure from doses response relationships observed in mice.

It seems that within the field of radiation, extrapolation between species had some success, but this differs from Pearl's conclusion that a fundamental law of mortality applying to various species does not exist. The reason lies on the environmental conditions of the animals being compared, because Pearl's studies were based on the comparison with species that experienced high levels of exogenous mortality, and the laboratory animals used in radiation studies came from controlled environments without predation and where infectious diseases were minimized. These environmental conditions are far more similar to the sheltered environment and medical attention received by humans, leading to a better comparison between species [5].

The modern development of biodemography originated with a series of articles published by Weiss and colleagues [25]. Weiss (1990) recognized that the field of genetic epidemiology could provide insights into the biological constraints influencing the shape of the mortality function in populations. Weiss's merging of the fields of demography and genetics and his subsequent elaboration using principles of evolutionary biology served as a launching point for the latest developments in the field of biodemography.

For most species survival beyond the age of reproduction is an extremely rare event with most deaths for a cohort occurring just after birth. At these ages the vast majority of deaths result from forces of mortality that are unrelated with senescence (e.g., predation or diseases). In this hostile environments, early reproduction has become an essential element in species' reproductive strategies ([26]). Consistent patterns of growth and development observed within species suggest that the reproductive biology of organisms alive today represents a genetic legacy of responses to environmental conditions that prevailed during early evolutionary history of each species. The modern evolutionary theory of senescence is based on the premise that selection is effective in altering gene frequencies until the time before the end of the reproductive period. When the normally high force of external mortality is controlled and survival beyond the end of reproductive period becomes a common occurrence, senescence and senescent-related diseases and disorders have the opportunity to be expressed. Because there are common forces (i.e., extrinsic mortality) responsible for molding species'reproductive strategies, a common pattern of intrinsic mortality, an evolutionary imprint, may become visible when species are compared on a biologically comparable time scale. Carnes et al. [27] have argued that the timing of genetically determined processes such as growth and development are driven by a reproductive biology, molded by the necessity of early reproduction, which in turn is driven by the normally high external force of mortality. If individual senescence is an inadvertent consequence of these developmental processes as predicted from the evolutionary theory of senescence, then age patterns of intrinsic mortality in a population should also be calibrated to some element(s) of a species's reproductive biology. These ideas have been introduced in various computational models.

Recent mortality schedules reveal a more *pure* biological influence because the external causes of death have been dramatically reduced by medical and technological advances and almost everyone now lives to his biological potential. At the same time, a greater understanding of biological processes has also allowed the modification of intrinsic mortality (e.g. medicine, treatments and operations) altering the survival trajectories of individuals whose intrinsic diseases have already been expressed. From this perspective, the biological life span of a specie is one based on a mortality schedule that would prevail in the absence of survival time manufactured by medical or pharmaceutical intervention of any kind - a view consistent with that of Raymond Pearl. When enough members of a population benefit from these medical interventions, it is possible that the life span of the population will exceed its biologically based limits. All past research on mortality suggests that Gompertz was right all along: there are biological reasons for why death occurs when it does, and a law of mortality for many species may very well exists. Which is the limit imposed by this law of mortality for humans, and the degree to which these limits can be manipulated is still subject of great interest [5].

The Gompertz equation was developed exclusively for human beings both as an empirical tool to describe the age pattern of death from all causes during a limited time frame, and as representing a law of mortality that arises from inherent biological processes. Gompertz never imagined that his equation would become a tool used in the analysis not only of failure time of organisms but also of failure time of mechanical devices and in the description of biological and tumour growth.

## 3.2 Tumour Growth Equations

A mathematical model of tumour growth is a mathematical expression of the dependence of tumour size in time. The common feature is that growth follows a sigmoid curve with three distinct phases: the initial exponential phase, the linear phase and the plateau. The most widely used framework is consider tumour growth as a dynamical system described by ordinary differential equations, although some growth models are formulated successfully also by partial differential equations.

The simple tumour growth model is described by a single, first order, autonomous differential equation:

$$y(t) = f(y) \qquad y(0) = y_0 > 0, \tag{3.1}$$

where $y(t) > 0$ is tumour size at time $t$ and $f(y)$ is a function describing the growth rate. The solution of (3.1) has the remarkably property of a monotonic ascending function of time when $f(y_0) > 0$, or a monotonic descending function of time when $f(y_0) < 0$. In the case of an ascending function, this implies that the stationary (critical) point corresponds to the maximum possible tumor size, $y_m$, achieved for $t \to \infty$. Similarly, in the case of a descending function, the stationary point achieved for $t \to \infty$ is $y_s \geq 0$. The model given by Eq. (3.1) describes continuous tumour growth which asymptotically approaches the finite value $y_m$ or infinity (that corresponds to the unattainable unrestricted growth). On the other hand, (3.1) can describe continuous tumour regression from size $y = y_0$ to extinction ($y = 0$) at some finite time or when $t \to \infty$. However, the solution of (3.1) can not describe oscillatory tumour growth with regressions and relapses. The solution $y(t)$ represents a sigmoidal ascending curve characteristic of tumour growth if a unique point of inflection exists. This condition can be achieved for some simple functions $f(y)$. It is conceivable that functions $f(y)$ exist which yield solutions with multiple inflection points resulting in "multisigmoidal" curves. Such curves would describe tumor growth with recurrent stagnation phases [28].

More complex models of tumour growth kinetics are described by systems of ordinary autonomous first-order differential equations:

$$\begin{cases} \frac{dy}{dt} &= f(y, x_1, \ldots, x_n), \\ \frac{dx_i}{dt} &= f_i(y, x_1, \ldots, x_n), \end{cases} \tag{3.2}$$

for $i = 1, \ldots, n$ and with initial conditions $y(0) = y_0 > 0$, $x_i(0) = x_0$. Here $x_i, \ldots, x_n$ are variables describing various factors responsible for tumour growth (e.g., levels of available nutrients, growth factor activity, size of quiescent cell population, etc.). The functions $f$ and $f_i$ and the variables $x_i$ are chosen to represent growth mechanisms of particular interest. Unlike the simple model given by Eq. (3.1), the system of two differential equations ($n = 1$ in Eq. (3.2)) can describe smooth oscillatory tumor growth [28].

There is no further advance without specifying model functions $f(y)$ that represent tumor growth mechanisms. The first approach is to consider the classical chemical kinetics paradigm, based on mass conservation. For tumour growth this paradigm can be expressed in its simplest form by:

$$y(t + \triangle t) = y(t) + G(y(t))\triangle t - D(y(t))\triangle t. \tag{3.3}$$

The tumor size (mass) at time $t + \triangle t$ is equal to the size at time $t$ enlarged by $G(y(t))\triangle t$ (generation of mass) during the small time interval $\triangle t$, and diminished by $D(y(t))\triangle t$ (degradation of mass) during the same time interval. The functions $G(y) > 0$ and $D(y) > 0$ are the growth and degradation rates respectively, assumed to depend on tumor size only. Within the limit of $t \to 0$, (3.3) becomes a differential equation:

$$\frac{dy}{dt} = G(y) - D(y), \qquad y(0) = y_0 > 0. \tag{3.4}$$

Necessary conditions for the establishment of a sigmoidal (ascending) growth curve includes:

- $G(y_0) > D(y_0)$;

- Only one solution $y_m > y_0$ of $G(y) = D(y)$ exists as does only one solution $y_i > 0$ of $\frac{dG(y)}{dy} = \frac{dD(y)}{dy}$, and
- $y_i < y_m$.

In the latter case, $y_m$ is the maximal tumor size achieved asymptotically and $y_i$ is the tumor size at the inflection point. The stated conditions can be met easily if both $G(y)$ and $D(y)$ are monotonic ascending functions. In a typical kinetics paradigm, these functions are given by the power function, $ky^n$, where $k$ is the rate constant and $n$ is the order of the process.

The second approach takes a fundamental idea: tumor growth results from exponential cell proliferation (often called "Malthusian growth") described by:

$$\frac{dy}{dt} = \alpha y, \qquad \alpha > 0. \tag{3.5}$$

This equation describes unrestricted growth leading to infinite tumor size, a notion not supported by observation. Initially tumor growth behaves approximately according to (3.5), but eventually it becomes stagnant due to restrictions within the tumor itself and those imposed by the environment. Thus, exponential growth must be modified to include terms that restrict growth. This can be achieved by multiplying $y$ on the right-hand side of (3.5) with a function $F(y) > 0$ satisfying $\lim_{y \to y_m} F(y) = 0$. The corresponding differential equation is:

$$\frac{dy}{dt} = \alpha y F(y). \tag{3.6}$$

Biologically, the function $F(y)$ can be interpreted as a growth function, i.e. as the ratio of proliferating cells in tumour versus total cell population, or more generally, the ratio of growing tumour mass versus total tumour mass. The consequence of this interpretation requires that $F(y) \leq 1$ and that parameter $\alpha$ be interpreted as the growth rate constant for the hypothetical unrestricted growth.

The maximal tumor size, $y_m$, predicted by the model is often designated as *carrying capacity*, $S > 0$, of the environment for tumors in vitro or of the host for tumors in vivo. It is useful to introduce $S$ explicitly into the growth fraction model:

$$\frac{dy}{dt} = \alpha y g\left(\frac{y}{S}\right), \qquad y(0) = y_0 > 0. \tag{3.7}$$

Mathematically, both considered approaches [yielding Eq. (3.4) or Eq. (3.7)] are equivalent and one can easily transform one equation into the other. However, on the vantage point of modeling and interpretation, the two approaches are quite different. The same differential equation can yield an intuitively acceptable interpretation in one approach, while it can lack a transparent interpretation in the other. The paradigms of mass conservation and growth fraction can obviously be used in development of more elaborated models yielding systems of equations Eq. (3.2) [28].

### 3.2.1 Exponential Growth

If the number of cells in a tumour at time $t$ is denoted by $y(t)$, then, at time $t + \Delta t$, the number of cells would be expressed as $y(t + \Delta t)$. The number of cells added to the tumour in the time interval $\Delta t$ can be found subtracting $y(t + \Delta t) - y(t)$, but this number is proportional to the duration of the time interval (i.e. more cells arrive in a long interval than in a short one) so:

$$\begin{aligned} y(t + \Delta t) - y(t) &= N\Delta t, \\ \frac{y(t+\Delta t) - y(t)}{\Delta t} &= N. \end{aligned} \tag{3.8}$$

Suppose that the increase in number of the cell population is due entirely to cells being born. As time progresses the division or birth rate may be altered so that more or less divisions occur, so the number of cells born in the interval $\Delta t$ may vary with time. Moreover, if there are more cells at time $t$, more divisions are likely to occur and $N$ will also depend on $y(t)$. Letting $\Delta t \to 0$, the left-hand side of Eq. (3.8) becomes the derivative of $y$ with respect to $t$, and we have:

$$\frac{dy(t)}{dt} = N\{t, y(t)\}, \qquad (3.9)$$

where we show explicitly the quantities on which $N$ depends. The expression $(1/y)(dy/dt)$ is known as the **specific growth rate**. Therefore, another way of describing Eq. (3.9) is to say that the specific growth rate is $N(t,y)/y$.

It is plausible to assume that, in a short time interval, there will be about twice as many births as in a time interval of half its length. Thus, one could expect that the number of births would be proportional to $y(t)\Delta t$ when $\Delta t$ is small. If the birth rate does not change in the time interval, $\Delta t$ can be expressed as $\alpha y(t)\Delta t$ with $\alpha$ a suitable constant. Then Eq. (3.9) becomes:

$$\frac{dy(t)}{dt} = \alpha y(t), \qquad (3.10)$$

which states that the specific growth rate is $\alpha$, the same for all times and all sizes of tumour. This equation has the same form of the expression found in Eq. (3.5) and its solution can be realized by the following procedure:

$$\begin{aligned}
\alpha \int_0^t dt &= \int_{y(0)}^{y(t)} \frac{dy}{y}, \\
\alpha t &= \ln\{y(t)/y(0)\},
\end{aligned} \qquad (3.11)$$

leading to:

$$y(t) = y_0 e^{\alpha t}, \qquad (3.12)$$

where $y_0$ is any constant that can be fixed by putting $t = 0$ in Eq. (3.12), and evidently is the size of the tumour at $t = 0$.

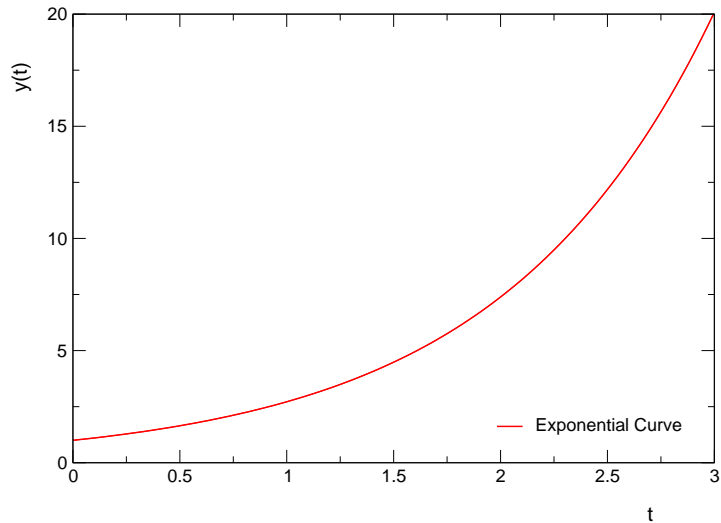

Figure 3.1: Exponential growth, with $y_0 = 1$, $\alpha = 1$. See Eq. (3.12)

The behaviour of a tumour, or a population, as time increases according to (3.12) is displayed in figure (3.1). The size grows steadily, and the increase becomes dramatic as time goes on. Of course, in any real situation, there will be a limit to the growth because of a shortage of essential supplies or insufficient food. Nevertheless, many organisms exhibit exponential growth in their initial stages [29].

Notice that Eq. (3.10) has been derived on the assumption that only births can occur. In the event that there are deaths but no births the same equation can be reached. However, $\alpha$ is now a negative number since the population or cell number decreases in the time interval $\Delta t$. It follows from (3.12) that the population decays exponentially with time from its size at $t = 0$.

More facets of the population problem can be incorporated in this equation. For instance, we may postulate that the number of deaths in the short time interval $\Delta t$ is $\beta y(t)\Delta t$. Similarly, individuals may enter the given area from outside, say $I(t)\Delta t$ immigrants in the interval $\Delta t$. Likewise, some may depart from the area giving rise to $E(t)\Delta t$ emigrants. We can model this population facets via the following equation:

$$y(t+\Delta t) - y(t) = \alpha y(t)\Delta t - \beta y(t)\Delta t + I(t)\Delta t - E(t)\Delta t, \quad (3.13)$$

leading to:

$$\frac{dy(t)}{dt} = (\alpha - \beta)y(t) + I(t) - E(t), \quad (3.14)$$

when $\Delta t \to 0$. More generally, $I$ and $E$ could be made to depend on $y$ so that Eq. (3.14) (often called **Verhulst´s differential equation**) can be difficult to solve. Notwithstanding, it is transparent that, if we hope to predict the size of a population at a given time, to find the solution of a differential equation will be an essential requirement [29].

### 3.2.2 Logistic Growth

A characteristic that must be taken into account is that the multiplication in cell numbers is restricted by crowding effects. Biochemically, these may be due to lack of nutrients, shortage of oxygen, change in pH or the production of inhibitors, for example. Whatever the cause, the cells are interacting between them. Since each cell can interact with $y$ others, there are $y^2$ possibilities in total. This suggests that, in Eq. (3.9), we should put:

$$N\{t, y(t)\} = \alpha y(t) - \beta y(t)^2, \quad (3.15)$$

where $\alpha$ and $\beta$ are positive constants. The term involving $\alpha$ is the same as before and takes into account the increase due to division. The term containing $\beta$ represents the inhibition on growth causes by crowding. With the substitution of Eq. (3.15) toward Eq. (3.9) we have:

$$\frac{dy}{dt} = \alpha y - \beta y^2, \quad (3.16)$$

which is called the **differential equation of logistics.** In the growth fraction paradigm Eq. (3.7), the equation equivalent to Eq. (3.16) is:

$$\frac{dy}{dt} = \alpha y(1 - y/S), \quad (3.17)$$

where $S = \alpha/\beta$.

If we integrate Eq. (3.16) from 0 to $t$, we obtain:

$$\begin{aligned}
\int_0^t dt &= \int_{y(0)}^{y(t)} \frac{dy}{\alpha y - \beta y^2}, \\
t &= \frac{1}{\alpha}\int_{y(0)}^{y(t)} \left(\frac{1}{y} - \frac{\beta}{\beta y - \alpha}\right)dy, \\
&= \frac{1}{\alpha}\ln\left(\frac{y}{\beta y - \alpha}\right)\Big|_{y(0)}^{y(t)}, \\
&= \frac{1}{\alpha}\ln\left(\frac{y(t)\{\beta y(0) - \alpha\}}{y(0)\{\beta y(t) - \alpha\}}\right).
\end{aligned} \quad (3.18)$$

Hence, solving for $y(t)$, we have:

$$y(t)\{\beta y(0) - \alpha\} = \{\beta y(t) - \alpha\} y(0) e^{\alpha t}$$
$$y(t) = \frac{\alpha y(0)}{\beta y(0) + \{\alpha - \beta y(0)\} e^{-\alpha t}}, \quad (3.19)$$

which is known as the **logistic law of growth**. In terms of the carrying capacity $S = \alpha/\beta$, Eq. (3.19) takes the next form:

$$y(t) = \frac{S y(0)}{y(0) + \{S - y(0)\} e^{-\alpha t}}. \quad (3.20)$$

The logistic curve is used to model a great variety of physical situations in which growth of a quantity is "self-limited", that is, the growth rate of the quantity depends on the size of the quantity in such a way that if the quantity grows beyond a certain level, the growth rate decreases. The logistic model nicely describes the behaviour of certain types of growth in business, economics, populations and sales forecasts [29].

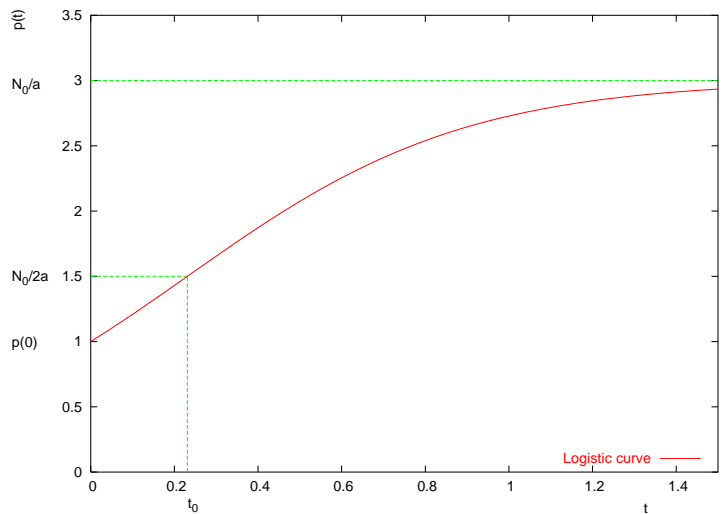

Figure 3.2: Logistic curve with $\alpha = 3$, $\beta = 1$, $y(0) = 1$.

The curve of logistic growth is shown in figure (3.2), assuming that $\alpha > \beta y(0)$. The curve rises steadily from the value $y(0)$ at $t = 0$ to an eventual value of $\alpha/\beta$, there being neither maxima nor minima in the curve. There is, however, a point of inflexion where the curve crosses its tangent at $t = t_0$ where:

$$t_0 = \frac{1}{\alpha} \ln\left(\frac{\alpha}{\beta y(0)} - 1\right), \quad (3.21)$$

and $y(t_0) = \alpha/2\beta$.

Notice that the final value $\alpha/\beta$ of $y$ does not involve $y(0)$, so that, no matter what the initial size of the population, its final size is always the same and does not depend on the starting size of the population.

In 1838, Verhulst proposed this model as a description of population growth. The model had been virtually forgotten until Pearl "rediscovered" it years later. Since then it has been often used as a point of departure for more advance population models. In 1945, Rashevsky, one of the founders of modern mathematical biology, arrived to the logistic model by considering tumor growth. Interestingly, the logistic model was used for fitting to tumor growth data much less frequently than the Gompertz model. On the other hand, the logistic model has been used in kinetics models describing immune response to tumor, where it has served as a mathematically simple description of immunologically unaffected tumor growth. Similarly, the logistic model has been used in models for chemotherapy optimization [28].

It is important to remember that the logistic law assumes that all cells divide at the same rate, and this is not always true. There are types in which some cells divide faster than others. Whether the logistic law can be applied still depends upon the differences between the various rates of division present. If the rates are not too far apart it is probably feasible to take $\alpha$ as their average. For greater deviations may be necessary to adopt a model in which the statistics of the number of cells of a given age and type at a given time play a part [29].

The immediate generalization of the logistic model Eq. (3.16) is:

$$\frac{dy}{dt} = \alpha y - \beta y^\nu, \qquad \nu > 1, \tag{3.22}$$

with solution

$$y = \left[k - \left(k - y_0^{1-\nu}\right) e^{\alpha(1-\nu)t}\right]^{\frac{1}{(1-\nu)}}, \qquad k = \frac{\alpha}{\beta}. \tag{3.23}$$

Function (3.23) is often designated as the Richard function. The solution of Eq. (3.22) has been thoroughly discussed by Fletcher. Interestingly, when this model was fitted to tumor growth data with $y_0$, $\alpha$, $\beta$, and $\nu$ as free parameters, in most cases it was found that $\nu \approx 1$. Clearly, $\nu$ cannot be exactly 1, because then (3.22) would describe unrestricted exponential growth. However, if (3.22) is reparametrized somewhat peculiarly as:

$$\frac{dy}{dt} = \left(a + \frac{b}{\nu - 1}\right) y - \frac{b}{\nu - 1} y^\nu = ay - by \frac{y^{\nu - 1} - 1}{\nu - 1}, \tag{3.24}$$

then in the limit $\nu \to 1$ one obtains the Gompertz model $dy/dt = ay - by \ln y$, using the general result: $\lim_{x \to 0} \frac{c^x - 1}{x} = \ln c$.. The result that fitting to data yielded $\nu \approx 1$ can be interpreted as a clear indication that the Gompertz model is a much more adequate description of tumor growth kinetics than is the logistic model [28].

### 3.2.3 Von Bertalanffy Growth

The combination of the chemical kinetics paradigm and the principle of allometry led von Bertalanffy to formulate the model of organismic growth represented by the equation:

$$\frac{dy}{dt} = \alpha y^\mu - \beta y^\nu, \qquad \mu > 0, \qquad \nu > 0. \tag{3.25}$$

It was shown that for any $\mu$ and $\nu$ the solution of Eq. (3.25) can not be expressed in terms of elementary functions, but in terms of the modified beta-function,

$$\beta(x, r, s) = \int_{1/2}^{x} (1 - u)^{r-1} u^{s-1} \, du \tag{3.26}$$

and its inverse.

The model characterized by $\mu = 2/3$ and $\nu = 1$ is based on the so called "surface rule", which is often named von Bertalanffy model. The underlying notion is that the anabolic growth rate is proportional to the surface area (expressed as $y^{2/3}$ where $y$ is interpreted as volume), and the catabolic growth rate is proportional to the volume itself. Another especial case of Eq. (3.25) is the generalized logistic model with $\mu = 1$, presented by Eq. (3.22), and its counterpart with $\nu = 1$:

$$\frac{dy}{dt} = \alpha y^\mu - \beta y, \qquad \mu < 1. \tag{3.27}$$

The solution of this equation is of the same form as Eq. (3.23), because Eq. (3.27) can be formally transformed into Eq. (3.22) by parameter redefinition, as clearly presented by Fletcher. Obviously, model Eq. (3.27) contains the von Bertalanffy "surface rule" model.

Returning to the general model Eq. (3.25), we wish to point out its not obvious relationship to the Gompertz model. Similarly to the generalized logistic model, Eq. (3.25) can be reparametrized into:

$$\frac{dy}{dt} = ay^\mu - \frac{1}{\varepsilon}by^\mu(y^\varepsilon - 1). \tag{3.28}$$

In the limit $\varepsilon \to 0$, one then obtains the so called "generalized Gompertz model":

$$\frac{dy}{dt} = ay^\mu - by^\mu \ln y, \tag{3.29}$$

which for $\mu = 1$ reduces to the original Gompertz model. In practice, this means that tumor growth data described by the generalized von Bertalanffy model with $\mu \approx 1$, $\nu \approx 1$, are described also by the Gompertz model [28].

### 3.2.4 Gompertz-Makeham Growth

In the paradigm of chemical kinetics (see Eq. (3.4)), the equation

$$\frac{dy}{dt} = \alpha y - \beta y \ln y, \qquad y(0) = y_0 > 0, \tag{3.30}$$

has the Gompertz growth formula as the unique solution. The growth rate $\alpha y$ reflects the Malthusian law with clear interpretation, but the degradation rate lacks any such interpretation.

In the growth fraction paradigm (Eq. (3.7)), the equation equivalent to Eq. (3.30) is obtained for $g(z) = -\ln z$, i.e.

$$\frac{dy}{dt} = -\gamma y \ln\left(\frac{y}{S}\right), \qquad y(0) = y_0 > 0. \tag{3.31}$$

Thus the growth fraction $g(z)$ is the simplest possible elementary transcendental function which obeys $g(z) \in [0,1)$ for $z \in (0,1]$ with $g(1) = 0$. Besides the simplicity argument, there is not an obvious interpretation of the growth fraction function. The solution of Eq. (3.31) and Eq. (3.30) reads:

$$\begin{aligned} y &= y_0 e^{(\alpha/\beta - \ln y_0)(1 - e^{-\beta t})} \\ y &= y_0 e^{\ln(S/y_0)(1 - e^{-\gamma t})} \\ &= S e^{-\ln(S/y_0)e^{-\gamma t}}. \end{aligned} \tag{3.32}$$

Comparison of (3.30) and (3.31) yields interesting relations among parameters:

$$\beta = \gamma, \qquad \alpha = \gamma \ln S. \tag{3.33}$$

These relations suggest that the inherent growth rate constant $\gamma$ (the rate constant for unrestricted growth, i.e., $S \to \infty$) is equal to the degradation rate constant $\beta$ and yet $\gamma$ is also proportional to the Malthusian growth rate constant $\alpha$. This indicates that the Gompertzian growth is regulated by the parameter $\gamma$ which controls both growth and degradation [28].

If we start from Eq. (3.31) and declare the growth fraction a new time dependent variable:

$$x = g\left(\frac{y}{S}\right) = \ln\left(\frac{S}{y}\right). \tag{3.34}$$

The solution (3.32) satisfies also the system of equations:

$$\begin{aligned} \frac{dy}{dt} &= \gamma x y, \\ \frac{dx}{dt} &= -\gamma x, \end{aligned} \tag{3.35}$$

with initial conditions $y(0) = y_0$ and $x(0) = \ln(S/y_0)$. From here, it is clear that the parameter $\gamma$ is at the same time the inherent growth rate constant and the rate constant for the temporal decrease of the growth fraction.

This certainly is a peculiarity of the Gompertz model which supports the idea that the single parameter $\alpha$ controls an inhibitory feedback mechanism operating in tumors. Beyond this and beyond the transparent structure of Eq. (3.35), that has a simple interpretation, other fundamental insights are not apparent. Another possibility to present the Gompertz model as a system of two differential equations is based on the introduction of the effective growth rate $x_1' = \gamma x$ as a variable:

$$\begin{cases} \frac{dy}{dt} = x_1 y, \\ \frac{dx_1}{dt} = \gamma x_1. \end{cases} \tag{3.36}$$

This system of equations is interpreted as describing exponential growth with exponential retardation. However, this can be inferred directly from Eq. (3.30).

### 3.2.5 Mathematical Properties and Comparison Between Logistic and Gompertz Growth

It is convenient to write equation Eq. (3.32) as:

$$y = c e^{-e^{a-bx}}, \tag{3.37}$$

in which $c$ and $b$ are essentially positive quantities. From Eq. (3.37) it is clear that as $x$ becomes negatively infinity $y$ will approach zero, and as $x$ becomes positively infinity $y$ will approach $c$. Differentiating Eq. (3.37) we have:

$$\frac{dy}{dx} = cbe^{a-bx}e^{-e^{a-bx}} = bye^{a-bx}, \tag{3.38}$$

and it is apparent that the slope is always positive for finite values of $x$, and approaches zero for infinite values of $x$. Differentiating again:

$$\frac{d^2y}{dx^2} = b^2 y e^{a-bx}(e^{a-bx} - 1), \tag{3.39}$$

and we obtain the point of inflection in:

$$x = \frac{a}{b}; \qquad y = \frac{c}{e}, \tag{3.40}$$

or approximately, when 37 % of the final growth has been reached. Therefore, when we desire to fit growth data which show a point of inflection in the early part of the growth cycle, we may use the Gompertz curve with the expectation that the approximation to the data will be good. Notice Figure 1, which shows the form of the curve for the case $c = 1$, $a = 0$, $b = 1$; there are also shown the logistic and the first derivative of the Gompertz curve [30].

The logistic possesses the same number of constants as the Gompertz curve, but has the point of inflection mid-way between the asymptotes. It is described by the following equation:

$$y = \frac{c}{1 + e^{a-bx}}. \tag{3.41}$$

It has been found useful to add a constant term to the logistic, giving it a lower asymptote different from zero:

$$y = d + \frac{c}{1 + e^{a-bx}}. \tag{3.42}$$

This procedure is equally applicable to the Gompertz curve giving:

$$y = d + c e^{-e^{a-bx}}. \tag{3.43}$$

The Gompertz curve and the logistic possess similar properties which make them useful for the empirical representation of growth phenomena. Each curve has three arbitrary constants, which corresponds essentially to the upper asymptote, the time origin, and the time unit or rate constant. In each curve, the degree of skewness, as measured by the relation of the ordinate at the point of inflection to the distance between the asymptote, is fixed [30].

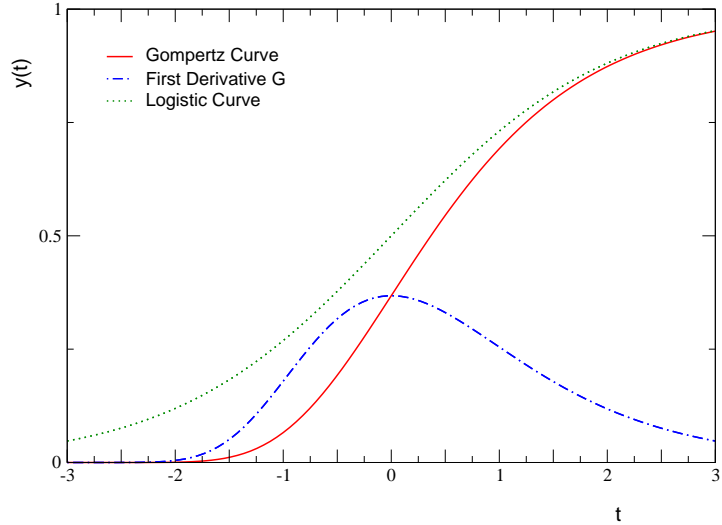

Figure 3.3: Gompertz curve and its first derivative, and the logistic curve, with $c = 1$, $a = 0$, $b = 1$.

To illustrate the mathematical properties of the Gompertz and logistic curves, the table from [30] has been reproduced on the next page.

The Gompertz equation is used as a predictive tool in demography [31]. However, the Gompertz law of exponential increase in mortality rates with ages is observed in many other biological species, such as rats, mice, fruit flies and flour beetles [32], not only on humans, and, therefore, some general theoretical explanation for this phenomenon is required. Furthermore, it often fits growth of organisms, organs and tumours. Despite numerous attempts, no consensus has been forged about the biological foundation of the broad applicability of the model [33].

| CURVE | GOMPERTZ | LOGISTIC |
| --- | --- | --- |
| Equation | $y = ce^{-e^{a-bx}}$ | $y = \frac{1}{1+e^{a-bx}}$ |
| Number of Constants | 3 | 3 |
| Asymptotes | $y = 0, y = c$ | $y = 0, y = c$ |
| Inflection | $x = \frac{a}{b}, y = \frac{c}{e}$ | $x = \frac{a}{b}, y = \frac{c}{2}$ |
| Straight line of equation | $\log\log\frac{c}{y} = a - bx$ | $\log\frac{c-y}{y} = a - bx$ |
| Symmetry | Assymetrical | Symmetrical about inflection |
| Growth rate | $\frac{dy}{dx} = bye^{a-bx} = by\log\frac{c}{y}$ | $\frac{dy}{dx} = \frac{b}{c}y(c-y)$ |
| Maximum growth rate | $\frac{bc}{e}$ | $\frac{bc}{4}$ |
| Relative growth rate as function of time | $\frac{1}{y}\frac{dy}{dx} = be^{a-bx}$ | $\frac{1}{y}\frac{dy}{dx} = \frac{b}{1+e^{-a+bx}}$ |
| Relative growth rate as function of size | $\frac{1}{y}\frac{dy}{dx} = b(\log c - \log y)$ | $\frac{1}{y}\frac{dy}{dx} = \frac{b}{c}(c-y)$ |

Table 3.1: Mathematical properties of Gompertz and logistic curves.

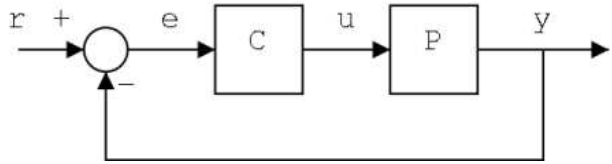

Figure 3.4: A simple feedback control loop.

# Appendix A: Control Theory Fundamentals

Control theory deals with the behaviour of dynamical systems over time. In a few words, is the mathematical study of how to manipulate the parameters affecting the behaviour of a system to produce the desired or optimal outcome. Control theory plays an important role in the design of manufacturing processes in industry, robotics, transportation, and biology, among other applications. Some of its basic concepts are the following:

**System**: set of elements that act in coordination to perform some objective.

**Plant, $P$**: is the physical element that one desires to control. Some examples are motors, ovens, navigation systems, bioreactors, etc.

**Output signal, $y(t)$**: is the variable that one wishes to control (position, velocity, pressure, temperature, etc). Is also called control variable.

**Reference Signal, $r(t)$**: is the desired value for the output signal to reach.

**Error, $e(t)$**: the difference between the reference signal and the real output signal.

**Control signal, $c(t)$**: is the signal produced by the controller $C$ in order to modify the control variable in such a way that the error decreases.

**Process**: steps that drive us to certain result.

**Perturbation**: a signal affecting the output of the system, deviating it from the desired value.

**Sensor**: device that turns the value of certain physical quantity (pressure, temperature, flow, etc.) into an electrical signal codified in analogic or digital forms.

**Closed-loop controller**: the output of the system $y(t)$ is compared to the reference value $r(t)$, through the measurement performed by a sensor. The controller then takes the difference between the reference and the output, the error $e(t)$, to change the inputs $u(t)$ to the system under control. Is known as feedback control.

**Open-loop controller**: the output signal $y(t)$ is not monitored to generate a control signal $c(t)$. There is no direct connection between the output of the system and its input $u(t)$. One of the main disadvantages of this type of controller is the lack of sensitivity to the dynamics of the system under control.

**Stability**: means that for any bounded input over any amount of time, the output will also be bounded. This is known as BIBO stability. If a system is BIBO stable then the output cannot diverge if the input remains finite.

The most simple closed-loop controller is a so-called single-input-single-output (SISO) control system, and is presented in Fig. 3.4. Examples where one or more variables can contain more than a value (MIMO, i.e. Multi-Input-Multi-Output - for example when outputs to be controlled are two or more) are frequent. In such cases variables are represented through vectors instead of simple scalar values.

If we assume the controller $C$ and the plant $P$ are linear and time-invariant (i.e.: elements of their transfer function $C(s)$ and $P(s)$ do not depend on time), we can analyze the system shown in the Fig. 3.4 by using the

Laplace transform on the variables. This gives us the following relations:

$$Y(s) = P(s)U(s) \tag{3.44}$$

$$U(s) = C(s)E(s) \tag{3.45}$$

$$E(s) = R(s) - Y(s) \tag{3.46}$$

Solving for Y(s) in terms of R(s), we obtain:

$$Y(s) = \left(\frac{P(s)C(s)}{1+P(s)C(s)}\right)R(s) \tag{3.47}$$

The term $\frac{P(s)C(s)}{1+P(s)C(s)}$ is referred to as the transfer function of the system. If we can ensure $P(s)C(s) >> 1$, i.e. it has very great norm with each value of $s$, then $Y(s)$ is approximately equal to $R(s)$. This means we control the output by simply setting the reference.

**Controllability** and **observability** are main issues in the analysis of system before decide the best control strategy to be applied. **Controllability** is related to the possibility to force the system in a particular state by using an appropriate control signal. If a state is not controllable, then no signal will ever be able to force the system to reach a level of controllability. **Observability** instead is related to the possibility to"observe", through output measurements, the system occupying a state. If a state is not observable, the controller will never be able to correct the closed-loop behaviour if such a state is not desirable.

Every control system must guarantee first the stability of the closed-loop behaviour. For linear systems, this can be obtained directly placing the poles. The behaviour of a non-linear system is not expressible as a linear function of its state or input variables, so non-linear control systems used instead specifical theories (normally based on Lyapunov Theory) to ensure stability without regard to inner dynamics of the systems. The possibility to fulfill different specifications varies from the model considered and/or the control strategy chosen.
Solutions to problems of uncontrollable or unobservable system include adding actuators and sensors.

An **observer** is an auxiliary dynamical system which uses the available measurement on the system in order to provide an estimate $\hat{x}$ of the state of the system. The dynamical nature of an observer means that the estimates of the state variable are provided on line. By an **adaptive scheme** we mean an observer that is able to provide an estimate state even in face of parameter uncertainties.

http://en.wikipedia.org/wiki/Control_theory

# Bibliography: Chapter 1

# Bibliography: Chapter 2

# Bibliography: Chapter 3

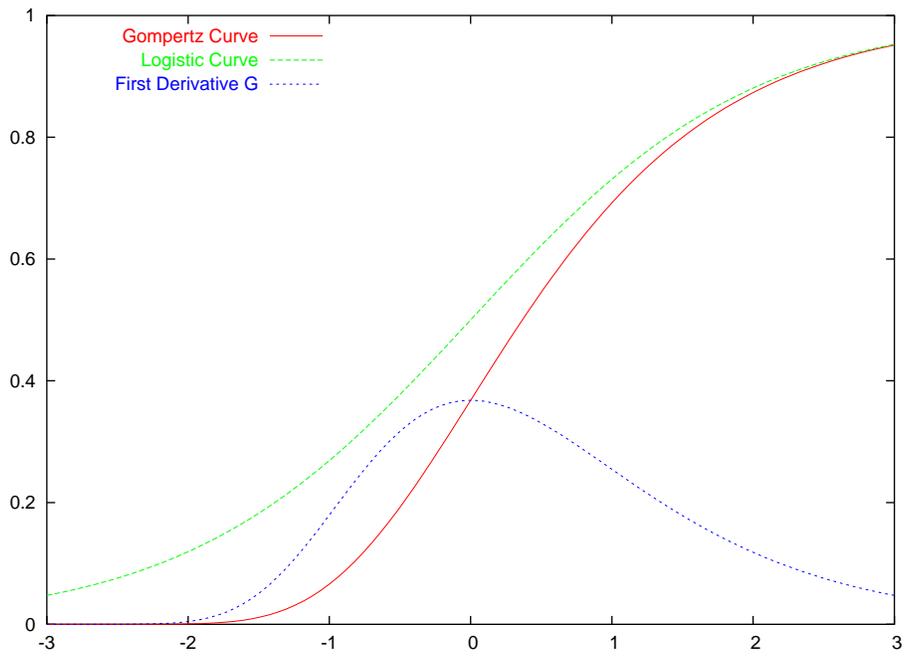

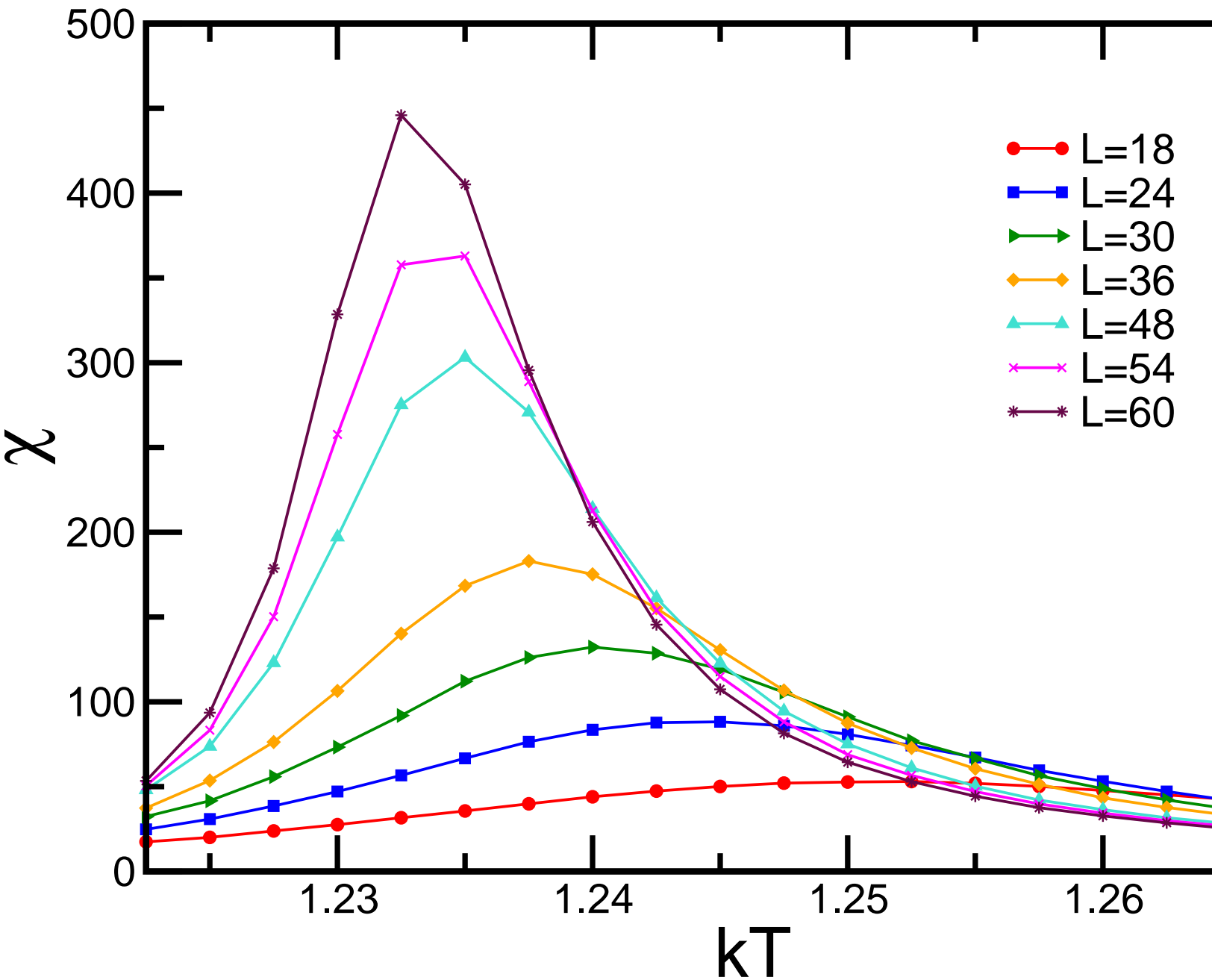

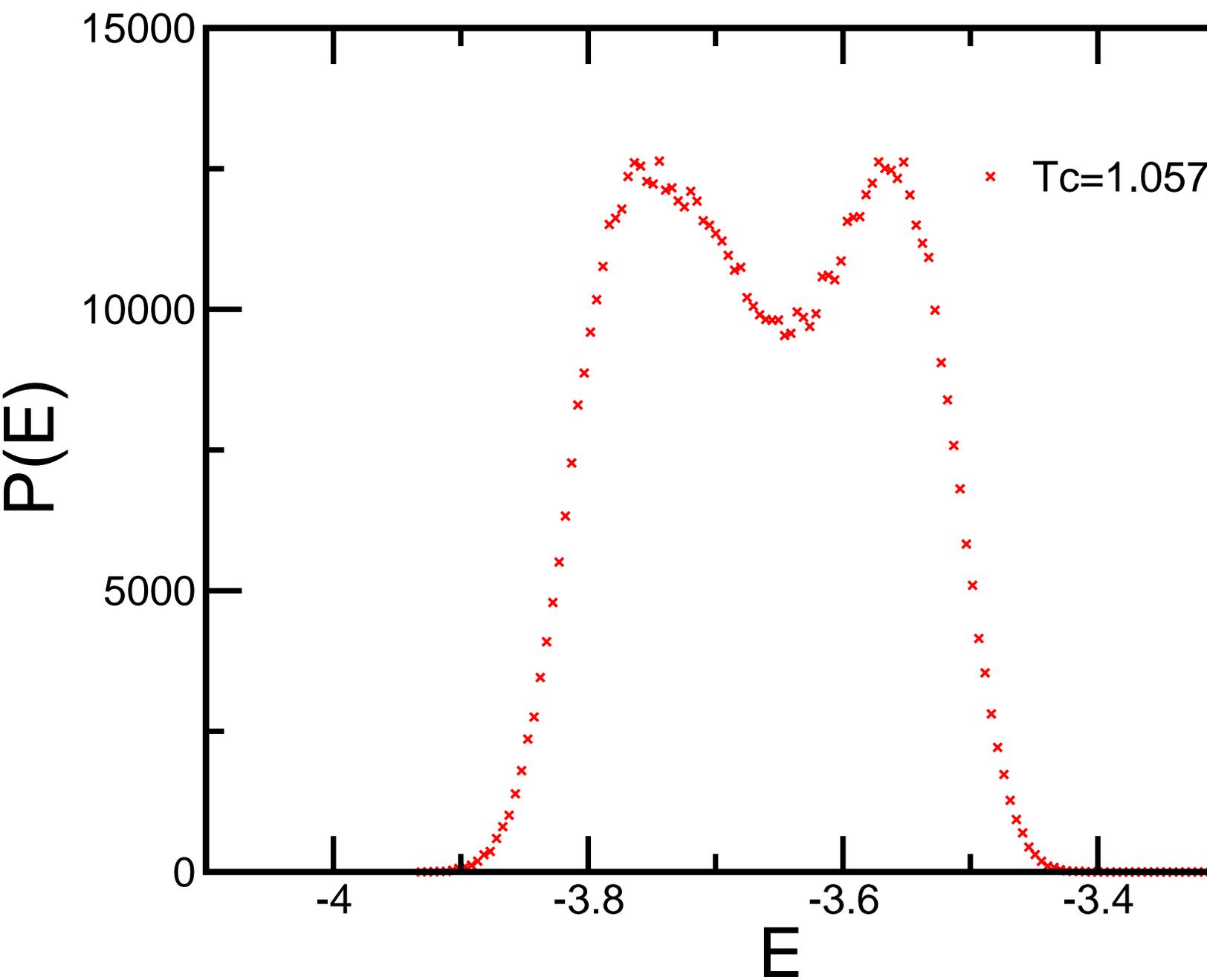